# Lignocellulosic Biomass: A Sustainable Platform for Production of Bio-Based Chemicals and Polymers

***Furkan H. Isikgor*[a], *C. Remzi Becer**[b]**

[a] Department of Chemistry, Boğaziçi University, Bebek, 34342 İstanbul, Turkey
[b] School of Engineering and Materials Science, Queen Mary University of London, Mile End Road, E1 4NS London, United Kingdom

The demand for petroleum dependent chemicals and materials has been increasing despite the dwindling of their fossil resources. As the dead-end of petroleum based industry has started to appear, today's modern society has to implement alternative energy and valuable chemical resources immediately. Owing to the importance of lignocellulosic biomass for being the most abundant and bio-renewable biomass on earth, this critical review provides insights into the potential of lignocellulosic biomass as an alternative platform to fossil resources. In this context, over 200 value-added compounds, which can be derived from lignocellulosic biomass by using various treatment methods, are presented with their references. Lignocellulosic biomass based polymers and their commercial importance are also reported mainly in the frame of these compounds. The review article aims to draw the map of lignocellulosic biomass derived chemicals and their synthetic polymers, and to reveal the scope of this map in today's modern chemical and polymer industry.

## 1. Introduction.

Modern industrial polymerization technologies make it possible to produce versatile polymers with highly tunable properties and broad range of applications. Depending on request, today's polymers can be stiff or soft, transparent or opaque, conducting or insulating, permeable or impermeable, stable or degradable. Some indispensable and irreplaceable applications of them include high strenght fibers, composites, construction materials, light weight engineering plastics, coatings, adhesives, packaging materials, microelectronics and novel materials for biomedical applications such as drug delivery systems, implants, membranes for artificial kidneys and water purification, dental fillings, wound dressing and artificial hearts. No other class of materials can have such diverse properties and versatile applicability. This means that modern life would be impossible without polymeric materials since they provide high quality of life for all humankind.[1-3]

On the other hand, industrial production of a wide range of chemicals and synthetic polymers heavily relies on fossil resources.[4] Dwindling of these resources together with their frightening environmental effects, such as global warming and littering problems, have started to threaten the future of polymer industry. In the early part of the 19th century, Henry Ford suggested that the implementation of a bio-based economy is a logical and necessary option for the growth of any civilization. This implementation was postponed because oil has always been cheaper than any other commodity products. However, the competitive price advantage of fossil fuels during the last century has disappeared.[5] After crossing the oil production peak, the dwindling of fossil resources will further boost the oil price and this situation will drastically impact the cost-effectiveness and competiveness of polymers. More importantly, mass consumption of petroleum based materials leaves devastating environmental problems, which are lethal threats for human beings. Growing concerns regarding these issues have inevitably started to enforce our society demanding for sustainable and green products. The European Union has already approved laws for the reduction of environmentally abusive materials and started to put greater efforts in finding eco-friendly materials based on natural resources. Hence, alternative solutions are sought to develop sustainable polymers from renewable natural resources for decreasing the current dependence on fossil resources and fixing the production rate of $CO_2$ to its consumption rate.[6, 7]

Biomass and biomass derived materials have been pointed out to be one of the most promising alternatives.[8, 9] These materials are generated from available atmospheric $CO_2$, water and sunlight through biological photosynthesis. Therefore, biomass has been considered to be the only sustainable source of organic carbon in earth and the perfect equivalent to petroleum for the production of fuels and fine chemicals with net zero carbon emission.[10, 11] In this context, lignocellulosic biomass, which is the most abundant and bio-renewable biomass on earth,[10] has a critical importance. Many studies have shown that lignocellulosic biomass holds enormous potential for sustainable production of chemicals and fuels. Besides, it is a renewable feedstock in abundance and is avaliable worldwide.[12, 13] Lignocellulosic biomass has been projected as an abundant carbon-neutral renewable source, which can decrease $CO_2$ emissions and atmospheric pollution. Thus, it is a promising alternative to limit crude oil, which can be utilized to produce biofuels, biomolecules and biomaterials.[14-16] Furthermore, the major component of lignocellulosic biomass; cellulose, is considered as the strongest potential candidate for the substitution of petroleum-based polymers owing to its ecofriendly properties like renewability, bio-compatibility and bio-degradability.[7]

Sustainability of the production of fuels and chemicals from biomass, on the other hand, has been greatly debated. As an example, there are cricital concerns regarding the sustainability of current production of bioethanol, which relies on starch and sugar crops. The limited supply of such crops can lead to competition with food production.[17] Lignocellulosic feedstocks have crucial advantages over other biomass supplies in this manner because they are the non-edible portion of the plant and therefore, they do not interfere with food supplies.[18] Moreover; forestry, agricultural and agro-industrial lignocellulosic wastes are accumulated every year in large quantities. Disposal of these wastes to the soil or landfill causes serious environmental problems, however; they could be utilized for the production of a number of value added products.[19] From economic point of view, lignocellulosic biomass can be produced quickly and at lower cost than other agriculturally important biofuels feedstocks such as corn starch, soybeans and sugar cane. It is also significantly cheaper than crude oil.[20]

On the other hand, the development of the conversion of lignocellulosic biomass to fine chemicals and polymers still remains a big challange.[10] Lignocellulose has evolved to resist degradation. This inherent property of lignocellulosic materials makes them resistant to enzymatic and chemical degradation.[16] For changing the physical and chemical properties of lignocellulosic matrix, the pretreatment of lignocellulosic biomass, which is an expensive procedure with respect to cost and energy, is essential.[21] Although lignocellulosic materials are abundant and usually low-priced, the crucial challenge in converting lignocellulosic biomass is to produce value-added chemicals at high selectivities and yields at economical costs.[22] Extensive research is currently being undertaken all over the world to address this problem.[23] Biorefinery and biofuel technologies are developed to refine biomass in analogy to petrochemistry for producing renewable oil and green monomers.[24] In addition, the number of biorefinery-related pilot and demonstration plants has been increasing.[25] For instance, Lignol, Verenium and Mascoma are promising companies, which aim to undertake the development of biorefining technologies for the production of advanced biofuels, biochemicals and biomaterials from non-food cellulosic biomass feedstocks.

In this review article, we report the ongoing activities in the field of lignocellulosic biomass for the production of value-added chemicals and polymers that can be utilized to replace petroleum-based products. After a description of the structure and sources of lignocellulosic biomass, different pre- and post-teratment methods for the degradation of lignocellulosic biomass into its components are summarized. Over 200 value-added compounds, which can be derived from lignocellulosic biomass by using various treatment methods, are presented with their references. Finally, detailed overviews of the polymers that can be produced mainly from these compounds are depicted. The current research studies and commercial product examples of these polymers are also explained in order to reveal their indispensable need in our modern society.

## 2. Structure and Sources of Lignocellulosic Biomass.

Lignocellulosic biomass is mainly composed of three polymers; cellulose, hemicellulose and lignin together with small amounts of other components, like acetyl groups, minerals and phenolic substituents (Figure 1). Depending on the type of lignocellulosic biomass, these polymers are organized in complex non-uniform three-dimensional structures to different degrees and varying relative composition. Lignocellulose has evolved to resist degradation and this robustness or recalcitrance of lignocellulose stems from the crystallinity of cellulose, hydrophobicity of lignin, and encapsulation of cellulose by the lignin-hemicellulose matrix.[16, 26, 27]

The major component of lignocellulosic biomass is cellulose. Unlike to glucose in other glucan polymers, the repeating unit of the cellulose chain is the disaccharide cellobiose. Its structure consists of extensive intramolecular and intermolecular hydrogen bonding networks, which tightly binds the glucose units (Figure 1). Since about half of the organic carbon in the biosphere is present in the form of cellulose, the conversion of cellulose into fuels and valuable chemicals has a paramount importance.[10, 12, 13]

Hemicellulose is the second most abundant polymer. Unlike cellulose, hemicellulose has a random and amorphous structure, which is composed of several heteropolymers including xylan, galactomannan, glucuronoxylan, arabinoxylan, glucomannan and xyloglucan (Figure 1). Hemicelluloses differ in composition too; hardwood hemicelluloses contain mostly xylans, whereas softwood hemicelluloses contain mostly glucomannans. The heteropolymers of hemicellulose are composed of different 5- and 6-carbon monosaccharide units; pentoses (xylose, arabinose), hexoses (mannose, glucose, galactose) and acetylated sugars. Hemicelluloses are imbedded in the plant cell walls to form a complex network of bonds that provide structural strength by linking cellulose fibres into microfibrils and cross-linking with lignin (Figure 1).[26, 28]

Finally, lignin is a three dimensional polymer of phenylpropanoid units. It functions as the cellular glue which provides compressive strength to the plant tissue and the individual fibres, stiffness to the cell wall and resistance against insects and pathogens.[29] The oxidative coupling of three different phenylpropane building blocks; monolignols: *p*-coumaryl alcohol, coniferyl alcohol, and sinapyl alcohol, forms the structure of lignin. The corresponding phenylpropanoid monomeric units in the lignin polymer are identified as *p*-hydroxyphenyl (H), guaiacyl (G), and syringyl (S) units, respectively (Figure 1).[30]

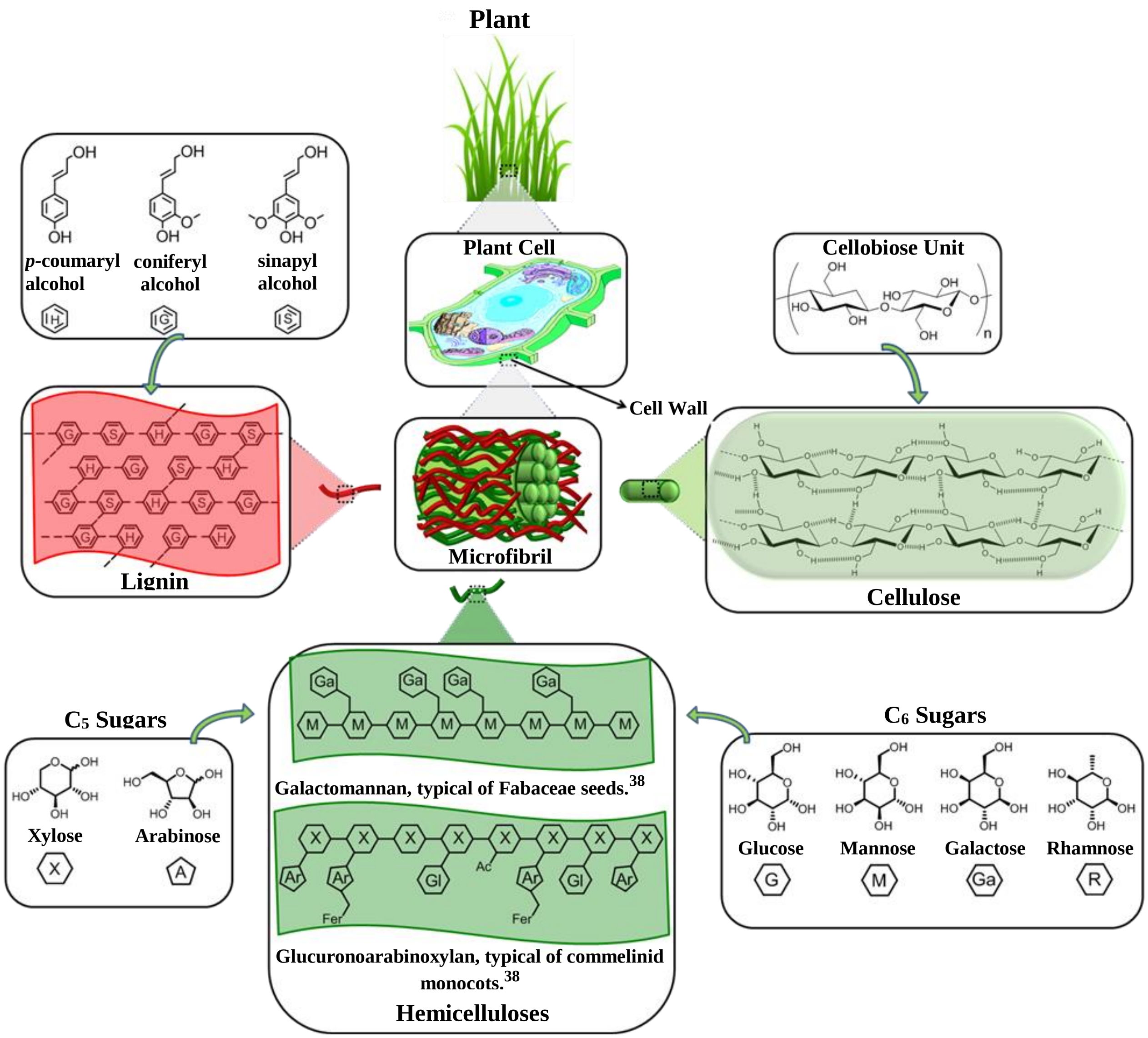

Figure 1. The main components and structure of lignocellulose. "Gl" represents glucuronic acid and "Fer" represents esterification with ferulic acid, which is characteristic of xylans in commelinid monocots.[28]

Cellulose, hemicellulose and lignin are not uniformly distributed within the cell walls. The structure and the quantity of these plant cell wall components vary according to species, tissues and maturity of the plant cell wall.[16] Generally, lignocellulosic biomass consists of 35–50% cellulose, 20–35% hemicellulose, and 10–25% lignin. Proteins, oils, and ash make up the remaining fraction.[21] Table 1 summarizes particular types of lignocellulosic biomass and their chemical composition.[17, 31]

## 3. Lignocellulosic Biomass Treatment Methods.

One of the most important goals of lignocellulosic biomass refining is to fractionate lignocellulose into its three major components; cellulose, hemicelluloses and lignin. Single step treatment methods, like pyrolysis, are not efficient. Although they render lower costs, deconstruction of the lignocellulosic biomass takes place since these methods generally rely on high temperatures. It is highly inconvenient and difficult to separate the targeted chemicals and fuels *via* single step methods because the produced bio-oil consists of a mixture of hundreds of compounds. For downstream and efficient separations, additional costs and various pretreatment methods are required. Application of the pretreatment methods changes the natural binding characteristics of lignocellulosic materials by modifying the supramolecular structure of cellulose–hemicellulose–lignin matrix. Hence, the pretreatment of lignocellulosic biomass, prior to other treatment methods, is an essential step in order to increase cellulose and hemicellulose accessibility and biodegradability for enzymatic or chemical action.[16, 22]

**Table 1: Types of lignocellulosic biomass and their chemical composition.**

| Lignocellulosic Biomass | | Cellulose (%) | Hemicellulose (%) | Lignin (%) |
|---|---|---|---|---|
| **Hardwood** | Poplar | 50.8-53.3 | 26.2-28.7 | 15.5-16.3 |
| | Oak | 40.4 | 35.9 | 24.1 |
| | Eucalyptus | 54.1 | 18.4 | 21.5 |
| **Softwood** | Pine | 42.0-50.0 | 24.0-27.0 | 20.0 |
| | Douglas fir | 44.0 | 11.0 | 27.0 |
| | Spruce | 45.5 | 22.9 | 27.9 |
| **Agricultural Waste** | Wheat Straw | 35.0-39.0 | 23.0-30.0 | 12.0-16.0 |
| | Barley Hull | 34.0 | 36.0 | 13.8-19.0 |
| | Barley Straw | 36.0-43.0 | 24.0-33.0 | 6.3-9.8 |
| | Rice Straw | 29.2-34.7 | 23.0-25.9 | 17.0-19.0 |
| | Rice Husks | 28.7-35.6 | 12.0-29.3 | 15.4-20.0 |
| | Oat Straw | 31.0-35.0 | 20.0-26.0 | 10.0-15.0 |
| | Ray Straw | 36.2-47.0 | 19.0-24.5 | 9.9-24.0 |
| | Corn Cobs | 33.7-41.2 | 31.9-36.0 | 6.1-15.9 |
| | Corn Stalks | 35.0-39.6 | 16.8-35.0 | 7.0-18.4 |
| | Sugarcane Bagasse | 25.0-45.0 | 28.0-32.0 | 15.0-25.0 |
| | Sorghum Straw | 32.0-35.0 | 24.0-27.0 | 15.0-21.0 |
| **Grasses** | Grasses | 25.0-40.0 | 25.0-50.0 | 10.0-30.0 |
| | Switchgrass | 35.0-40.0 | 25.0-30.0 | 15.0-20.0 |

Pretreatment methods are divided into different categories such as mechanical, chemical, physicochemical and biological methods or various combinations of these.[16] Various pretreatment options were reported to fractionate, solubilize, hydrolyze, and separate cellulose, hemicellulose, and lignin components.[21] Some of them include milling, irradiation, microwave, steam explosion, ammonia fiber explosion (AFEX), supercritical $CO_2$ and its explosion, $SO_2$, alkaline hydrolysis, liquid hot-water pretreatment, organosolv processes, wet oxidation, ozonolysis, dilute- and concentrated-acid hydrolyses, and biological pretreatments.[19, 21] The common goal of these methods is to reduce the biomass in size and open its physical structure. Each of these methods has been reported to have distinct advantages and disadvantages.

Through research and development, pretreatment of lignocellulosic biomass has great potential for improvement of efficiency and lowering cost of production.[32] The integration of various biomass pretreatment methods with other processes like enzymatic saccharification, detoxification, fermentation of the hydrolyzates, and recovery of products will greatly reduce the overall cost of using lignocellulose for practical purposes.[21] Hence, the future achievements of lignocellulosics conversion at commercial scale is expected to depend on the improvements in pretreatment technologies, cellulolytic enzymes producing microorganisms, fullest exploitation of biomass components and process integration.[33]

# 4. Valuable Chemicals from Lignocellulosic Biomass.

Lignocellulosic biomass has higher amount of oxygen, and lower fractions of hydrogen and carbon with respect to petroleum resources. Owing to this compositional variety, more classes of products can be obtained from lignocellulosic biorefineries than petroleum based ones. Nevertheless, a relatively larger range of processing technologies is needed for the treatment of lignocellulosic biomass. In fact, most of these technologies are still at pre-commercial stage.[25, 34] Nowadays, only production of bio-ethanol from biomass feedstocks is well established and it is turning to be a mature technology. Fermentation of glucose to lactic acid is also established on the market and it is commercially avaliable.[25, 35]

Higher oxygen content in biofuels reduces the heat content of the products and prevents their blending with existing fossil fuels. Therefore, in terms of transportation fuels and chemicals, lignocellulosic biomass needs to be depolymerized and deoxygenated. For the production of the other value-added chemicals, the presence of oxygen often provides valuable physical and chemical properties to the product. Thus, the production process requires much less deoxygenation.[25, 36] Oxygenation or deoxygenation of biomass feedstocks result in completely different products. Besides these, various sources of lignocellulosic biomass need to be considered separately since they have different compositions of cellulose, hemicelluose and lignin. Against all odds, depolymerization process of the lignocellulosic biomass is a common goal for all different feedstocks for the production of all sorts of chemicals.

## 4.1. $C_5$ and $C_6$ Sugar Production From Lignocellulosic Biomass

The first platform chemicals in the biorefinery can be sugar compounds obtained from non-food biomass.[37] Efficient release of the $C_5$ and $C_6$ sugars (Figure 1) with lower energy consumption has a critical importance because the generation of the further degradation products depends on that step.

Glucose is the sugar degradation product of cellulose. The depolymerization of hemicellulose, on the other hand, results in formation of both glucose as well as the other five (xylose, arabinose) and six (mannose, galactose, rhamnose) membered sugars (Figure 1). According to Zviely, concentrated HCl-driven hydrolysis is currently the most powerful and industrially proven technology for conversion of lignocellulosic biomass to low-cost fermentable sugars.[38] However, recovery of the acid still remains as a key limitation to any concentrated acid hydrolysis. Continued research is needed for addressing the use and separation of mineral acids, increasing the concentration of product streams, and improving product separations. In that respect, a roadmap was presented by Wettstein *et al.* to address current challenges and future prospects of lignocellulosic biomass conversion to sugars, fine chemicals and fuels.[39]

### 4.2. Lignocellulosic Sugar and Lignin Derivable Chemicals

Figures 2-5 depict the map of lignocellulosic sugars and lignin derivable chemicals under 16 platforms with their references. These chemicals can be produced *via* biological or chemical conversions. The 16 building blocks can subsequently be converted to a number of valuable chemicals or materials. The chemistry of these conversions was described in detail in the corresponding references. As such, the report of the US Department of Energy (DOE) describes twelve sugar derivable building block chemicals, which can be transformed into new families of useful molecules. These $C_5$ and $C_6$ sugar derived platform chemicals include 1,4-diacids (succinic acid, fumaric acid, malic acid), 2,5-furan dicarboxylic acid (2,5-FDCA), 3-hydroxy propionic acid (3-HPA), aspartic acid, glucaric acid, glutamic acid, itaconic acid, levulinic acid, 3-hydroxybutyrolactone (3-HBL), glycerol, sorbitol and xylitol/arabinitol.[40] Gallezot has also reviewed the synthesis of chemicals by conversion of platform molecules obtained by depolymerisation and fermentation of biopolymers.[41] Successful catalytic conversion of these building blocks into intermediates, specialties and fine chemicals was examined in detail. Unlike the DOE report, Gallezot has considered 5-hydroxymethyl furfural (5-HMF) as a separate building block because its derivates, such as 2,5-FDCA, were identified as very promising chemical intermediates. In the same review, xylose and furfural were considered as a $C_5$ platform. A variety of chemicals that can be obtained from that platform were also presented.

Since the original DOE report, considerable progress in biobased product development has been made. As an example, ethanol was omitted from DOE's original list because it was categorized as a supercommodity chemical due to its expected high production volume. On the other hand, Bozell and Petersen revisited that report and presented an updated group of candidate structures. They included ethanol as a platform chemical because bio-based ethanol and related alcohols such as bio-butanol are promising precursors to the corresponding olefins *via* dehydration.[42] In this manner, acetone can also be considered as a platform chemical because its conversion products include important monomeric chemicals. Acetone, butanol and ethanol (ABE) fermentation process is largely studied in scientific and industrial community.[43] Their conversion products such as ethylene, ethylene glycol, butadiene, propene and vinyl chloride have a great impact in polymer chemistry. Lactic acid also was not indicated as a building block in the DOE report. However, lactic acid is the most widely occurring carboxylic acid in nature. Owing to its biofucntionality, it can be converted into a variety of reaction compounds such as acetaldehyde, acrylic acid, propanoic acid, 2,3-pentanedione and dilactide.[44]

Lignin conversion also has a promising potential. The unique structure and chemical properties of lignin allow the production a wide variety of chemicals, particularly aromatic compounds. Hence, lignin can be considered as the major aromatic resource of the bio-based economy. Different approaches and strategies that have been reported for catalytic lignin conversion were comprehensively reviewed by Zakzeski *et al.*[45]
[46-200]

## 5. Sustainable Polymers from Lignocellulosic Biomass.

There are three main routes for the production of polymeric materials from lignocellulosic biomass. The least common route is to revaluate of the waste stream of the biorefinery or paper mill, which consists of lignin, cellulose, monomeric sugars and various extractives. The second route includes the cell wall polymers themselves in which cellulose, hemicellulosic polysaccharides and lignin are isolated, processed, and converted to end products.[201] It can be more convenient to directly make use of these readily avaliable bio-polymers because the process of polymerization is done perfectly by nature itself. But the challenge for lignocellulosic biomass remains in the efficient separation of celllulose, hemicellulose and lignin from each other. In this respect, lignocellulosic biomass related research studies are also focused on that particular concept. Recently, promising studies on isolation of cellulose, hemicellulose and lignin from various lignocellulosic biomass resources have been reported.[202-205] Once isolated, cellulose, hemicellulose and lignin can be converted and/or incorporated into a wide range of materials. The production strategies and specific applications of functionalized polymers of cellulose, hemicellulose and lignin were already highlighted in the recent literature.[201, 206, 207] That is why; this review article focuses on the third route and in fact the most important route which relies on the deconstruction of the cell wall polysaccharides into monomeric hexose and pentose sugars. These sugars are then converted into a wide range of value-added chemicals and bio-based polymers.

### 5.1. Sugar Containing Polymers

Lignocellulosic biomass can become as a major feedstock for sugar containing polymers by providing $C_5$ (xylose, arabinose) and $C_6$ (glucose, mannose, galactose, rhamnose) monosaccharides (Figure 1), and their many functionalized derivatives including glucaro-□-lactone, methylglucoside and glucuronic acid. These sugars and their derivatives can be either incorporated in polymer backbone or be used as pendant groups. The latter type is identified as glycopolymers; synthetic macromolecules with pendant carbohydrate moieties (Figure 6a).[208] Such polymers have gained a special interest since they

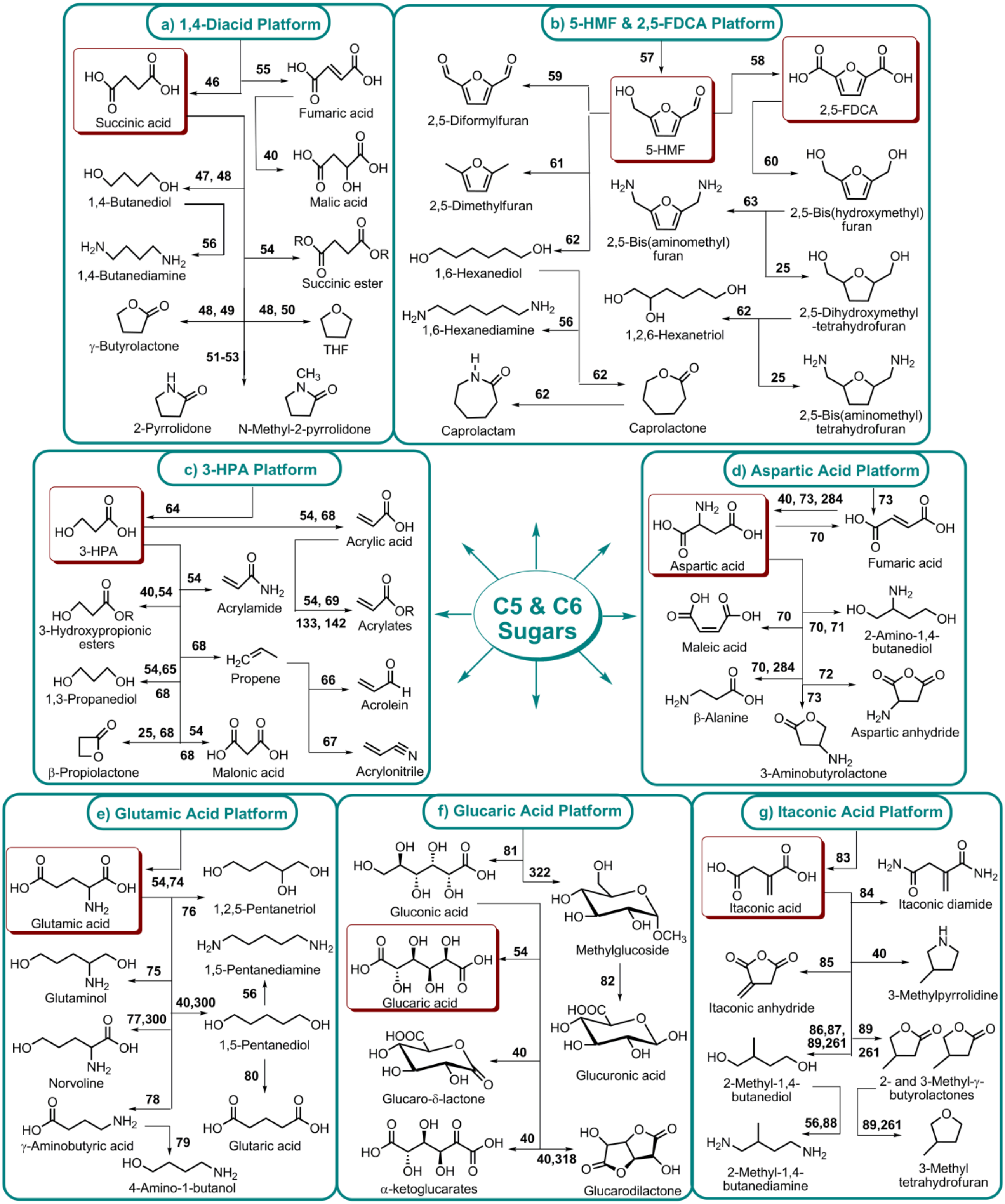

Figure 2. a) 1,4-diacid, b) 5-HMF and 2,5-FDCA, c) 3-HPA, d) aspartic acid, e) glutamic acid, f) glucaric acid, and g) itaconic acid platform chemicals.

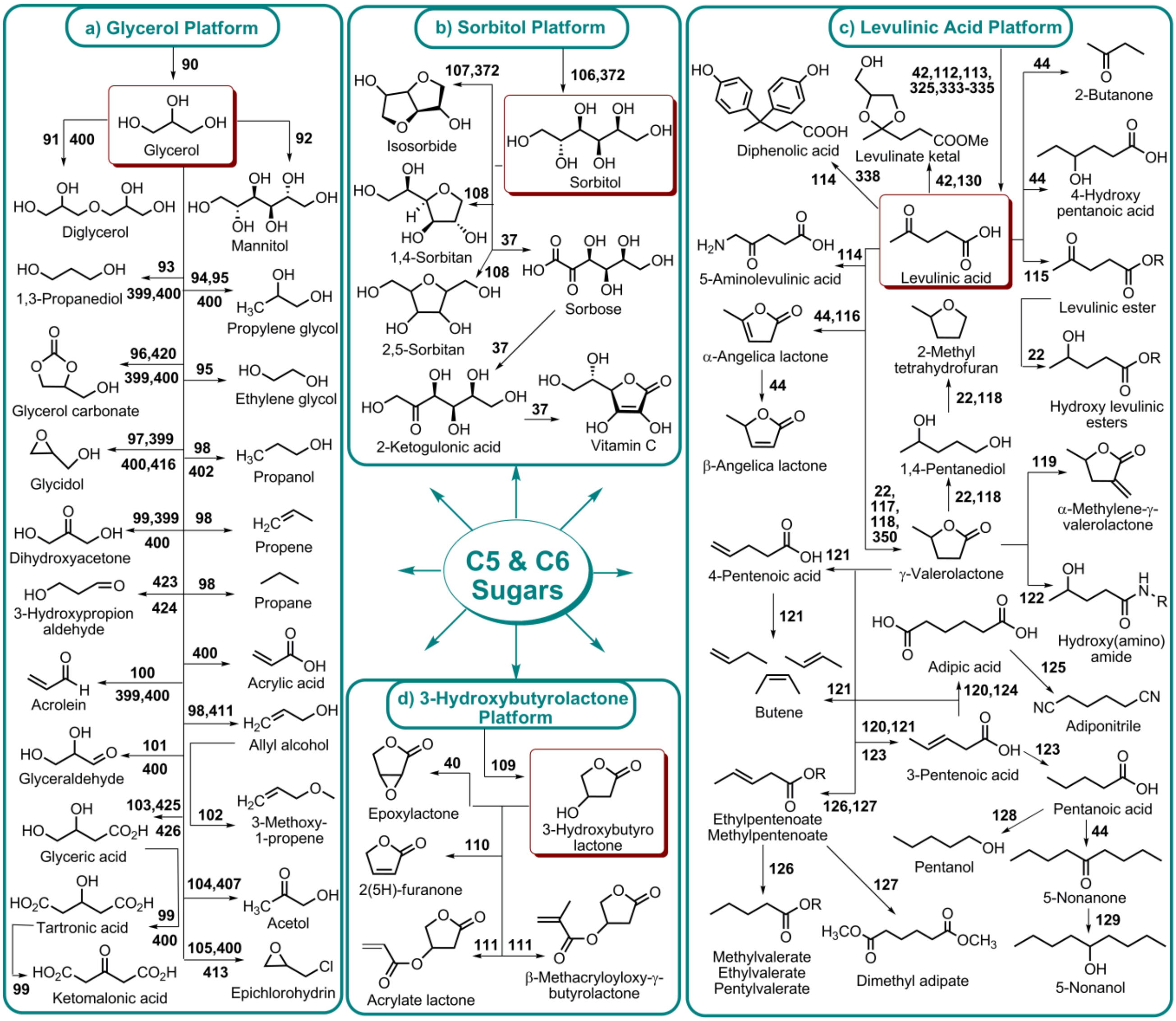

Figure 3. Glycerol, sorbitol, 3-hydroxybutyrolactone and levulinic acid platform chemicals

.

are promising in mimicking structural and functional responsibilities of glycoproteins.[209] Their multivalent character, which is displayed by the large number of repeating carbohydrate units, can represent "the cluster glycoside effect"; high affinity derived from multivalency in oligosaccharide ligands. The cluster glycoside effect greatly enhances carbonhydrate-lectin binding events.[210, 211] Therefore, different glycopolymer architectures have been developed with a specific effort to design novel drug and gene delivery systems. The current research on glycopolymeric drugs mainly focuses on millions of people affecting diseases such as influenza hemagglutinin and neuraminidase inhibitors, HIV, and Alzheimer's disease. However, the majority of these studies are still in preliminary level and extensive research is required to for clinical trials of glycopolymer-based drugs.[212, 213]

Recent advances in polymer chemistry have enabled the living polymerization of glycopolymers having various architectures.[214] In that respect, Miura reviewed the design and synthesis of glycopolymer incorporating systems such as block copolymers, micelles and particles, star polymers, bioconjugated polymers, hyperbranched polymers, and polymer brushes. For the synthesis of such systems, even when the monomer structure of the glycopolymer is bulky, various living radical polymerization (LRP) techniques, such as nitroxide-mediated polymerization (NMP), atom-transfer radical polymerization (ATRP), and reversible addition–fragment chain transfer (RAFT) polymerization, were proven to be very promising.[215-224] Ring-opening metathesis polymerization (ROMP) and polymerization with the saccharide-reactive group approaches also provide a living manner. For these living polymerization methods, several monomer structures of glycopolymers were developed (Figures 6b and 6c).[225]

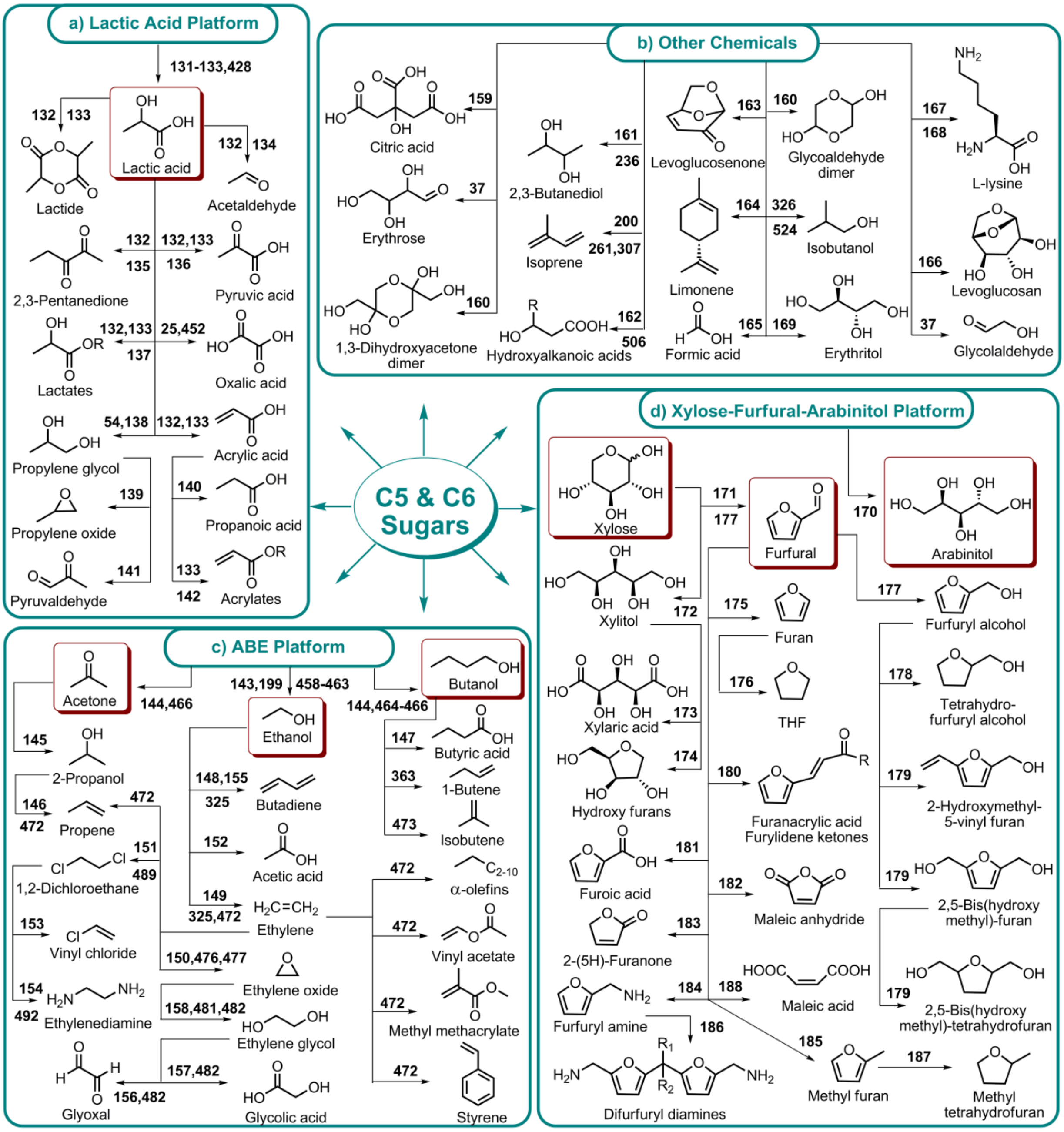

Figure 4. Lactic acid, ABE, xylose-furfural-arabinitol platforms and other lignocellulosic chemicals.

Controlled/living polymerization of glycopolymers with a combination of click chemistry has attracted attention in the last years.[23, 226-232] Controlled polymerization can give well-defined and fine-tuned glycopolymers. In addition, click chemistry can provide highly efficient and facile click of monosaccharide unit to the monomer itself or to the polymer backbone. Slavin *et al.* reviewed the combination of different polymerization techniques, i.e. RAFT polymerization, NMP, ATRP, cobalt catalyzed chain transfer polymerization (CCCTP) and ring opening polymerization (ROP), with selected click reactions including copper catalyzed azide-alkyne cycloaddition (CuAAC), p-fluoro-thiol click, thiol-ene click, thio-halogen click, and thiol-yne click.[23] Three different click chemistry and controlled polymerization combining synthetic routes have been developed so far. The first route is called glycomonomer approach. In this strategy, carbohydrate-bearing monomers can be polymerized to yield glycopolymers. The second route is based on post-polymerization modifications of clickable polymers with carbohydrates. In the third route, glycopolymers are synthesized *via* a one-pot process with simultaneous click reaction and living radical polymerization. Examples of these three synthetic routes to glycopolymers using ATRP and copper catalyzed azide-alkyne cycloaddition (CuAAC) reactions were summarized in Figure 7.[23] Apart from providing $C_5$ and $C_6$ monosaccharides, lignocellulosic biomass is also a huge feedstock for linear derivatives of these monosaccharides. Lignocellulosic biomass derivable alditols (isosorbide, erythritol, mannitol, sorbitol, xylitol, arabinitol), aldaric acids (glucaric acid, xylaric acid, □-ketoglucarates), and aldonic acids (gluconic acid, 2-ketogluconic acid) (Figures 2-4) can offer outstanding advantages for the preparation of bio-based polymers. These linear carbohydrate-based polymers typically display enhanced hydrophilicity. They are more prone to be non-toxic and biodegradable than petrochemical-based polycondensates. Therefore, they offer a wide possibility of applications in food packaging and medical devices. The –OH and –COOH rich functionality of the linear carbohydrates makes them particularly useful in polycondensation reactions. However, group protection and activation procedures are generally required before their polycondensation process.[233, 234] Munoz-Guerra reported that linear polycondensates, such as polyesters, polyamides, polycarbonates, polyureas, polyanhydrides, and polyurethanes, are obtained by the end group functionalization and/or blocking or removal of the secondary –OH groups (Figure 8). A variety of polycondensates have been derived from tetroses, pentoses and hexoses. Their structures have been examined, and some of their more relevant properties comparatively evaluated.[234] Apart from polycondensation reactions, linear carbohydrates are also employed as pendant groups. In that respect, the common strategy is to anchor a linear sugar derivative to a polymerizable double bond through ester, amide, ether and C-C linkages. As an example, Narain *et al.* described the polymerization of an C-C connected monomer, namely 4-vinylphenyl-□-gluco(d-manno)hexitol, which is derived from □-gluconolactone.[235]

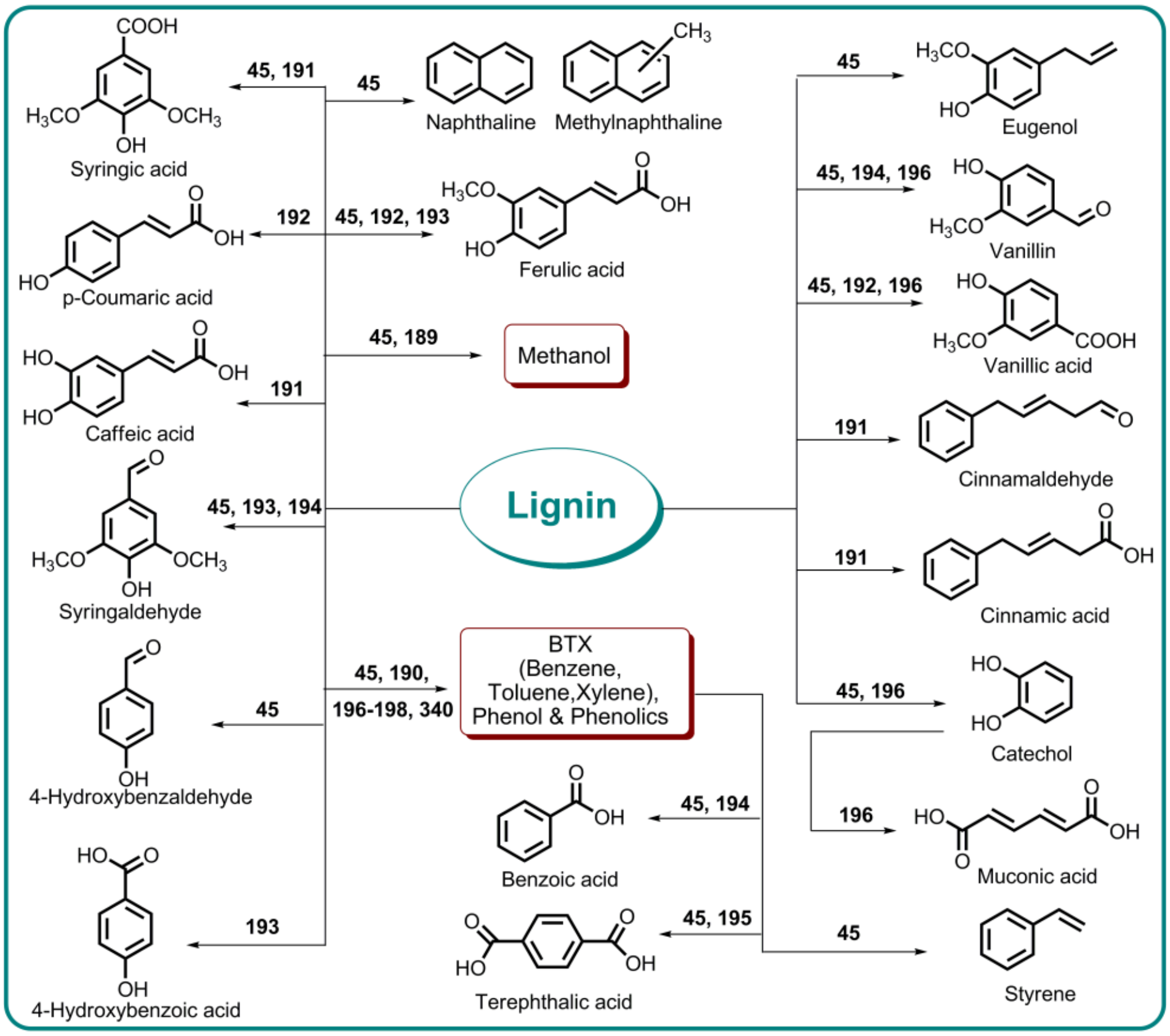

Figure 5. Lignin derived chemicals.

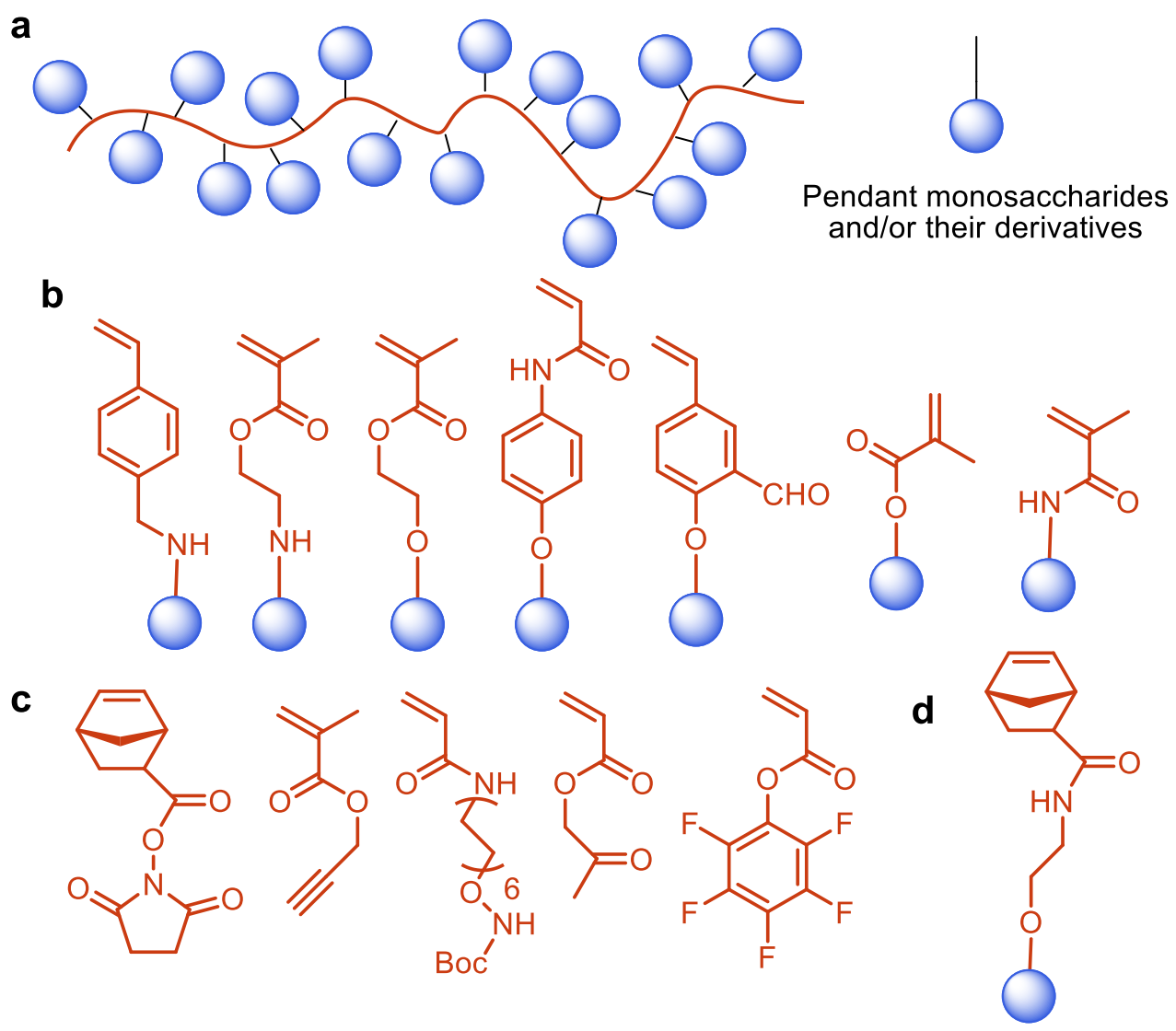

Figure 6. (a) Glycopolymer structure. Monomer structures of glycopolymers used for living polymerization: monomers for (b) LRP, (c) polymerization with the saccharide-reactive group, (d) ROMP.

For production of commodity materials, protection/deportection as well as further functionalization processes of sugar monomers are required. These processes add up to the manufacture costs and thus, decrease cost competitiveness of sugar containing polymers. Hence, polymer research in this filed is much more directed on specialty products instead of commodity materials. Particularly, controlled/living polymerization of sugar monomers with the help of click chemistry has the highest contribution to the systhesis of sugar containing specialty polymers. Nonetheless, extensive research is required since the majority of these studies are in a preliminary level for bio-applications.

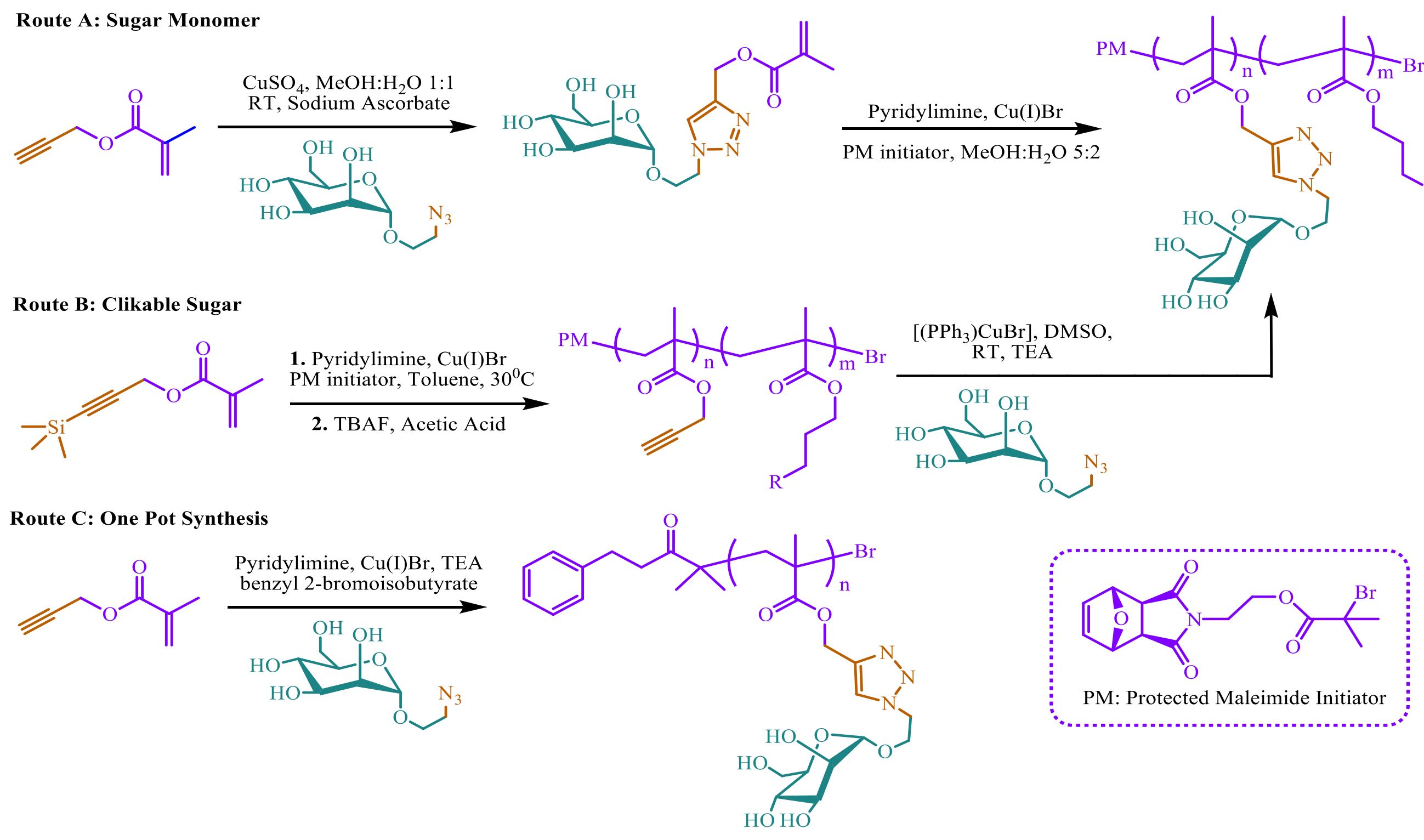

Figure 7. Various synthetic routes to glycopolymers using ATRP and CuAAC employed by Haddleton *et al.*[23]

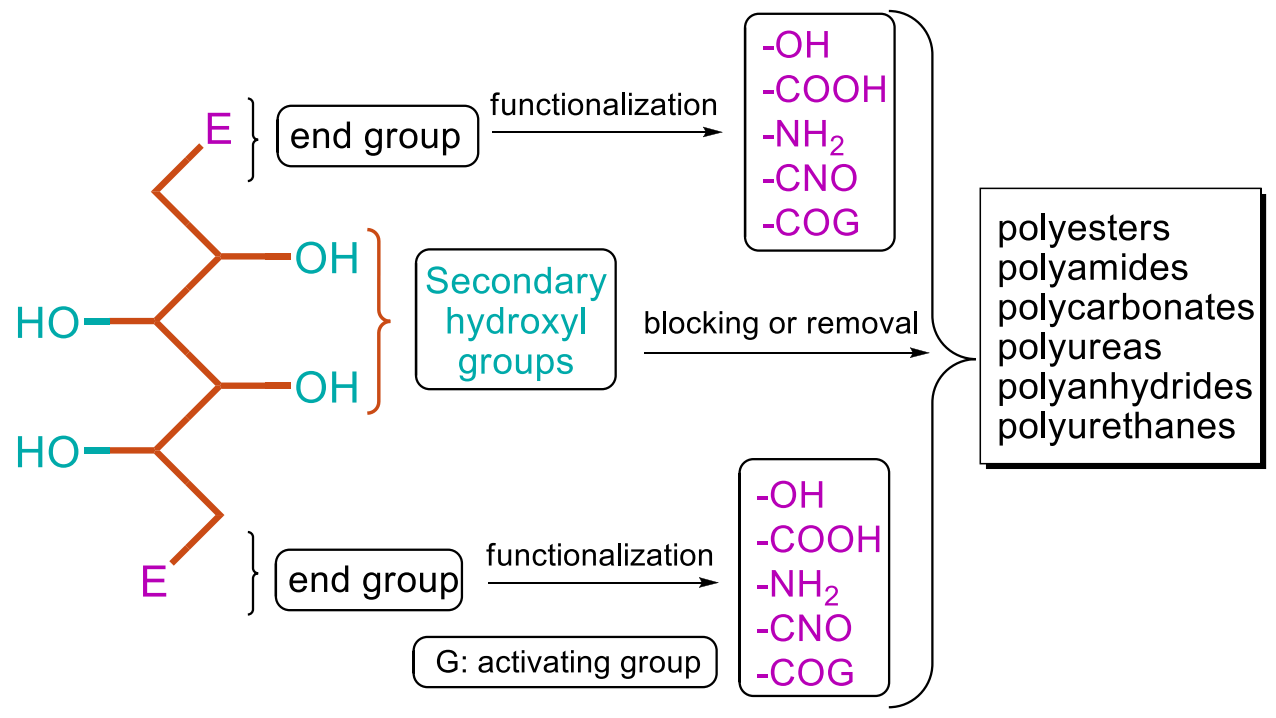

Figure 8. Linear polycondensates from carbohydrates.

### 5.2. 1,4-Diacid Platform Based Polymers

As the microbial production of succinic acid (Figure 2a) has been turning to be a mature technology,[46] several companies have started to develop renewable polymers from succinic acid and its derivatives. As an example, NatureWorks and BioAmber recently formed an alliance (AmberWorks) to investigate the production of completely renewable polyester copolymers of succinic acid and 1,4-butanediol (1,4-BDO).[236] Various polyamides (PA), polyesters (PE), and poly(ester amide)s (PEA) can be produced *via* the condensation reaction of succinic acid or succinic acid diesters with diamines or diols. In terms of polyamides, the literature generally describes polyamides which are either on the basis of 1,4-butanediamine (PA 4n) or succinic acid (PA n4).[237] Certain types of succinic acid based polyesters, particularly poly(alkylene succinate)s (PAS), can also be manufactured. Poly(ethylene succinate) (PES), poly(propylene succinate) (PPS) and poly(butylene succinate) (PBS) are the mostly studied polyesters of succinic acid. From these set of polyesters, PES and PBS are succussfully commercialized due to their relatively high melting temperatures, controllable biodegradation rate and good processability. They have polyethylene (PE) like crystallization behavior, and their elongation at break and tensile strenght are comparable with those of polypropylene (PP) and low-density polyethylene (LDPE).[238] PBS and its copolymers have been receiving a special interest in today's polymer market.[239] The commercially avaliable products of PBS are presented in Table 2, which summarizes the most important commercial products of 1,4-diacid platform, their large scale manufacturers and specific application areas.

Through hydrogenation of succinic acid; 1,4-BDO,[47, 48]□-butyrolactone (GBL)[48, 49] and THF[48, 50] are obtained (Figure 2a). Luque *et al*. proved that the selectivity of the formation of these products can be tuned by the proper choice of the catalyst.[48] The production of polybutylene terephthalate (PBT) (Table 2) is the second largest use of 1,4-BDO. Recently, there are over 50 companies engaged in the production of PBT. Among these companies, Toray has recently manufactured a partially bio-based PBT using 1,4-BDO made with Genomatica's bio-based process technology. PBT exhibits excellent thermal stability, high strenght, high chemical resistance and good durability. It is also easily made into yarn which has a similar natural stretch similar to Lycra®. These characteristics of PBT polyester resins and yarns lead to its many uses. A great deal of efforts have also been devoted to further broaden its current applications by changing the crystallinity and crystallization rate of it.[240]

**Table 2: Important commercial products of 1,4-diacid platform, their large scale manufacturers and specific application areas.**

| Polymer Structures | Product Names and Manufacturers | Major Applications |
|---|---|---|
| 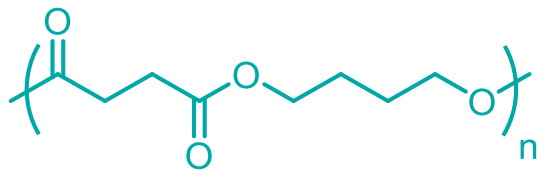 PBS and its copolymers | GS Pla® (Mitsubishi Chemical), Bionolle® (Showa), Lunare SE® (Nippon Shokubai), Skygreen® (SK Chemical) and etc. | Disposable goods, films (compost bag, shopping bag, packaging film and etc.), agriculture or horticulture (mulch film, plant pot, rope and etc.), fishing gear (fishing net, fishing trap, fishing line and etc.), containers (tray, food containers, bottles and etc.) |
| 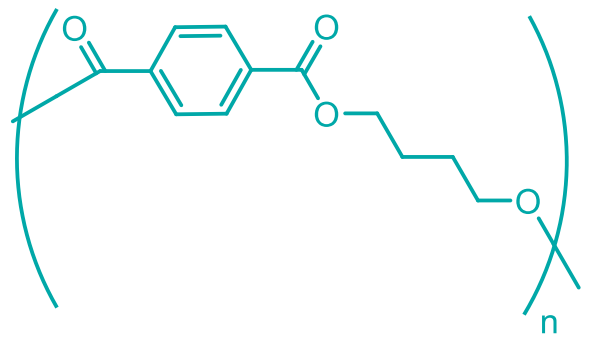 PBT resins and yarns | Arnite® (DSM), Crastin® (DuPont), Ultradur® (BASF), Advanite™ (SASA), Celanex® (Ticona), Toraycon® (Toray), Valox® (SABIC) and etc. | Automotive parts, electrical (as an insulator in the electrical and electronic industries), footwear, recreation equipment, appliances, furniture and etc. PBT yarn: sportswear, underwear, outerwear, car interior textiles and etc. |
| 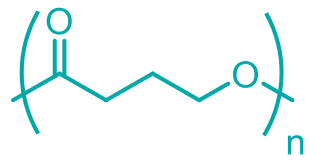 P4HB and its copolymers | PHA4400 (Tepha Inc.) | Medical applications and Devices: cardiovascular, wound healing, orthopedic, drug delivery and tissue engineering. Ex: tissue engineered heart valve, vascular grafts, stents, patches, sutures and etc. |
| 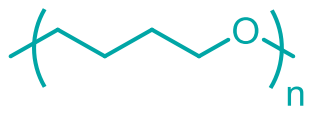 PTMEG, its copolymers and elastic fibers | Terethane (Invista), (PolyTHF) BASF | Stretchable fabrics, artificial leather, apparel and clothing, compression garments, home furnishing. |
| 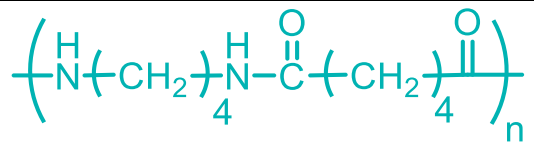 PA-4,6 | Stanyl (DSM Engineering Plastics) | Electrical & electronics, outdoor power equipments, automotive |
| 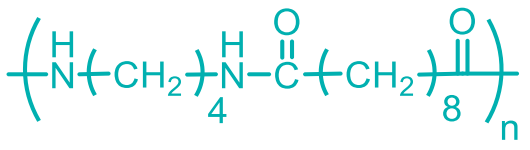 PA-4,10 | EcoPaXX (DSM Engineering Plastics) | |

In terms of ring opening polymerization (ROP), poly(4-hydroxybutyrate) (P4HB) (Table 2) is produced from GBL. However, the polymerization reaction gives low molecular weight products.[241] Nonetheless, the copolymers of GBL with L-lactide, glycolide, □-propiolactone, □-valerolactone and □-caprolactone were reported. It was stated that the use of GBL as a monomer for the synthesis of biodegradable polymer materials offers distinct advantages.[242, 243] P4HB polyester belongs to polyhydroxyalkanoates (PHAs) family (Section 5.16). So, another and more efficient production route for the production of P4HB is fermentation process, in which a much higher molecular weight, strong and flexible plastic can be derived. For instance, Tepha Inc. currently produces PHA4400, the commercial product of P4HB, for medical applications using a proprietary transgenic fermentation process.[244] Unlike GBL, the ROP of THF has attracted much more attention. The resultant linear polymer of the ring opening process is poly(tetramethylene oxide) (PTMO). The most important difunctional derivative of PTMO is PTMEG (Table 2); $\alpha,\omega$-dihydroxy-terminated polymer. PTMEG has turned to be a large-scale commercial product and the main use of this polymer is to make thermoplastic elastomers.[245] In this manner, it can be either reacted with diisocyanates to produce thermoplastic polyurethane elastomers, or with diacids/their derivatives to produce thermoplastic polyester elastomers. Thermoplastic elastomers contain two different types of blocks or chain segments to form a hard and a soft segment.[246] In particular, PTMEG is often applied as polyol because comparing to other soft segments, it renders low temperature flexibility, low content of extractable substances, microbial resistance and hydrolytic stability.[247] For instance, the main use of PTMEG is to make spandex (Elastane, Lycra), a highly elastic synthetic fiber. PTMEG based spandex is typically produced by first making a prepolymer of PTMEG and methylene diphenyl diisocyanate (MDI). Later on, the prepolymer is extended *via* a diamine extension process in which ethylenediamine is often preferred.[248] Other examples of PTMEG based commercial thermoplastic elastomers include Arnitel (DSM), Hytrel (DuPont) and Pibiflex (SOFTER). Common applications of these products were summarized in Table 2.

Through reductive amination of succinic acid by using amine, ammonium or ammonia and optionally alcohol, 2-pyrrolidone can be synthesized (Figure 2a).[51] 2-pyrrolidone can undergo an anionic ROP. The corresponding polymer, PA-4, has been known to be difficult to be synthesized and processed due to its thermal unstability. Its melting and decomposition points are very close and when melted it reverts to its thermodynamically stable monomer.[249] Although it has not been commercially produced due to its insufficient thermal stability, PA-4 membranes show interesting dialysis properties which are comparable to those of commercial cellophane and cellulose acetate membranes.[238]

Either by employing metal catalysts or biocatalytic methods, 1,4-BDO can be transformed to its diamine product; 1,4-butanediamine (putrescine) (Figure 2a). In the presence of metal catalysts, the primary alcohol firstly gives the corresponding carbonyl compound and hydrogen and then, the imine product. Subsequent hydrogenation of the imine leads to the desired amination product.[56] Putrescine has been extensively used in the production of polyamides. DSM Engineering Plastics currently uses putrescine to produce PA-4,6 and PA-4,10 (Table 2), which have been marketed as Stanyl™ and EcoPaXX™, by using adipic acid and sebacic acids as co-monomers, respectively.[236] These polyamides have led to the emergence of novel technologies. For example, EcoPaXX™ has started to be used in the engine cover of Mercedes Benz A-Class engine. Moreover, Bauser, an injection molding company located in Wehingen, Germany, has successfully introduced a production process for automotive gears using Stanyl™. PA-4I (I: isophthalic acid), PA-4T (T: terephthalic acid), and PA-4,2, are the other polymers of putresine which were studied.[250]

In addition to succinic acid, the other diacid members of this platform; fumaric and malic acids, can be obtained through fermentation which overproduces $C_4$ diacids from Krebs cycle pathways (Figure 2a).[40, 55] Homopolymers of these diacids, either alone or in combination with other polymers, have been used for general biomedical applications. Unlike succinic acid, having a double bond in its structure can allow fumaric acid to form cross-linked, degradable polymer networks with tunable material properties. Among them poly(propylene fumarate) (PPF) has been widely investigated because PPF-based polymers can be crosslinked in situ to form solid polymer networks for injectable applications. Recently, Kasper *et al.* reported a novel protocol for the synthesis of PPF by a two-step reaction of diethyl fumarate and propylene glycol for its potential utilization in the fabrication of orthopedic implants, scaffolds for tissue engineering, controlled bioactive factor delivery systems and cell transplantation vehicles.[251] By providing high water solubility, poly(malic acid) was specifically studied for the design of targeted drug delivery systems.[252]

Recently, bio-succinic acid costs higher than its petroleum based counterpart. Although this is the case, microbial production of succinic acid has been attracting a lot interest mainly due to three factors: (i) The estimated $115.2 million market value of succinic acid in 2013 is expected to reach $1.1 billion by 2020.[253] So, bio-based feedstocks are needed to compensate this demand. (ii) The succinic acid produced through traditional chemical routes using petroleum-based starting materials cannot give high purity and yield without the usage of high temperature, high pressure, and catalysts.[254] On the other hand, microbial succinic acid production provides milder and environmental friendly conditions as well as high purity and yields. (iii) As the microbial production of succinic acid becomes a mature technology, the current bio-succinic acid production costs will further decrease. It is highly expected that petroleum-based succinic acid will lose its cost advantage in the near future. Microbial production of succinic acid also opens new oppurtunities to bio-based production of 1,4-BDO, THF, putrescine, and polymers of these such as PBS, PBT, PTMEG, PA-4,6 and PA-4,10. Hence, bio-versions of these products will expand together with bio-succinic market.

### 5.3. 5-HMF & 2,5-FDCA Platform Based Polymers.

5-HMF can easily be converted to various 2,5-disubstituted furan derivatives as shown in Figure 2b. These compounds have been called the “sleeping giants” of renewable intermediate chemicals for having enormous market potential.[255] In fact, the “sleeping giants” have already woken-up, and started to claim its place in today’s market. As an example, Avantium has developed 100% bio-based polyethylene-furanoate (PEF) (Table 3) bottles, fibers and films under the brand name of YXY by using plant sugar derived 2,5-FDCA.

5-HMF has leaded its derivatives to become potential building blocks for step-growth polymers. In addition to this, the development of the vinyl polymers inexpensively from biomass-derived 5-HMF can introduce new substituents for commodity polymers. For instance, Yoshida *et al.* synthesized novel biomass-based vinyl polymers from 5-HMF. They efficiently converted HMF or its methylated derivative, 5-(methoxymethyl)furfural (5-MMF), to the vinyl monomers by the Wittig reaction in a solid-liquid phase transfer process, followed by free radical polymerization (Figure 9).[256]

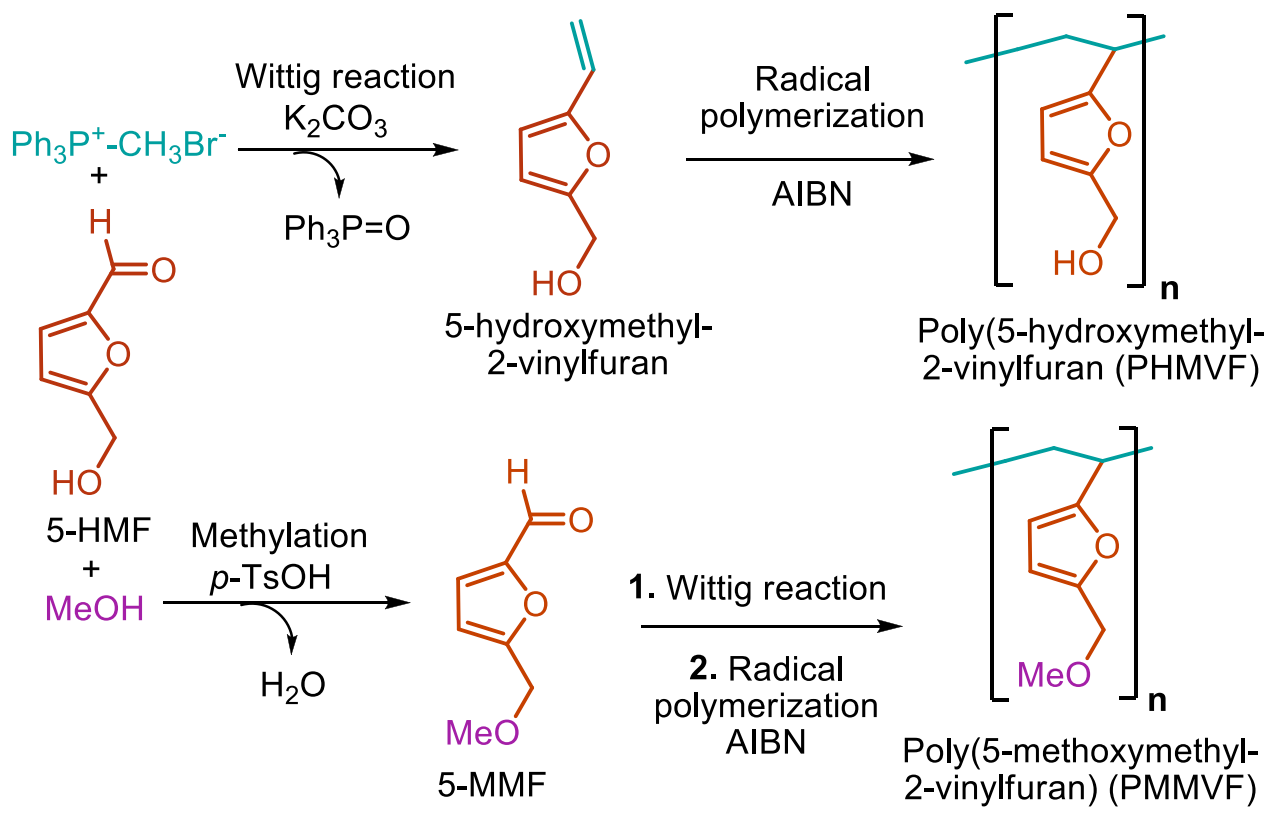

Figure 9. Synthesis of PHMVF and PMMVF

Table 3 summarizes some of the most important furanoic homo- and copolymers mainly in the frame of furan ring based monomers (Figures 2b and 4d). The presented furan-based polymers can offer distinct advantages and special features. As

**Table 3: Furan parent ring based monomers and their corresponding homo- and copolymers.** A[257-262]

| Furan Based Monomers | Furanoic Homo- and/or Copolymers | Ref. |
|---|---|---|
| 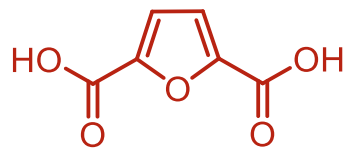 2,5-FDCA | 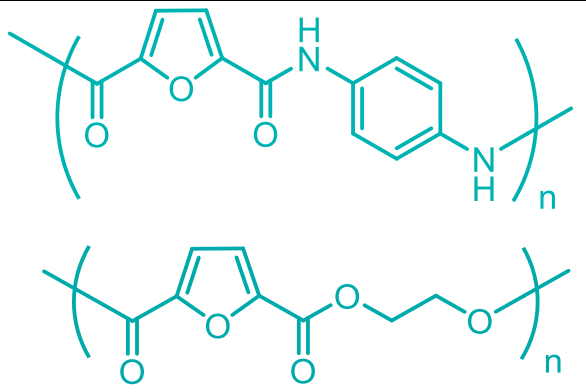 | 257 258 263 |
| 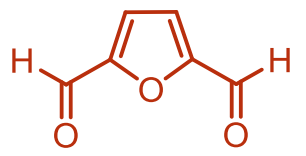 2,5-Diformyl furan | 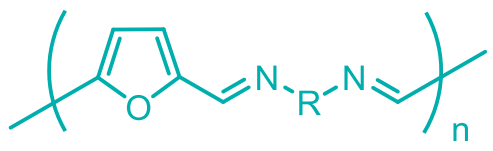 | 259 |
| 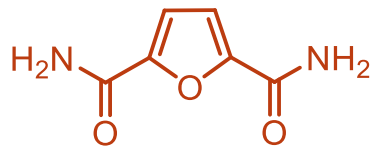 2,5-Bis(aminomethyl) furan | 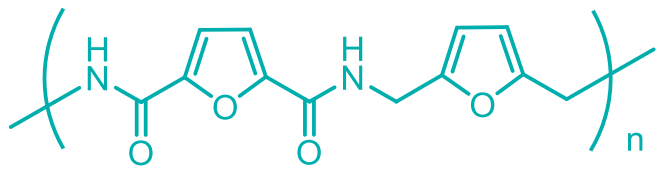 | 260 |
| 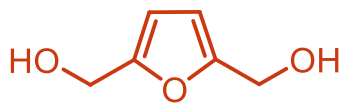 2,5-Bis(hydroxy methyl)furan | 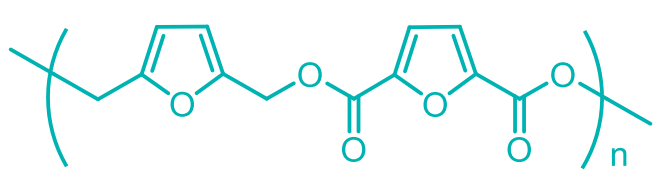 | 260 |
| 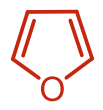 Furan | 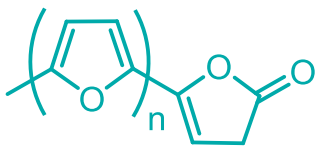 | 261 262 |
| 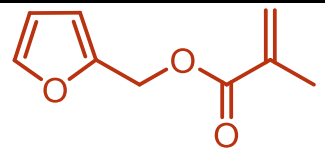 Furfuryl methacrylate | 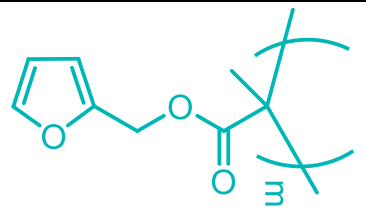 | 261 |
| 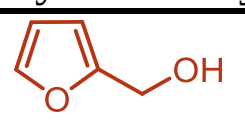 Furfuryl alcohol | 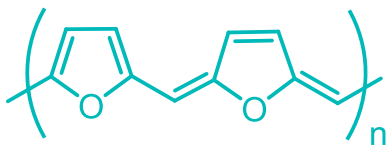 | 257 259 |
| 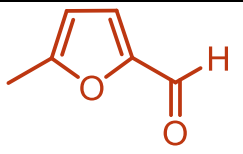 5-Methylfurfural | 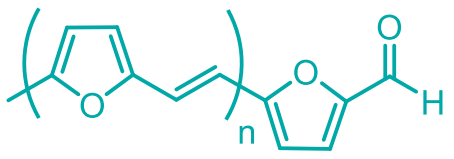 | 257 |
| 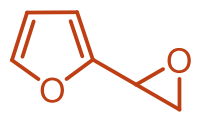 2-Furyl oxirane | 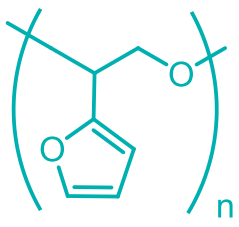 | 257 |
| 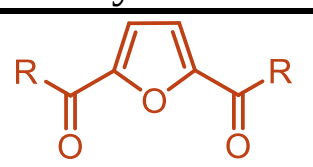 R=$OCH_2CH_2OH$ | 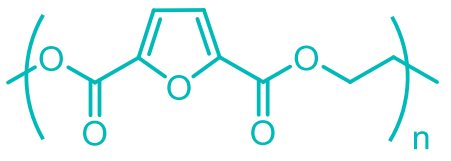 | 260 |
| 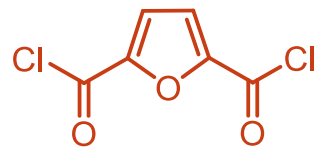 2,5-FDCA dichloride | 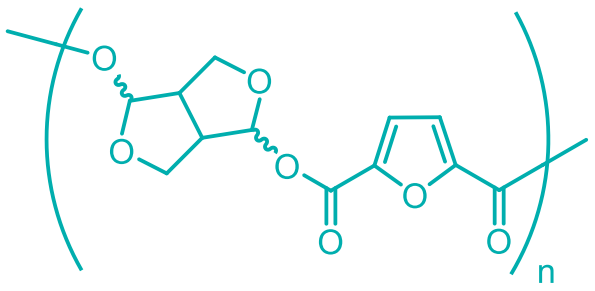 | 260 |
| 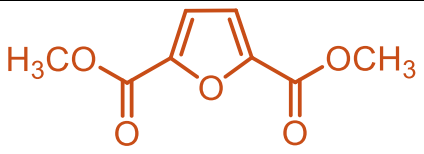 Dimethyl-2,5-furan dicarboxylate | 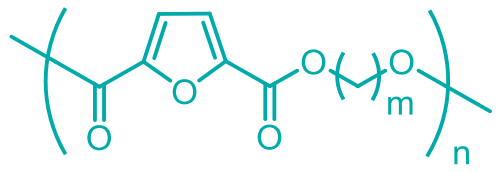 m= 3, 6, 12, 18 | 260 |

such, aromatic nature of furan ring allows the synthesis of conjugated polymers, especially for optoelectronic applications. Recent progress in the synthesis, characterization, and physical properties of 5-HMF derived furanoic polymers including poly-Schiff-bases, polyesters, polyamides, polyurethanes, polybenzoimidazoles and polyoxadiazoles reviewed in detail by Amarasekara.[260] Also, Gandini discussed a wide range of furanmonomers and their corresponding polymers.[257] Amongst these monomers, a particular attention has been recently paid to 2,5-FDCA. The oxidation of HMF to FDCA was mainly catalyzed by Pt, Pd, and Au-based catalysts. The polyamide, which is produced *via* polycondensation of 2-5-FDCA and *p*-phenylenediamine (PPD), can render high strenght by forming many inter chain hydrogen bonds as in the case of its entirely aromatic counterpart Kevlar.[257] But more importantly, 2,5-FDCA shows comparable properties with terephthalic acid (TA) in a range of polyesters. Hence, Avantium has been developing next generation 2,5-FDCA based bioplastics, called “YXY building blocks”, to replace oil-based polyesters. For instance, Avantium aims to replace oil-based PET with its furanoic counterpart; polyethylene furanoate (PEF) (Table 3). For PEF to become a successful replacement polymer, it has to compete with PET in terms of price and performance as well as it has to deliver a better environmental footprint. Regarding the physical properties, PEF has higher glass transition temperature (Tg), heat deflection temperature (HDT) and tensile strenght but it has lower melting temperature (Tm) and elongation to break than PET. PEF bottles exhibit superior barrier properties than PET; for both $H_2O$ and $CO_2$ more than two times and for $O_2$ six times better. Also, recyclability of PEF should be excellent. It was shown that both mechanical and chemical recycling of PEF is feasible. Moreover, PEF production can reduce the non-renewable energy usage and green house gas emissions by 40-50% and 45-55%, respectively, compared to PET. It is expected that the price of 2,5-FDCA will drop below € 1000 once its production capacity exceeds 300kT/y and then, it will be competitive with purified terephthalic acid (TA) produced at the same scale. All these findings suggest that PEF can compete with PET in terms of price and performance particularly in bottling applications, and it can provide a significantly better environmental footprint.[258, 263]

Recently, one of the most important motivations to study furan-derived monomers is their utilization in Diels-Alder (DA) reactions for new polymeric materials. Both the DA and retro-DA reactions between furan and maleimide, which are straightforward and not affected by side reactions, have been well investigated over the past several years (Figure 10). The DA reaction can be employed to produce both linear and branching polymers by simple modification on the number of maleimide funtionality.[261]

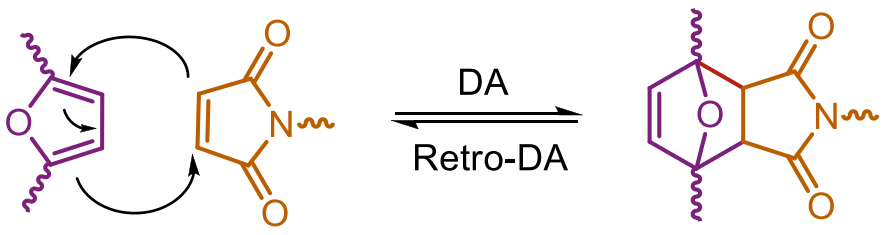

Figure 10. DA and retro-DA reactions between furan and maleimide.

Sequentially, dihydrofuran and/or tetrahydrofuran based monomers can be obtained *via* hydrogenation of furan ring based

**Table 4: THF and Furanone rings based monomers and their corresponding polymers.**

| Monomers | Homo- and/co-polymers | Ref. |
|---|---|---|
| 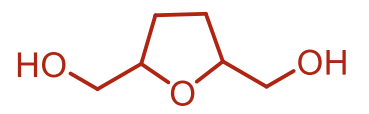 2,5-Dihydroxymethyl -tetrahydrofuran | 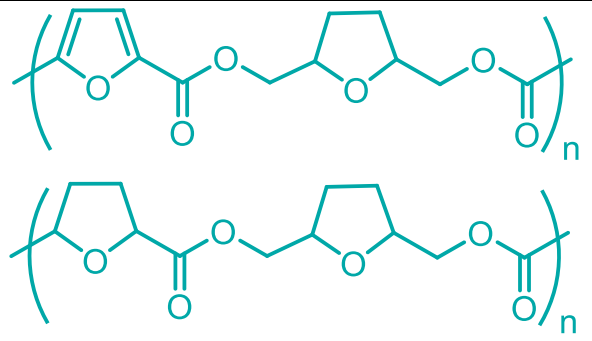 | 264 |
| 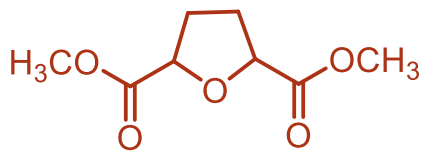 Dimethyl tetrahydrofuran-2,5-dicarboxylate | 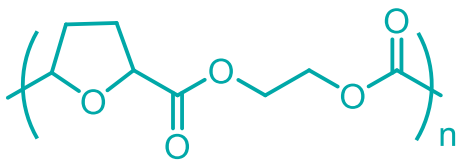 | 264 |
| 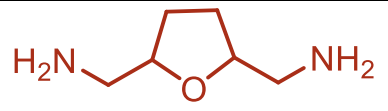 2,5-Bis(aminomethyl) tetrahydrofuran | 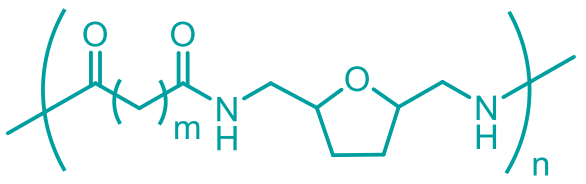 m= 4, 8, 10 | 265 |
| 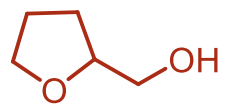 Tetrahydrofurfuryl alcohol | 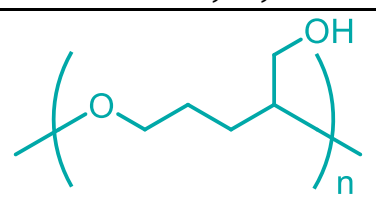 Linear saturated 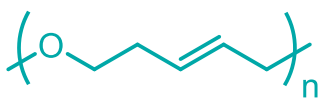 Linear unsaturated 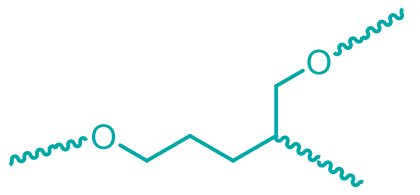 Branched | 266 267 |
| 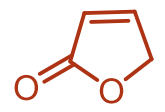 2(5H)-Furanone | 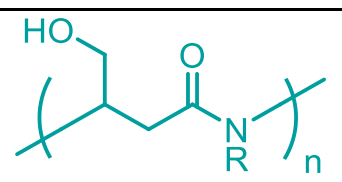 R: 1-50 carbon atoms | 268 |

compounds as shown in Figures 2b and 4d. The resulted new monomers' behaviors are quite dissimilar to that of their furan precursors because the aromatic nature of furan ring is now broken down in the process. Through step-growth polymerization method, THF ring bearing polyesters[264] and polyamides[265] are reported and some of their literature examples are summarized in Table 4. In terms of ROP, another derivative of THF, tetrahydrofurfuryl alcohol, can give cationic-ROP for the production of its homo- or hyperbranced-polymer architectures.[266, 267]

Buntara et al. proposed four different reaction pathways for conversion of 5-HMF into 1,6-hexanediol (1,6-HDO). These pathways include; (i) the direct hydrogenation of 5-HMF to 1,6-HDO. (ii) a two step process through 2,5-THF-dimethanol (THFDM) intermediate. (iii) a three-step synthesis through THFDM and 1,2,6-hexanetriol (1,2,6-HT) intermediates, respectively. (iv) a four-step synthesis via THFDM, 1,2,6-HT, and tetrahydro-2H-pyran-2- ylmethanol intermediates.[62] The first route is much more desirable since it avoids the usage of intermediate steps and chemicals. Once obtained, the double diamination of 1,6-HDO yields 1,6-hexanediamine (1,6-HDA) (Figure 11).[56] The conversion of 5-HMF to 1,6-HDO, and subsequently to 1,6-HDA is particularly important since these products have a last-born and very large polymer market. Successful commercialization of their polymers mainly depends on their fairly long hydrocarbon chain, which provides hardness and flexibility to their incorporating polymers, and high reactivity of their terminal alcohol or amine groups. Ube Industries, Ltd. currently produces 1,6-HDO based polycarbonatediols under the trend name of Eternacoll® (Figure 11), particularly for the production of polyurethanes as coatings, elastomers and adhesives.[269] For the case of 1,6-HDA, more than 95 % of it is reacted with adipic acid for the manufacture of PA-6,6.[270, 271] Recently, the world's largest producer of this polyamide is Invista and it is commercialized as Torzen® for general use in industrial, textile, and automotive applications (Figure 11). 1,6-HDA can be effectively converted to its diisocyanate derivative, hexamethylene-diisocyanate (HDI) by employing a phosgene-free synthetic approach.[272] The resultant new monomer is especially useful for the synthesis of polyurethanes (PU).

5-HMF derived 1,6-HDO can be alternatively oxidized to ε-caprolactone. In this process, 1,6-HDO is oxidized to its corresponding monoaldehyde, which cyclizes spontaneously to the lactol, and then the dehydrogenation of the lactol yields ε-caprolactone. Further amination of the ε-caprolactone gives caprolactam (Figure 11), which is the monomer for PA-6, a widely used synthetic polymer with an annual production of about 4 million tons. This 5-HMF-based bio-route may become an alternative to the traditional caprolactam synthesis process. The bio-route has only four steps; on the other hand, the traditional caprolactam process requires six steps starting from benzene and ammonia.[62] In this respect, the bio-route is more feasible but it still requires process development in terms of product yields and cost competitiveness.

Although the great majority of ε-CL has been used for the commercial production of caprolactam, it is also converted to its commercially avaliable homopolymer, polycaprolactone (PCL), *via* the ROP of it using a variety of anionic, cationic and co-ordination catalysts. Perstorp (CAPA), Dow Chemical (Tone) and Daicel (Celgreen) are the major producers of this polyester. Woodruff and Hutmacher reviewed the important applications of PCL as a biomaterial over the last two decades with a particular focus on medical devices, drug delivery and tissue engineering.[273] Another significant industrial application of ε-CL is its utilization in the production of polyglecaprone 25, the copolymer of glycolide and ε-CL. This copolymer is formed into filaments for use as sutures and has trademarked by Ethicon as Monocryl®.[274] For caprolactam, approximately 90% of it is processed to PA-6 filament and fiber (Perlon), and most of the remaining 10% is used for the manufacture of plastics.[275] This aliphatic polyamide is trademarked as Capron®, Ultramid® and Nylatron®. It is exceptionally useful whenever high strenght and stiffness are required. The applications and properties of PA-6 and PA-6,6 are described in detail by James E. Mark in "Polymer Data Handbook".[270] Also, important commercial polymers of 5-HMF & 2,5-FDCA platform are depicted in Figure 11.

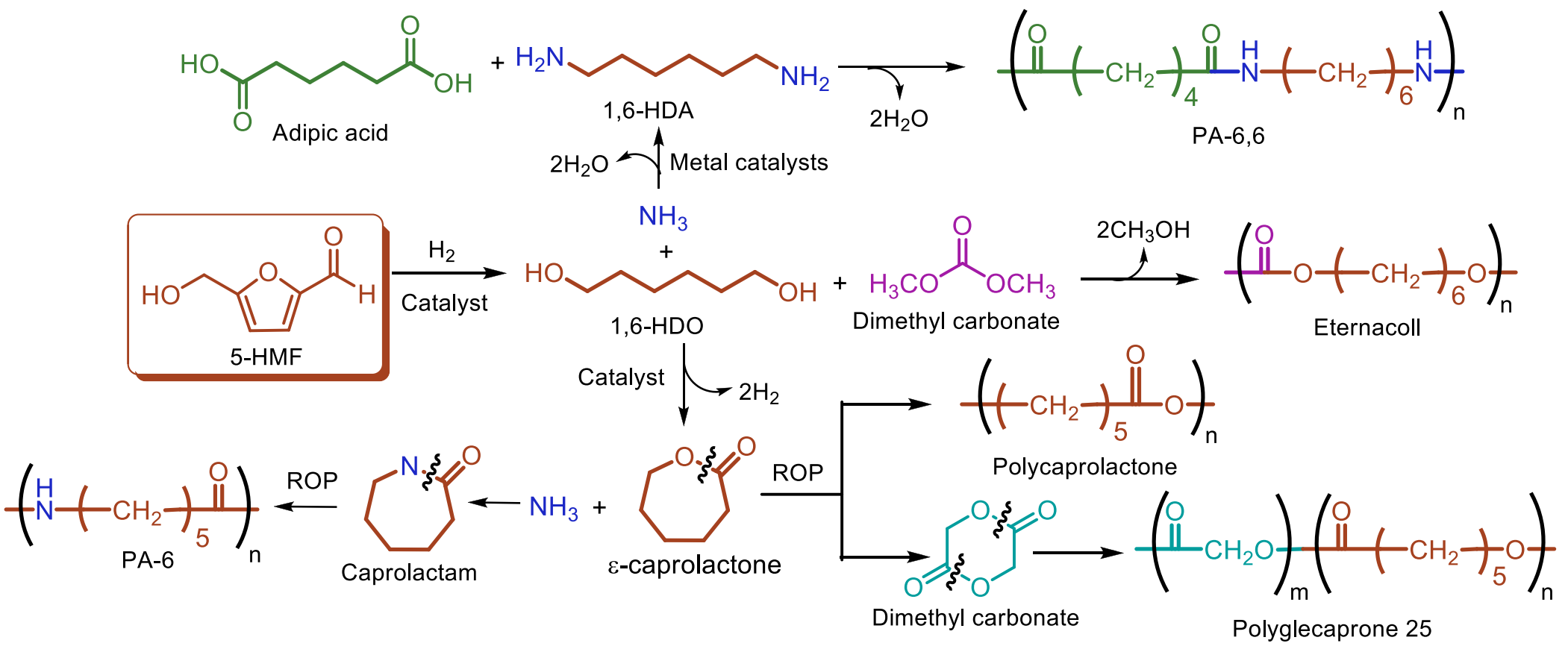

Figure 11. Commercial polymer products of 1,6-hexanediol, 1,6-hexanediamine, □-caprolactone and caprolactam.

### 5.4. 3-HPA Platform Based Polymers.

3-HPA is a key building block for production of high value products such as acrylic monomers, 1,3-propanediol (PDO), propene, malonic acid, □-propiolactone, and 3-hydroxypropionic esters (Figure 2c).[68] It can be chemically produced (i) through oxidation of PDO in the presence of Pd-containing supported catalyst (ii) through oxidation of 3-hydroxypropionaldehyde in the presence of Pd-containing supported catalyst, or (iii) through hydration of acrylic acid in the presence of acid catalysts. However, these chemical routes are too costly and thus, a biosynthetic pathway is required for large scale production of 3-HPA. In this consideration, cheap sugar feedstocks should be employed to obtain 3-HPA in a commercially efficient way. Even so, no known organisms can produce 3-HPA as a major metabolic end product from sugars. Hence, large scale production of 3-HPA can be realized through fermentation of sugars by employing genetically modified microorganisms. The process requires an improved microbial biocatalyst as well as reduction in the production cost.[64, 68]
3-HPA and its various derivatives, including 3-hydroxypropanal, □-propiolactone or its oligomers and esters of 3-HPA, can be used for the production of 1,3-propanediol (1,3-PDO) through hydrogenation process (Figure 2c).[54, 65, 68] In fact, direct production of PDO through the fermentation of sugars is more feasible and thus, the current industry on large scale production of 1,3-PDO has been shifting to this direction. As such, DuPont and Tate & Lyle has recently formed a joint venture for the production of bio-PDO from fermentation of corn sugar by using a genetically modified strain of *E. Coli*.[276] It was claimed that the bio-PDO process results in 40% less energy consumption and 20% reduction of greenhouse gas emissions. Based on the corn derived bio-PDO, DuPont has commericialized Sorona® family of poly(trimethylene terephthalate) (PTT); the polyester of 1,3-propanediol and terephthalic acid (or dimethyl terephthalate).[277]

3-HPA can be directly converted to acrylic acid by thermal dehydration at reduced pressure using homogeneous acid catalysts in the presence of copper powder. This method provides yields of acrylic acid around 80%. Higher yields are attained when heterogeneous catalysts, such as $NaH_2PO_4$ supported on silica gel, are employed.[54] As an alternative route, 3-HPA can be converted to propene through a dehydration reaction coupled with a decarboxlation process (Figure 12). But this reaction is yet to be realized. Once propene is obtained, the vapour phase oxidation of propene in the presence of oxygen to acrylic acid through acrolein intermediate is the most widely accepted commercial process. So, as shown in Figure 2c, the dehydration of 3-HPA, with or without further modification of its carboxylic unit, can open a completely new window to the world of acrylic monomers and their corresponding polymeric materials.[54, 66-69, 133, 142] The realization of technology in the end will depend either depend on either (i) the cost competitive production of 3-HPA and its efficient dehydration to acrylic monomers, or (ii) the cost competitive production of 3-HPA derived bio-propene with respect to the petroleum derived propene.
Production of acrylic monomers starting from 3-HPA is quite important owing to the giant market size of acrylic polymers. Almost all commercially produced acrylic polymers contain acrylic acid and/or methacrylic acid at some extent. Hence, homo- or copolymers of acrylic (Figure 12) and methacrylic acids have a myraid of applications. Many companies have involved in the production of these polymers. For polyacrylic acid, some of them include The Dow Chemical Company (Acrysol™, Acumer™, Acusol™, Dualite™), AkzoNobel (Alcogum®, Alcosperse®, Aquatreat®), Lubrizol (Carbopol®) and BASF (Sokalan®). Swift summarized common applications of polyacrylic acid and polymethacrylic acid under their specific types including emulsion, water-soluable, alkali-soluble, gel, block and graft polymers.[270, 278] In terms of acrylates, methyl acrylate and ethyl acrylate were the first produced derivatives of acrylic acid. Physical properties of acrylic ester polymers (Figure 12) greatly depend on their R ester groups. For instance, the glass-transition range can vary widely among the acrylic ester polymers from -65°C for 2-ethylhexyl acrylate ($R = C_4H_9$) to 103°C for acrylic acid (R = H). Since the R side-chain group conveys such a wide range of properties, acrylic ester polymers are used in various applications varying from paints to adhesives, concrete modifiers and thickeners. More detailed information regarding many different acrylic esters and their correcponding

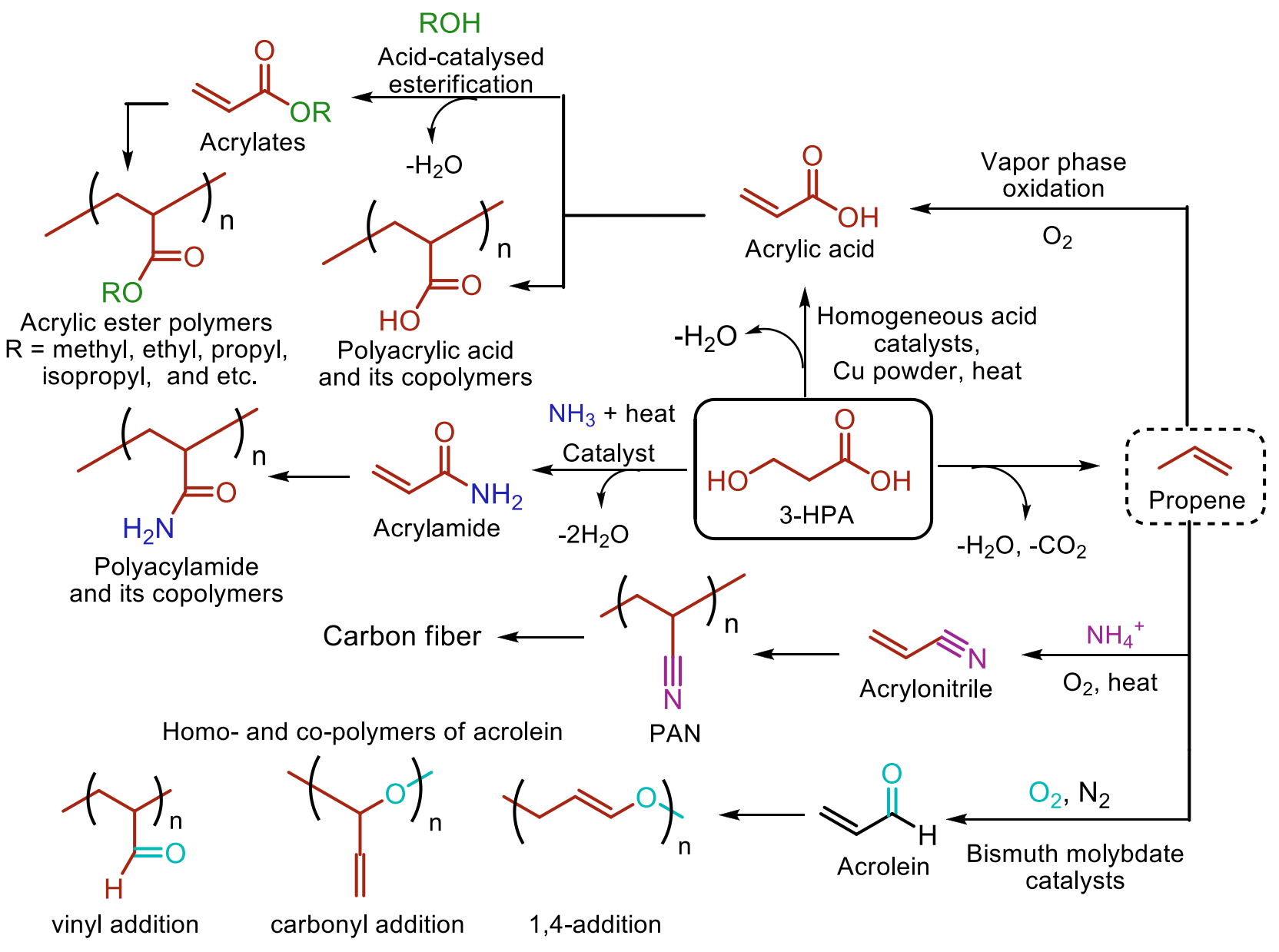

Figure 12. Industrially important 3-HPA derivable acrylic polymers.

polymers can be found in Slone's article.[279]

Acrylamides are also produced from 3-HPA by simply heating mixtures of 3-HPA and an amine without or with a catalyst which enhances the rate of the dehydration reaction (Figure 12).[54] Although acrylamide is toxic, its homopolymer; polyacrylamide, is harmless and it has a wide range of applicability. Huang *et al.* have extensively reviewed a very large number of applications of acrylamide-containing polymers. Some major applications of these polymers include flocculants in water treatment, paper manufacture, mining, oil recovery, absorbents, and gels for electrophoresis.[270, 280]

Acrylonitrile and acrolein are the other two induatrially important acrylic derivatives of 3-HPA. These monomers can be obtained from 3-HPA derived bio-propene (Figure 12). Acrylonitrile is produced commercially *via* SOHIO's catalytic vapor-phase propylene ammoxidation process in which propylene, ammonia, and air are reacted in the presence of a solid catalyst at 400–510°C.[67] On the other hand, propene is selectively oxidized to acrolein in the presence of bismuth molybdate catalysts and nitric acid.[66]

Acrylonitrile has unalterable status in today's fiber industry. It is the major monomer of DuPont's first acrylic fiber, Orlon. Moreover, polyacrylonitrile (PAN) fibers are the chemical precursor of high-quality carbon fiber. Carbon fibers are uniquely qualified for use in both high-tech and common daily applications since they have the inherent combination of high strength, high stiffness and light weight.[281] Wu stated that acrylonitrile easily copolymerizes with electron-donor monomers and there are over 800 acrylonitrile copolymers that have been registered. Some of the important copolymers include styrene-acrylonitrile (SAN), acrylonitrile-butadiene-styrene (ABS), acrylonitrile-styrene-acrylate (ASA), acrylonitrile-butadiene (ANB) and acrylonitrile-methyl acrylate (AN/MA).[270, 282]

Acrolein can polymerize in three different manners; by 1,2-addition at the vinyl group, or at the carbonyl group, or by 1,4-addition as a conjugated diene across the carbonyl and □,□ unsaturation (Figure 12). Acrolein homo- and co-polymers generally have been used as biocidal and biostatic agents. Moreover, a commercial product having both pendant aldehyde and carboxylic acid groups is obtained by oxidative co-polymerization of acrylonitrile with acrylic acid. The material has the Degussa trend name POC and it is particularly useful as a sequestering agent for water treatment. Schulz reviewed radical, anionic and cationic polymerizations of acrolein and more detailed information regarding acrolein as a monomer and its' polymers can be found in this article.[283]

## 5.5. Aspartic Acid Platform Based Polymers.

Large scale production of aspartic acid has recently been realized from ammonia and fumaric acid by employing immobilized aspartase from *E. coli* or suspended cells of *Brevibacterium bravum*. Its direct production from sugars is more desirable for cutting-off intermediate steps. However, this is not possible with the current techonology because a favorable equilibrium of the aspartase-catalyzed reaction requires toxic concentrations of intracellular ammonia and fumaric acid.[284] Thus, bio-based fumaric acid is required to industrially produce aspartic acid from renewable feedstocks.

As a chemical building block, aspartic acid leads to the formation of many valuable compounds including fumaric acid, maleic acid, 2-amino-1,4-butanediol, □-alanine, aspartic anhydride and amino-□-butyrolactone (Figure 2d).[40, 70-73, 284] Also, aspartic acid based polymers, poly(amino acids) with free carboxylic groups, have attained the greatest commercial success in terms of water soluable biodegradable polymers. Poly(aspartic acid)s (PASA) can be produced by different routes. Thermal polyaspartate (TPA) production is the simplest and oldest approach. In this process, powdered L-aspartic acid is firstly heated to high enough temperature, at least 185°C, to initiate a condensation reaction. Then, the temperature is increased to 225°C and maintained at that temperature until at least 80 % conversion of L-aspartic acid to polysuccinimide (PSI) is achieved. The subsequent step includes the alkaline hydrolysis of the intermediate PSI polymer to give the resultant TPA polymer (Figure 13). TPAs are functionally equivalent to poly(acrylic acid) and highly linear TPAs are fully biodegradable. These distinctive features of TPAs make them one of the best target polymers for use in three global markets; performance chemicals, diapers, and agriculture, which have a global market potential of $20 billion.[285, 286]

There are certain disadvantages of the TPA process. First of all, a copolymer is always obtained at the end of the polymerization. In addition, nearly complete racemization occurs during the thermal polycondensation reaction and the alkaline hydrolysis may proceed at both carbonyls of the PSI. Hence, the resultant copolymer consists of not only D- and L-isomers, but it may also contain □- or □-peptide bonds in the main chain (Figure 13).[286] Control over the repeating unit isomers, D- or L- PASA with pure α- or β-links, can be achieved by enzymatic preparation of PASA. For instance, Soeda *et al.* polymerized diethyl L-aspartate by employing a bacterial protease from *Bacillus subtilis* to yield only □-linked poly(ethyl L-aspartate) having an *M*w of up to 3700. The best polymerization result was obtained by using 30% protease *Bacillus subtilis* in acetonitrile containing 4.5 volume percent of water at 40°C for 2 days (Figure 13).[287] Although the enzymatic preparation route can give only D- or L- PASA with pure α- or β-links, it is not a suitable approach for large scale production. The TPA process is relatively much cheaper and more suitable for commercial manufacture of PASA.

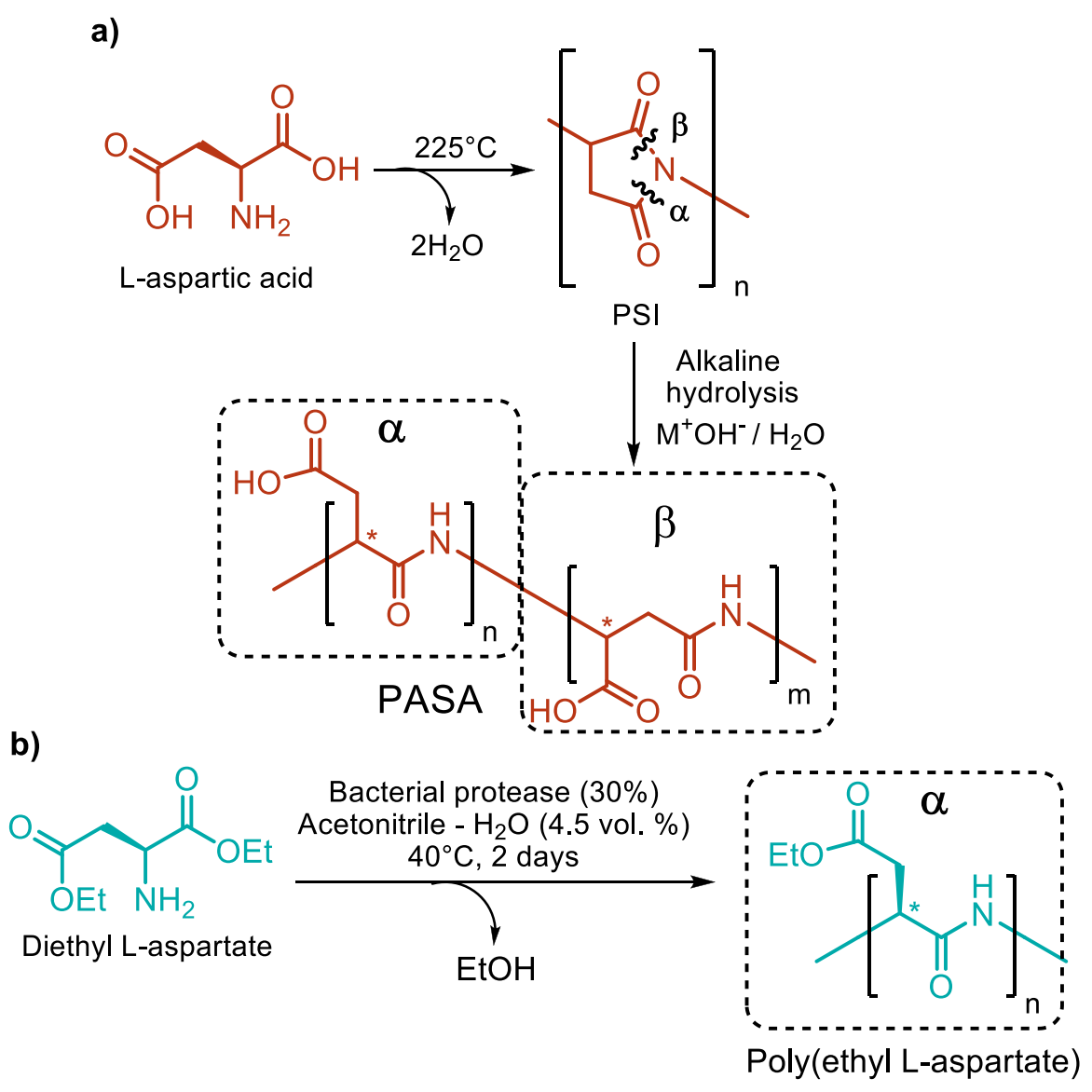

Figure 13. PASAs obtained via a) the TPA, b) the bacterial protease methods.

Nonetheless, the enzymatic process is quite useful for specialty polymer applications in which the isomeric purity of PASA really matters.

PSI has also been used as a polymer precursor for the production of polyaspartamide (PAA) based polymers apart from being an intermediate polymer for the production of commercial TPA. Recent literature studies regarding PAA derived polymers mainly focuses on its use as drug carriers and stimuli-responsive polymeric materials.[288, 289] As a recent example, Xu *et al.* synthesized a novel polyaspartamide derivative composed of partially tetraethylenepentamine grafted poly(aspartic acid) (PASP-pg-TEPA) nanoparticles as a potential intracellular delivery system (Figure 14).[290]

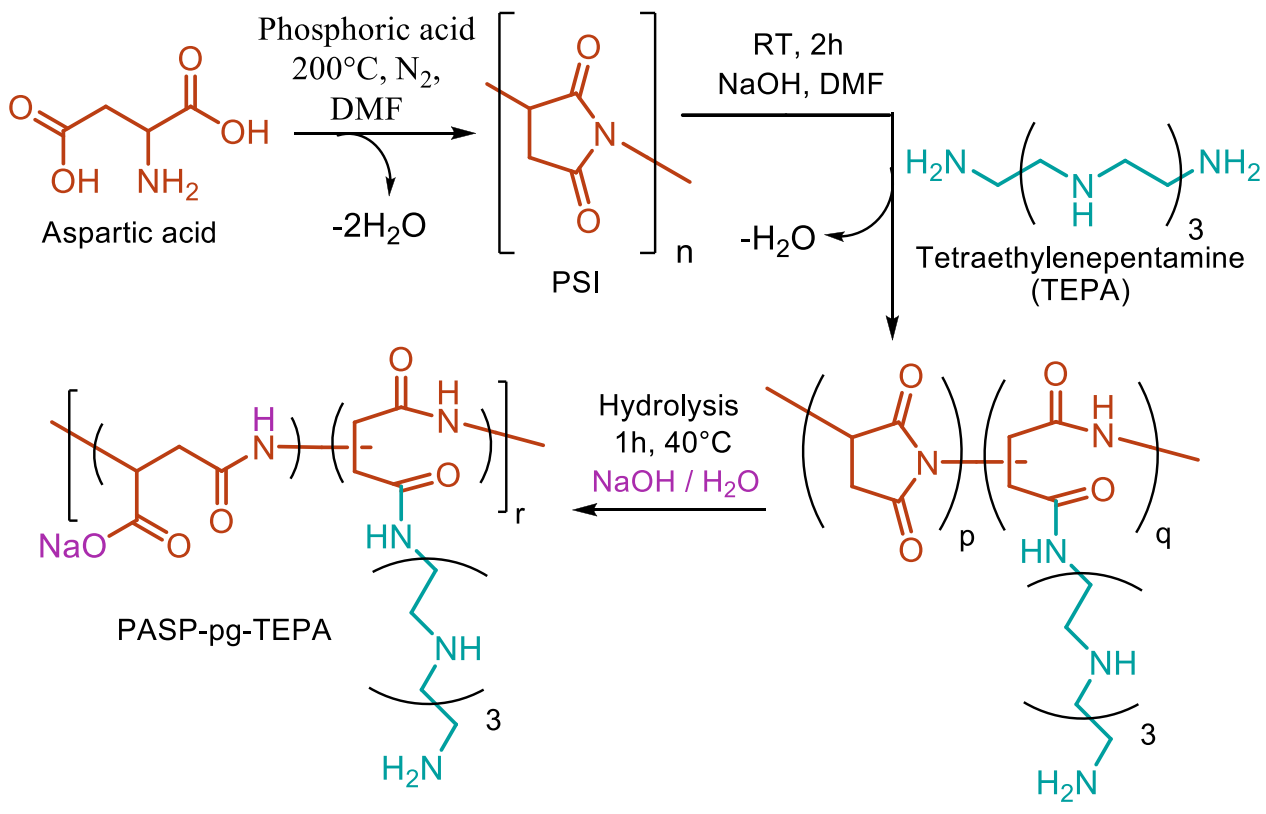

Figure 14. Synthesis of the zwitterionic polyaspartamide derivative composed of PASP-pg-TEPA.

α-Decarboxylation of aspartic acid results in the formation of β-alanine (Figure 2d).[70, 284] On the other hand, it is highly difficult to achieve this reaction through chemical methods because □-carboxyl group of aspartic acid can also undergo a decarboxylation reaction. The perfect conversion can be realized using *E. coli L-aspartate α-decarboxylase*. However, this enzymatic method still requires process development mainly due to the occurring enzyme inactivation during the catalysis.[284] Solution of this problem is critical for large scale enzymatic synthesis of □-alanine. Once cost competitive □-alanine is industrially available, it may open new oppurtunities for production of nitrogen containing base chemicals such as acrylamide and acrylonitrile. Besides, the direct condensation of □-alanine, or its esters, is used for the preparation of PA-3 (poly(□-alanine), which is used as stabilizer for polyoxymethylene, and polyacetal resin. It is a also an excellent formaldehyde scavenger.[270] Apart from these, Liu *et al.* have revealed the intriguing biological properties of PA-3. They introduced a new family of PA-3 polymers (poly-β-peptides) and these polymers were shown to display significant and selective toxicity toward *Candida albicans*, the most common fungal pathogen among humans.[291] In their another study, it was described that PA-3 backbone based polymers selectively support in vitro culture of endothelial cells but they do not support the culture of smooth muscle cells or fibroblasts. There are very few reported synthetic material examples which can display selective cell growth. Hence, these findings of Liu *et al.* regarding selective encouragement of the growth of specific cell types are very valuable for the engineering of complex tissues.[292]

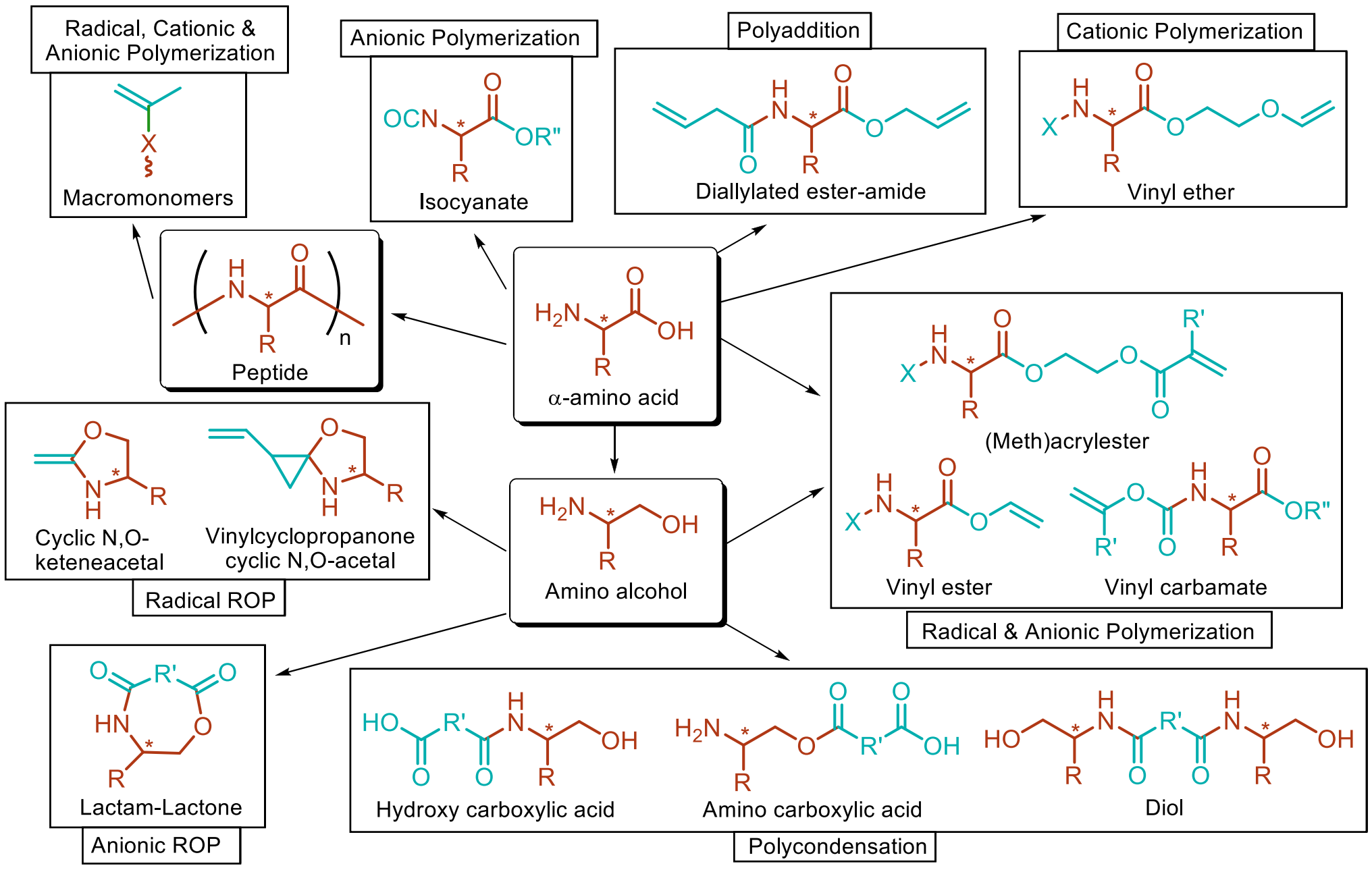

Figure 15. The design of novel functional monomers based on amino acids and their corresponding polymerization methods.

Amino acid-based polymers of aspartic acid and □-alanine, and also other lignocellulosic biomass derived amino acids such as glutamic acid, □-benzyl glutamate, norvoline, □-aminobutyric acid (Section 5.6), 5-aminolevulinic acid (Section 5.9) and lysine (Section 5.18) may offer many different advantages. These polymers can be modified further to introduce new functions such as imaging and molecular targeting. Their degradation products are expected to be non-toxic and readily excreted from the body. Drugs can be chemically bonded to these polymers. Moreover, these polymers can enhance biological properties like cell migration, adhesion and biodegradability, and improve on mechanical and thermal properties. Three major applications of them include drug delivery, gene delivery and tissue engineering. Various types of polymerization techniques, particularly different polycondensation and ROP reactions, can be employed to produce polymers consisting of amino acid moieties in the main chain. Besides, radical, anionic and cationic polymerizations are generally used to obtain polymers consisting of amino acid moieties in the side chain. On the basis of these techniques, it has been reviewed that polyamides, polyesters, polyurethanes, polyacetylenes, polymethacrylamides, polyanhydrides, polyphosphazines, polysulfides, poly(amide-ester-imide)s, poly(amide-imide)s, poly(amide-imide-urethane)s, poly(ester-amide)s, poly(ester-amide-sulfide)s, poly(ester-imide)s and poly(imide-urethane)s are produced with different architectures. Amino acid based polymers were reviewed in detail in the recent literature.[293-296] Thus, no further detail is provided in this article. Nonetheless, the design of novel functional amino acid based monomers and their corresponding polymerization methods are briefly described in Figure 15.[296] Although the figure covers up □-amino acid derived monomers and their polymerization methods, it can be extended to □-, □, □-amino acids such as □-alanine, □-benzyl glutamate, □-aminobutyric acid (Section 5.6), and 5-aminolevulinic acid (Section 5.9).

## 5.6. Glutamic Acid Platform Based Polymers.

Currently, the global production capacity of glutamic acid is more than 200 kt per year and most of this production has been achieved through fermentation. Variuos species of *Brevibacterium* and *Corynebacterium* can produce glutamic acid from different cheap carbon sources such as glucose, ethanol and glycerol. Although glutamic acid is produced from these cheap feedstocks, there are certain limitations in its large scale production process. The main limitation stems from its complex processing which consists of a number of downstream treatment schemes like precipitation, conventional filtration, acidification, carbon adsorption and evaporation. All these treatments are essential to obtain high purity glutamic acid but they greatly increase the production cost in the end. Thus, a new process is required to produce glutamic acid in a more eco-friendly and economically manner. In this consideration, a membrane-based processing is envisioned to eliminate the need for separate purification units and to reduce the overall production cost.[74] Thus, a shift from the current complex processing to a membrane-based technology might be observed in the glutamic acid market in the near future.

Industrial production of glutamic acid is quite important for production of the $C_5$ compounds (Figure 2e) and their corresponding polymers. So far, the most successful commercial polymer form of glutamic acid is poly-□-glutamic acid (□-PGA), an anionic biopolymer formed *via* the gamma-amide linkage. □-PGA is isolated particularly from various strains of *B.*

*licheniformis* and *B. subtilis.* It can be produced from a variety of carbon sources such as glutamic acid, glucose, fructose, sucrose, citrate and glycerol. The proposed pathway for □-PGA synthesis from L-glutamic acid in *B. subtilis* IFO 3335 is shown in Figure 16a. The recent pilot trial findings of Zhu *et al.* have suggested that the environmental friendly and efficient production of *γ*–PGA can also be achieved by using the low-cost and renewable lignocellulosic biomass.[297] □-PGA is water-soluable, edible, biodegradable, biocompatible and non-toxic for humans and the environment. Hence, □-PGA and its derivatives have been of interest in the past few years, especially in food, cosmetics, medicine, and water treatment industries. A wide range of unique applications of them in these industries were reveiwed by Shih and Van.[298]

A structurally different polymer of glutamic acid is poly-□,L-glutamic acid (□,L–PGA). It is generally synthesized starting from poly(□-benzyl-L-glutamate) (PBLG). At first, □-benzyl-L-glutamate is reacted with a carbonyl dichloride compound, such as phospene, to give the corresponding NCA derivative. Later on, the triethylamine-initiated polymerization of the NCA of □-benzyl-L-glutamate yields poly(□-benzyl-L-glutamate). Finally, removing the benzyl protecting group of poly(□-benzyl-L-glutamate) with the use of HBr in the presence of $CO_2$ results in formation of □,L–PGA (Figure 16b).[299] On the other hand, a better method, most probably a microbial method, is necessary because the above-mentioned route requires the synthesis of the intermediate NCA compound as well as the usage of highly toxic carbonyl dichloride compounds, and protection/deprotection steps. □,L–PGA shows the general characteristics of □-PGA and additionally, its carboxylic group provides functionality for drug attachment. Thus, anticancer drug conjugates of □,L–PGA have been studied extensively. □,L–PGA based anti-cancer drug conjugates were reviewed in detail by *Li*.[299]

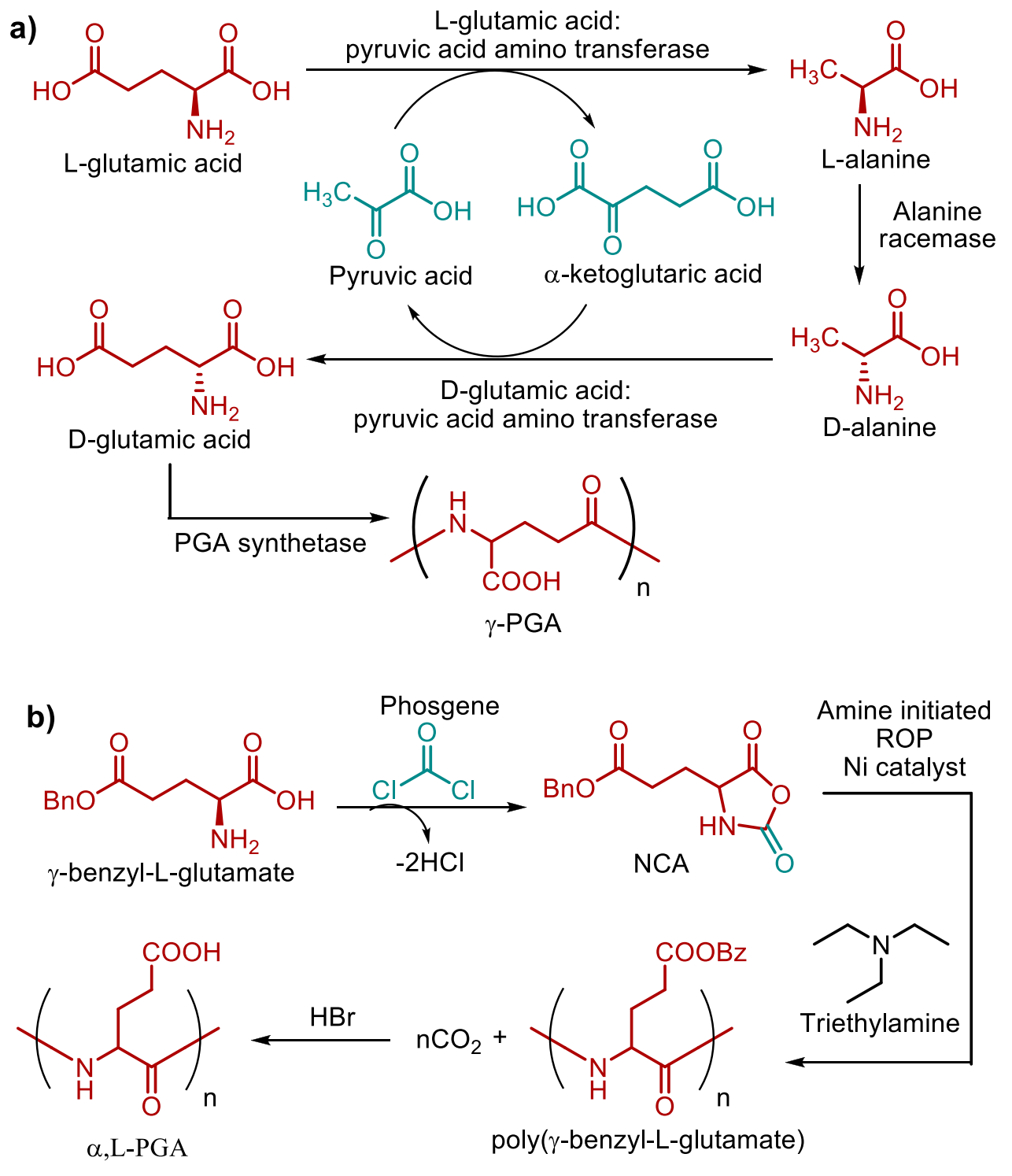

Figure 16. a) Proposed pathway for □-PGA synthesis from L-glutamic acid in *B. subtilis* IFO 3335 (The pathway starting from glucose to L-glutamic acid was not indicated) b)Reaction scheme for synthesis of □,L–PGA.

Norvoline and □-aminobutyric acid (GABA) are the other two amino acids of this platform in addition to glutamic acid (Figure 2e).[77, 78, 300] Through chemical methods, it is quite difficult to control selective reduction of □-carboxyl or selective decarboxylation of □-carboxyl groups of glutamic acid for production norvoline and GABA, respectively. Hence, biosynthetic routes are more convenient for production of these amino acids from glutamic acid. As such, the glutamate decarboxylase catalyzed reaction of glutamic acid to GABA is superior to conventional chemical methods owing to its higher catalytic efficiency and environmental compatibility, and milder reaction conditions.[78] Norvoline, GABA and other lignocellulosic biomass derivable monomers are quite useful for synthesis of novel biodegradable and bio-compatible amino acid containing polymers. As an example, Galán *et al.* studied poly(ester amide)s derived from glycolic acid and GABA or □-alanine for non-coated monofilament and flexible sutures.[301] GABA or □-alanine were specifically chosen as co-monomers because they can enhance enzymatic degradation and improve intermolecular interactions of their incorporating polymers. Moreoever, they can bring functionalizable groups where molecules with a pharmacological activity could be attached. Although GABA and □-alanine were the only studied amino acids, the study can be extended for other amino acids as shown in Figure 17. Amino acid derived monomers and their corresponding polymerization techniques are highlightened in more detail in Section 5.5.

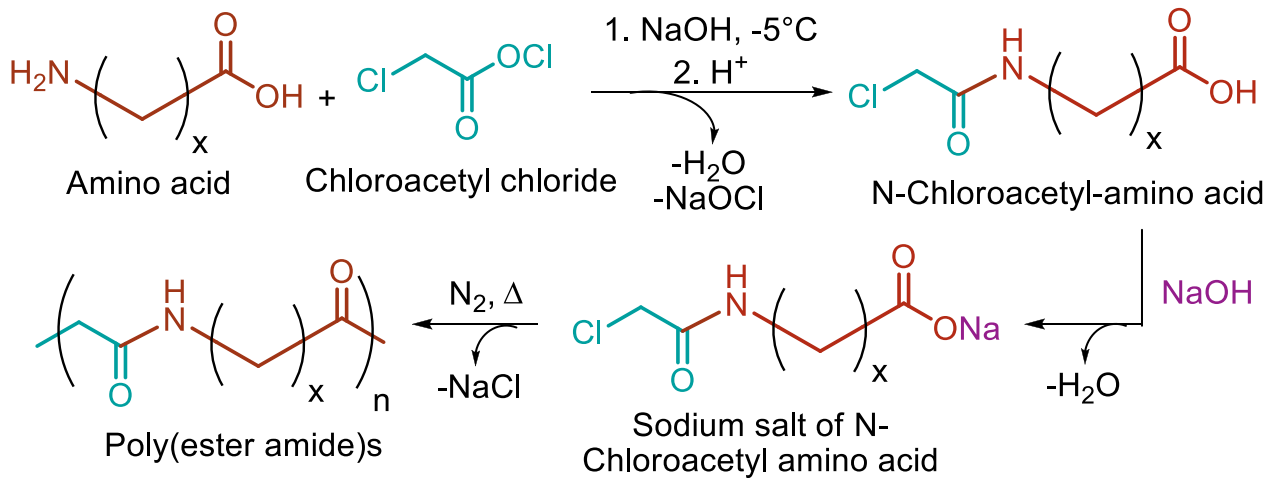

Figure 17. Synthesis of poly(ester amide)s based on glycolic acid and amino acids for suture applications.

GABA can be effectively reduced to 4-amino-1-butanol with a conversion of more than 99.9% and a yield of 91.9% through a hydrogenation process over Rh-MoOx/$SiO_2$ heterogeneous catalyst (Figure 2e).[79] 4-amino-1-butanol monomer has started to be widely used for the production of poly(□- or □-amino ester)s. Recently, there has been a great increase in the number of publications related to poly(amino ester)s, especially after the promising results of these polymers in gene delivery applications. For instance, Li *et al.* stated that cationic poly(β-amino ester)s have superior transfection efficiency and less toxicity with comparison to commercial reagents such as Lipofectamine 2000;

transfection reagent produced by Invitrogen.[302] Many 4-amino-1-butanol containing poly(amino ester) systems have been reported to deliver DNA, siRNA and proteins for gene threrapy applications.[302-305] Different libraries of structurally related poly(amino ester)s can be produced by the systematic modification of the polymer backbone, side chain, and end group. Sunshine *et al.*[306] introduced a new library of 320 poly(□-amino ester)s by implementing this method (Figure 18). In the process, the primary amino alcohols were reacted with diacrylates by Michael addition to yield the corresponding diacrylate-terminated base polymers. The reaction was carried out in the presence of excess diacrylate with no solvent stirring at 90°C for 24 h. Subsquently, the base polymers were dissolved in DMSO and then, allowed to react with an end-capping amine at RT for 24 h. As can be seen from the figure, other lignocellulosic biomass derivable alkanolamines such as 1,3-propanolamin and 1,5-pentanolamin, which are reduction products of □-alanine and 5-aminolevulinic acid, respectively, can also be employed for the introduction of novel poly(amino ester) libraries.[304, 306]

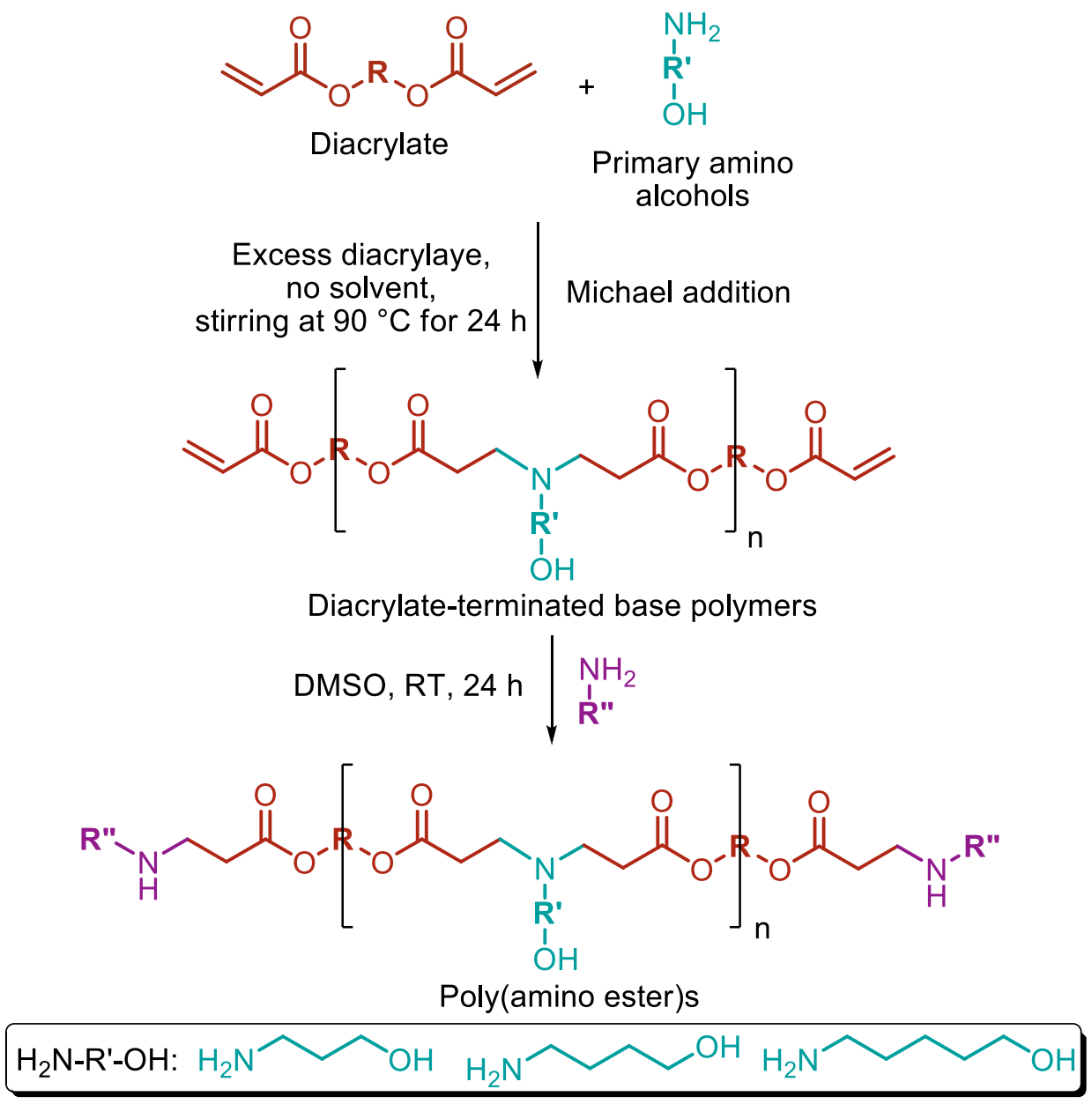

Figure 18. Synthesis of the poly(□-amino ester) library.[306]

By providing amine functionality together with a carboxyl, or an alcohol group, glutamic acid platform chemicals can boost up manufacture of amino acid based polymeric materials, once their commercial production routes are optimized. Additionally, their conversion to di-functional products such as 1,5-pentanediamine, (or cavaderine), glutaric acid and 1,5-pentanediol (1,5-PDO) (Figure 2e) opens a completely new window to the industry of polyamides and polyesters. Among these chemicals, cadaverine can be produced by the diamination process of glutamic acid derived 1,5-PDO,[56] or as a different route, it can be effectively exported from *E. coli* or *C. glutamicum* cells as the decarboxylation product of L-lysine.[236] Various PA-5 series can be produced by combining cadaverine with diverse diacids. For example, PA-5,10, PA-5,5 and PA-5,4 are obtained by polymerizing cadaverine with sebacic acid, glutaric acid and succinic acid, respectively (Figure 19).[307] According to Kind and Wittmann, polycondensation of microbial biosynthesis derived cadaverine with these bio-blocks provides completely bio-based PAs. Cadaverine incorporating PAs possess desirable properties such as high melting points and low water absorption. Hence, they are proposed to be bio-based alternatives to several petroleum-based PAs.[308] In fact, Carothers invented PA-5,10 even before PA-6,6;[309] a keystone in today's polymer industry. However, PA-5,10 is more expensive to make even though it has superior properties. So, industrial shifts from conventional PAs to bio-based PA-5,10 products might be observed in the future depending on the cheaper and more efficient production of cavaderine from biomass resources than its petroleum-based production routes.

Like cadaverine, glutaric acid (Figure 2e) is also a natural L-lysine degradation product. *Pseudomonas putida* uses L-lysine as the sole carbon and nitrogen source to produce glutaric acid by having 5-aminovaleric acid (5-AMV) as the key intermediate.[310] Thus, reconstruction of the 5-AMV pathway in a L-lysine over-producing host can bring renewable routes to PA-5 and PA-5,5 polymers (Figure 19). PA-5, as being homopolymer of 5-AMV, possesses close properties to PA-4,6 and could be a suitable substitute.[236] PA-5 based products have also been commercially manufactured by Lipo Chemicals Company under the trend name of Orgasol Caresse® for skin care and sun care applications.[311]

Glutamic acid platform can provide bio-based monomers for PAs as shown in Figure 19. In addition to this, glutaric acid can be extensively employed in polyester industry. Numerous polyesters polyols comprising about 50 to 70% by weight of glutaric acid were patented by Altounian and Grenier.[312] More importantly, its hydrogenation product; 1,5-PDO (Figure 2e),[300] has already been used as a common plasticizer in today's modern polymer industry owing to its fairly long carbon chain.[276] Flynn and Torres recently patented bio-derived 1,5-PDO plasticizer for biopolymers which improves the flexibility of the polymers while not adversely affecting their modulus.[313]

### 5.7. Glucaric Acid Platform Based Polymers.

The main contribution of glucaric acid platform (Figure 2f) to the polymer chemistry would be in the design and synthesis of novel sugar containing polymers. In terms of linear monosaccharide derivatives, D-gluconic acid has an estimated market size of 60 kt/year. It is usually produced by enzymatic oxidation of D-glucose which is facilitated by the enzyme glucose oxidase and glucose dehydrogenase.[81] D-gluconic acid is considered as an interesting source of comonomer to synthesize functionalized polyesters for biomedical and pharmaceutical applications. Marcincinova-Benabdillah *et al.* reported novel heterocyclic 1,4-dioxane-2,5-dione monomers derived from D-gluconic acid through selective protection of the 3-, 4-, 5-, and 6-hydroxyl groups. In the process, □-gluconolactone is reacted with methoxypropane and then, treated with NaOH and HCl, respectively, to yield 3,4:5,6-di-*O*-isopropylidene gluconic acid. Further reaction of the protected gluconic acid with bromoacetyl chloride in the presence of triethylamine, and subsequently with NaHCO3 gives the corresponding cyclic dilactone. Finally, the ROP of the cyclic dilactone with lactide yields novel degradable polyesters (Figure 20) with higher glass transition temperatures than poly(lactic acid-*co*-glycolic acid) polymers, which are routinely used for sutures, bone prostheses, and drug delivery systems.[314] However, the polyesters are not suitable for large scale production because several laborious and costly steps are required to synthesize the dilactone monomer.

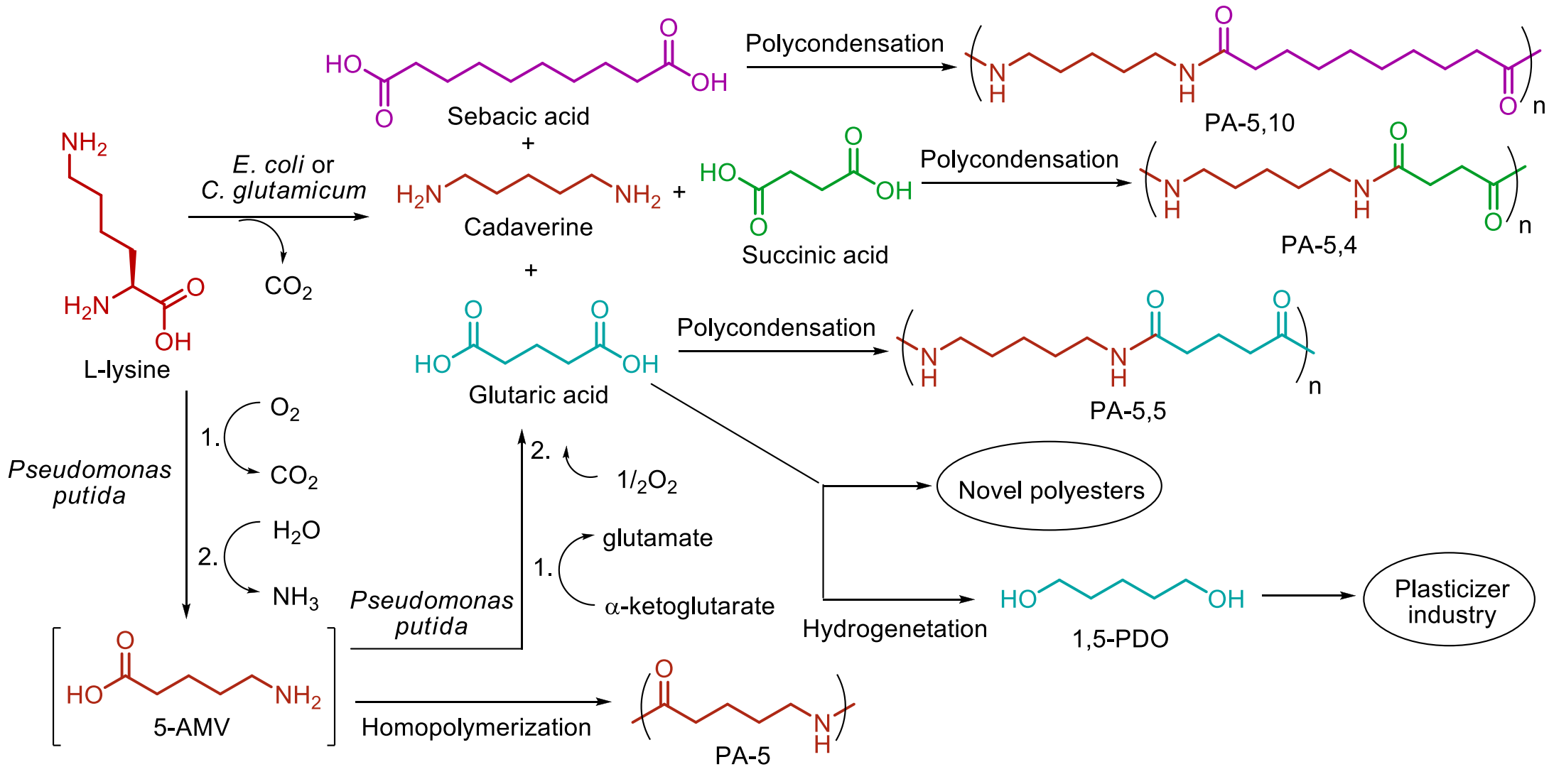

Figure 19. Glutamic acid platform based PAs and PEs.

As another linear monosaccharide derivative of this platform, glucaric acid can be obtained by oxidation of the primary hydroxy group of gluconic acid on Pt/C catalysts (Figure 2f).[54] The recent literature on polymers of glucaric acid is mainly based on poly(D-glucaramidoamine)s (PGAAs) (Figure 20); efficient and degradable gene delivery vehicles that consist of three main functionalities: carbohydrate groups, secondary amines, and amide bonds. As a promising example, Liu *et al.* synthesized a series of new PGAAs by reacting esterified D-glucaric acid comonomer with the amine-containing comonomers; diethylenetriamine, triethylenetetramine, tetraethylenepentamine, and pentaethylenehexamine, at room temperature in methanol. The PGAAs were proven to exhibit high gene delivery efficiency without toxicity with BHK-21 cells.[315] More detailed information regarding PGAA and other carbohydrate-based nucleic acid delivery vehicles were reviewed by Ingle *et al.*[316] By applying the same synthetic approach; the polycondensation of esterified glucaric acid, Nobes *et al.* synthesized and tested a number of poly(glucaramides); hydroxylated PAs (Figure 20), for applications requiring water resistance. In the study, the poly(glucaramides) were used as an additive in fiber-reinforced pressed panels. The water resistance of the panels increased dramatically with increasing amounts of poly(glucaramide) in the panels.[317] Another route of producing hydroxylated PAs is ring-opening polyaddition of glucaric acid derived glucarodilactone[318] (Figure 2f) with several alkylenediamines. Hashimoto *et al.* synthesized various hydroxylated PAs through this route even at room temperature and with no catalyst.[319] The same group also carried out the polyaddition of glucarodilactone with hexamethylene diisocyanate and methyl (S)-2,6-diisocyanatocaproate to obtain the corresponding polyurethanes (Figure 20) by using dibutyltin dilaurate as a catalyst.[320] Very recently, Gallagher *et al.* reported the synthesis of a new glucarodilactone based dimethacrylate monomer; glucarodilactone methacrylate (GDMA). Thermally initiated free radical polymerization of GDMA in the bulk yields a highly cross-linked thermoset network. The thermoset polymer is reported to have comparable mechanical properties with respect to commercially available stiff poly(dimethacrylates).[321]

For cyclic monosaccharide derivatives, their versatile usage as a monomer or as an initiator has been reported. Such a compound is methylglucoside which can be obtained through the methanolysis of cellulose in methanol in the presence of an acid catalyst at 200°C.[322] However, the yield of this process is low (between 40–50%) and thus, a process development is required to further increase the yield (Figure 2f). Park *et al.* utilized □-methylglucoside for the chemoenzymatic synthesis of sugar-containing biocompatible hydrogels (Figure 20) by a two step reaction. In the first step, lipase-catalyzed esterification of □-methylglucoside was performed with acrylic acid, methacrylic acid, vinyl acrylate or vinyl methacrylate. In the subsequent step, the esterified □-methylglucoside monomers were used to synthesize poly(□-methylglucoside acrylate) and poly(□-methylglucoside methacrylate) by free-radical polymerization of the monomers with and without ethylene glycol dimethacrylate (EGDMA) as crosslinker. The results of the study suggest that the obtained polymers; not crosslinked with EGDMA, are highly biocompatible and they are readily available for various biomedical applications.[323] A different utilization of □-methylglucoside was reported by Suriano *et al.* They synthesized amphiphilic A3B mikto-arm copolymers by using a t-butyl-diphenyl silyl-based methylglucoside derivative. The methylglucoside derivative was used as an initiator for the polymerization of □-CL to obtain star-shaped poly(□- caprolactone) macroinitiator (Figure 20). ATRP of the macroinitiator in the presence of galactose methacrylate allowed the formation of A3B mikto-arm copolymers with different compositions and molecular weights. Finally, selective deprotection of sugar protecting groups generated amphiphilic mikto-arm copolymers. Such polymeric micelles offer the advantage to trap drugs, such as anticancer agents, e.g., paclitaxel and doxorubicin, in their hydrophobic core.[324]

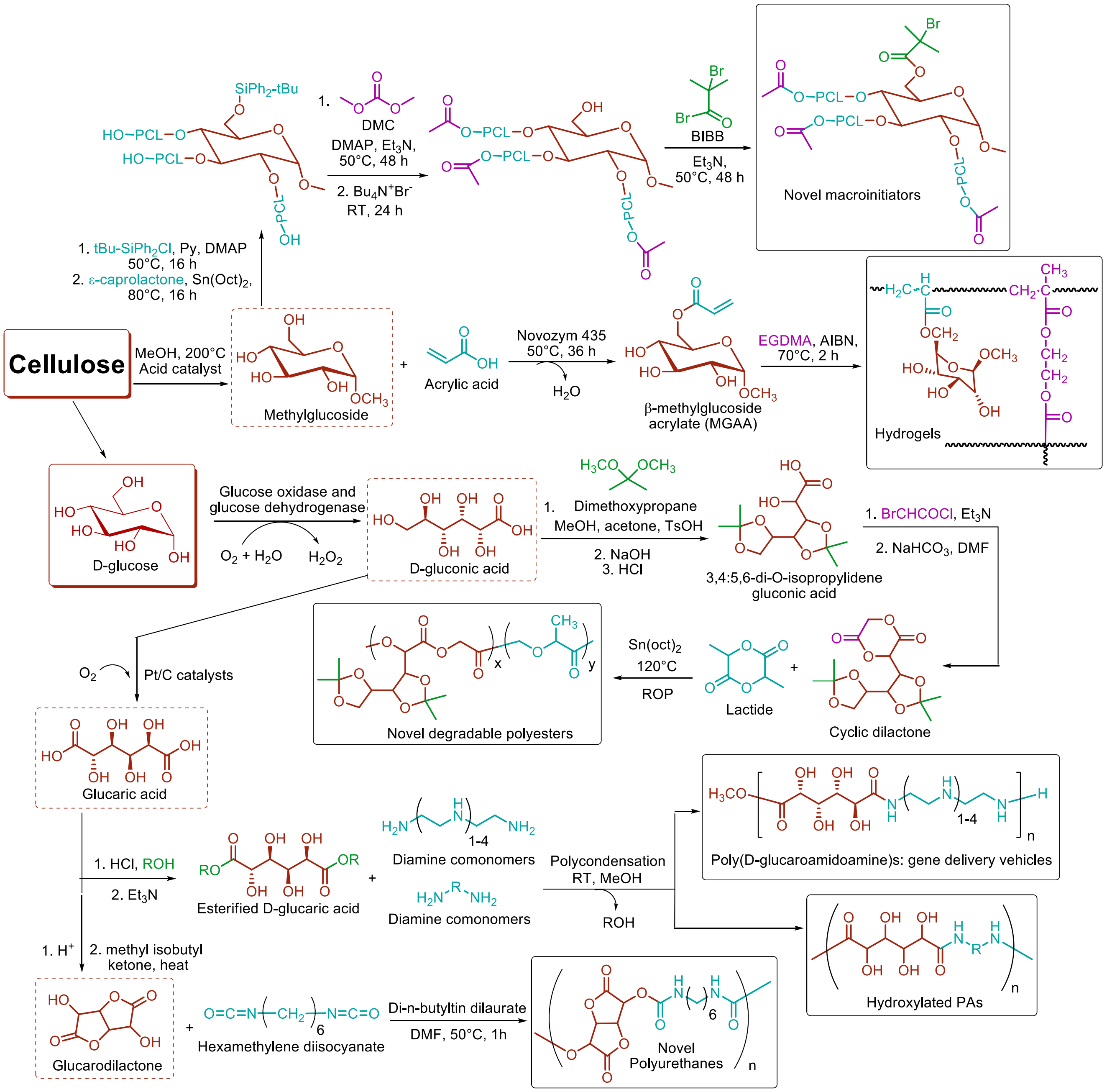

Figure 20. The versatile utilization of glucaric acid platform based monomers in production novel polymers.

Figure 20 shows the versatile utilization of glucaric acid platform based monomers in production of novel polyesters, polyurethanes, hydrogels, hydroxylated PAs, and novel initiators. As shown in the figure, applicability of polymers of glucaric acid platform is quite versatile. Their research and industrial utilization can extend from composites to cosmetic and biomedical materials. Since glucaric acid platform consists of monosaccharides and their derivatives, more detailed information regarding incorporation of sugars and their derivatives either in polymer backbone or as pendant groups is provided in Section 5.1.

## 5.8. Itaconic Acid Platform Based Polymers.

Itaconic acid (ITA) (Figure 2g) is one of the most attractive monomers for producing bio-based polymers by providing two functional acid groups as well as a vinyl functionality. It is produced industrially *via* fermentation of carbohydrates such as glucose by fungi and its recent market volume is about 15 k t/year. Chemical synthesis of ITA starting from different compounds, like citric acid, is also known but it is not economically and ecologically convenient for large scale production. According to Yao and Tang, ITA may become a replacement monomer for petrochemical-based acrylic or methacrylic acid owing to its similarity in structure.[261, 325] Current production of petroleum-based acrylic and methacrylic acid requires several chemical synthesis steps. In this consideration, ITA has process and ecological advantages since it is a direct fermentation product of cheap carbohydrates. On the other hand, its market penetration requires a cost competitive production with respect to other acrylic monomers.

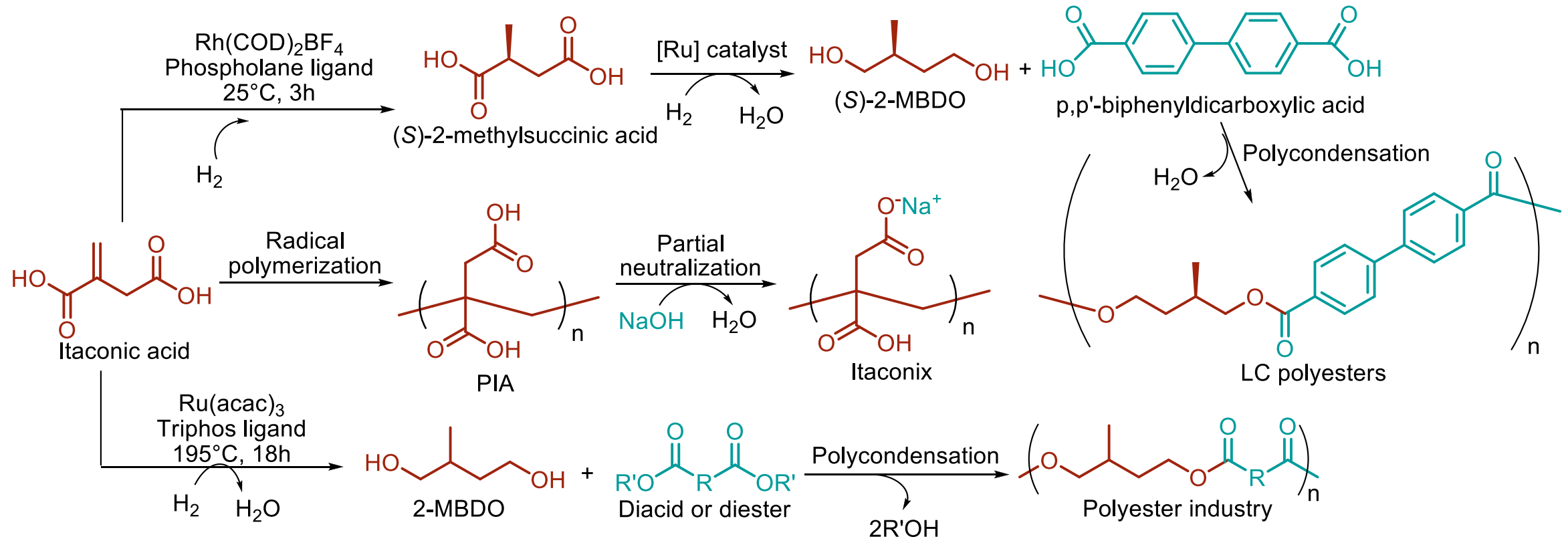

Figure 21. Itaconic acid and 2-MBDO based polymers.

ITA has already been used in the manufacture of synthetic fibers, coatings, adhesives, thickeners and binders.[261] For instance, it is primarily used as comonomer in styrene butadiene polymers to provide dye receptivity in fiber industry. Moreover, its homopolymer; polyitaconic acid (PIA), is commercially avaliable. Itaconix Corporation, USA Company, produces linear PIA partially neutralized with sodium salt for everyday applications (Figure 21).[326] The company uses bio-based ITA which is produced by fermentation of carbohydrates using *Aspergillus terreus*. Hence, 100% sustainable polymers from bio-based itaconic acid are already in the market since all of the carbon comes from renewable biomass resources. PIA has also been envisioned as a replacement polymer to polyacrylic acid. For this to happen, its current production cost (~$3/kg) must reach below $1.5/kg and this can be achieved via an integrated production infrastructure starting from carbohydrates to final PIA polymer.[327] Versatile utilization of ITA based polymers in resin, lattice, fibre, detergent and cleaner industries are summarized by Willke and Vorlop.[328]

Apart from being a valuable monomer itself, ITA also serves to be a parent molecule for production of other monomers (Figure 2g). 2-methyl-1,4-butanediol (2-MBDO) is a promising example. It can be obtained through ruthenium-catalyzed hydrogenation of ITA in the presence of triphos ligand at 195°C. The reaction proceeds in excellent yields up to 93%.[89] 2-MBDO can have a great contribution to the current polyester industry owing to its structural resemblance to 1,4-butanediol. Its usage in production of polyesters still awaits exploration. Besides, selective hydrogenation of ITA to R- or S-2-MBDO holds significance because entiomerically pure 2-MDBO is a promising monomer for synthesis of liquid crystalline (LC) polyesters. In this consideration, Almena *et al.* showed that the vinyl group of ITA can be effectively reduced with an enantiomeric excess (ee) of 97% with the help of a chiral phospholane ligand and $Rh(COD)_2BF_4$ catalyst.[86] Further reduction of the two carboxyl groups can give chiral 2-MBDO compounds. Recently, liquid crystalline (LC) polyesters of (S)-2-MBDO with p,p'-biphenyldicarboxylic acid or p,p'-biphenyldicarboxylate (Figure 21), have been attracting interest. As an example, Uchimura *et al.* incorporated 9,10-diphenylanthracene moieties into the LC polyesters consisting of (S)-2-MBDO, p,p'-biphenyldicarboxylate and 1,6-hexanediol units. The circular dichroism spectra showed that the polymer with 1 mol % of anthracene moiety exhibited the chiral smectic C phase. On the other hand, only a smectic A phase was formed in the other polymers.[329]

Upon a dehydration reaction under acidic conditions and at high temperatures, carbohydrate derived ITA can be converted to itaconic anhydride (ITAn). However, a faster synthesis process, which can proceed under milder conditions, is needed for this conversion. Such a process was reported by Robert *et al*.[85] They showed that reacting ITA with dimethyl dicarbonate in the presence of chromium (III)-based salen complex perfectly gives ITAn with 100% product yield (Figure 22). This conversion is quite valuable because it allows the formation of ITAnh monomer starting from carbohydrates in just two steps. ITAnh can be employed both in ROP and radical polymerization reactions owing to its cyclic anhyride and vinyl functionalities, respectively. It is one of the unique monomers for designing cyclic anhyride copolymers, which can exhibit biodegradable characteristics and are derived from renewable resources. Such copolymers have been drawing attention because they can further be modified for target applications. As an example, Shang *et al.* synthesized comb-like random copolymers of ITA and stearyl methacrylate (SM) by radical polymerization. These copolymers were then easily converted to ionomers (Na, Ca or Zn carboxylates) *via* the partial neutralization of the copolymers (Figure 22a). Both of the copolymers and their ionomers showed crystallinity of stearyl side-chains, by having a melting point about 30°C. Since the melting point is just below human body temperature (37°C), these materials were suggested to be potential candidates for biomaterials such as scaffolds, drug delivery systems, and prosthetics.[330] As another approach, ITA can be first utilized in polycondensation reactions, and then, the remaining vinyl functionality of its ring-opened copolymers can be further used in post-functionalization reactions. For instance, Okuda *et al.* performed the Sn-catalyzed ROP of ITA with poly(L-lactic acid) (PLA) to produce ITA-PLA macromonomers. The intact vinyl functionality of the macromonomers were then radically copolymerized with n-butyl methacrylate (BMA), n-butyl acrylate (BA), methyl methacrylate (MMA), and ethyl methacrylate (EMA) to efficiently give the corresponding graft copolymers (Figure 22b).[331]

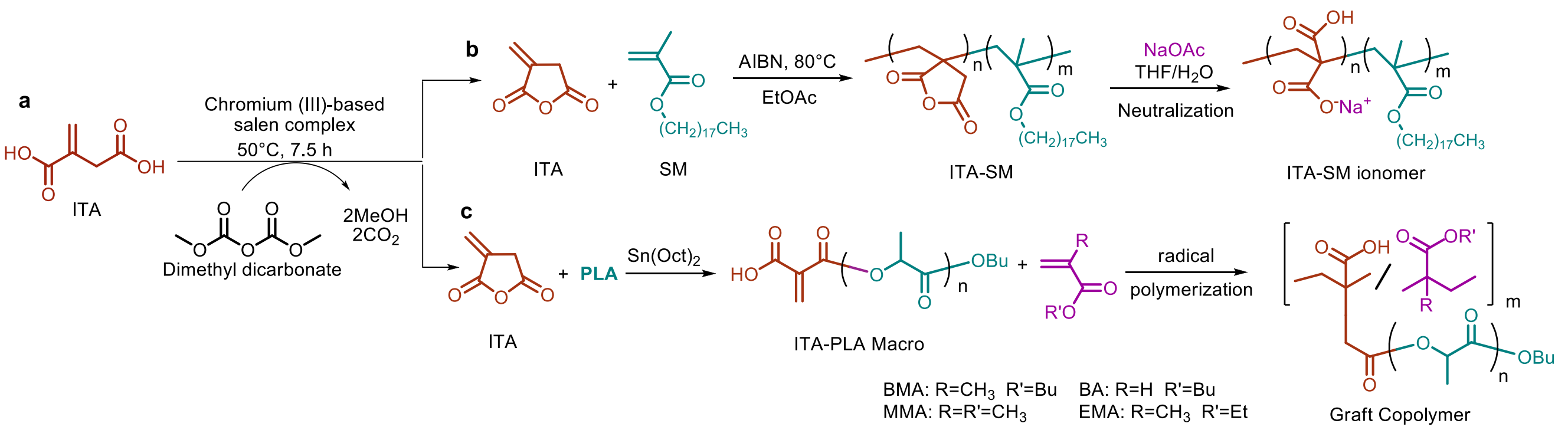

Figure 22. (a) Dehydration of ITA to ITAnh. b)Synthesis of ITA-SM copolymers and their conversion to the ITA-SM ionomer. (c) Synthesis of ITA-PLA graft copolymers.

Hydrogenation of the vinyl group of itaconic acid leads to formation of hydroxyl acids. Then, intramolecular esterification of the hydroxyl acids yields two regioisomeric monomers; 2-methyl butyrolactone (2-MGBL) and 3-methyl butyrolactone (3-MGBL) (Figure 2g). Polymerization of these butyrolactones is rarely investigated because □-lactones (5-membered) are thermodynamically stable. Nonetheless, it was stated that these monomers are worthy for more investigation.[261] Further reduction and dehydration of these lactones result in formation of 3-methyltetrahydrofuran (3-MTHF) (Figure 2g). 3-MTHF has a THF like cationic polymerization, however; its comperatively low value of activation entropy prevents its polymerization above 4°C.[261, 332] Literature on the polymerization of the other derivatives of ITA (Figure 2g); itaconic diamide, 2-methyl-1,4-butanediamine and 3-methylpyrrolidine, is scarce. Depending on the large scale production of ITA from lignocellulosic biomass, these derivatives can have a larger contribution to polymer research and industry.

### 5.9. Levulinic Acid Platform Based Polymers.

Amongst the lignocellulosic biomass derived monomers, levulinic acid (LevA) (Figure 3c) has a critical importance because its large scale production from lignocellulosic biomass has already been achieved. For instance, Maine BioProducts announced “Biofine Process” for the commercial production of LevA. The process is based on acid catalyzed dehydration of lignocellulosic feedstocks in a two-stage process.[325] Moreover, Avantium’s YXY technology has developed a tunable process to convert plant-based carbohydrates into high purity methyl levulinate, which can be easily converted to LevA through a straightforward process.[333] Although these processes for the first time provide cost competitive and high volume access to levulinics, the current LevA prices remaining in between 5-8 \$/kg must be further decreased to 1 \$/kg or below. Thereupon, the realization of LevA as a giant commodity chemical can be achieved. For further reduction of the LevA prices, its direct production from cheap cellulosic biomass feedstocks is more convenient and this issue is under intense investigation.[42, 325, 334, 335]

LevA is the precursor of the levulinic family and it branches to many valuable new compounds having novel applications (Figure 3c). Hence, it is a significant precursor to many pharmaceuticals, plasticizers, and additives.[336] Various companies recently launched production of novel bio-polymers based on LevA or its derivatives with a motivation towards producing cost competitive and sustainable polymers. As an encouraging example, Segetis recently launched the “Levulinic Ketal Technology” to investigate the direct utilization of LevA based ketals in polyurethane and thermoplastic applications.[325] In the process, esters of LevA, as degradation products of carbohydrates, are combined with alcohols derived from vegetable oils for synthesis of levulinic ketals (Figure 23). Leibig *et al.* evaluated over 25 levulinic ketal derivatives in terms of their plasticizer performance in PVC. A number of levulinic ketals exhibited superior performance with comparison to the commercial phthalate plasticizers.[337] In addition to giving rise to the development of novel polymer materials, levulinic monomers have the potential for substitution of petrochemical building blocks. For instance, there is a growing concern regarding the sustainability of bisphenol A (BPA) in consumer products and food containers since it has a pseudo-hormonal effect on the body.[338] The Food and Drug Administration (FDA) has already ended its authorization of the use of BPA in certain products.[339] LevA derived diphenolic acid (DPA) is considered as sustainable replacement for BPA. It can be easily produced by reacting levulinic acid with two moles of phenol at 100°C in acid conditions (Figure 23).[114, 338] Both of these starting chemicals can be produced from lignocellulosic biomass.[112, 113, 340] So, petroleum based BPA can be replaced with fully renewable and sustainable DPA, depending on its cheaper production from lignocellulosic biomass. DPA has also been attracting interest in the synthesis of poly(arylene ether ketone) (PAEK) and poly(ether ether ketone) (PEEK) based membranes for fuel cell applications. As a recent study, Zhou and Kim produced sulfonated PAEK (SPAEK) by polymerizing DPA with 4,4'-difluorobenzophenone and 5,5'-carbonylbis(2-fluoro benzenesulfonate) in the presence of DMSO and $K_2CO_3$. The intact carboxylic groups of the polymerized DPA served as pendant moieties so that they were then converted to unsaturated groups to obtain a series of cross-linked SPAEK (CSPAEK) membranes. The CSPAEK membrane with 80% sulfonation and 20% cross-linking degree exhibited much lower low methanol permeability and a comparable proton conductivity with respect to its commercial competitor; Nafion® 117.[341] Moreover, DPA is quite a useful monomer in the synthesis of hyperbranched polymers by providing tri-functionality; having two phenol and a carboylic acid groups. Foix *et al.* made use of this characteristic of DPA. They prepared a new hyperbranched-linear-hyperbranched polymer in a one pot process *via* polycondensation of DPA with poly(ethylene glycol) using DCC as coupling agent and 4-(N,N-dimethylamino)pyridinium-p-toluenesulfonate as catalyst (Figure 23).[342]

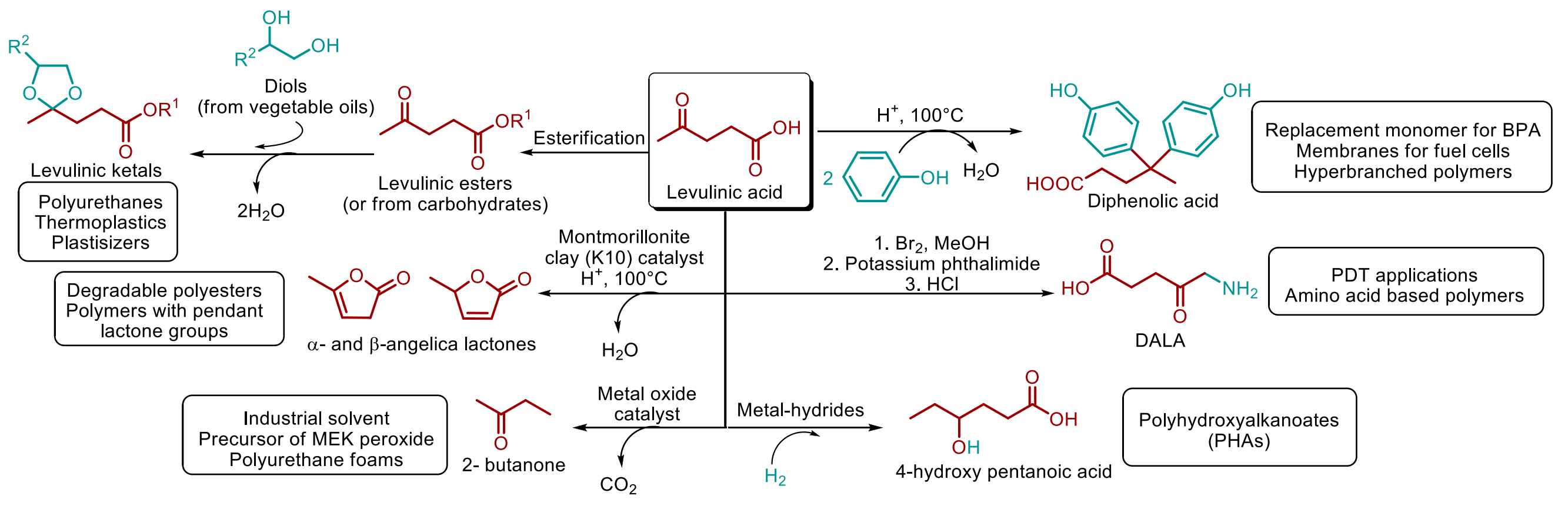

Figure 23. Synthesis and common applications of direct derivatives of LevA.

In addition to DPA and levulinic ketals, LevA can be directly converted to levulinic esters,[115] 5-aminolevulinic acid (□-aminolevulinic acid, DALA),[114] angelica lactones (ANLs),[44, 116] 2-butanone,[44] 4-hydroxy pentanoic acid or its esters,[22, 44] and gamma-valerolactone (GVL)[22, 117, 118] (Figure 3c). Common applications of these direct derivatives of LevA are summarized in Figure 23.

Amongst these derivatives, synthesis of DALA requires formation of a C–N bond at the C5 carbon. Although various amination methods have been successful on laboratory scale,[114] a process development is required for its industrial production. Particlularly, the laborious amination processes based on the bromination (Figure 23) must be replaced with a low cost and facile method. DALA is frequently used as an agent for photodynamic therapy. Fotinos *et al.* reviewed the use of DALA derivatives in photodynamic therapy (PDT) and fluorescence photodetection (FD) from a chemical, biochemical and pharmaceutical point of view. Two derivatives of DALA; methylaminolevulinate and hexylaminolevulinate, are marketed under the trade names of Metvix and Hexvix, respectively.[343] Macromolecules based on DALA were also reported for PDT applications. Battah and co-workers synthesized novel DALA containing well-defined dendritic molecules, in which, the DALA moieties were attached to the periphery by ester linkages which are hydrolyzable in cellular conditions.[344, 345] As being an amino acid itself, DALA can provide novel oppurtunities in amino acid based polymers. Such polymers were described in more detail in Section 5.5.

As being dehydration products of LevA (Figure 23), ANLs can potentially provide a route for producing functionalized aliphatic polyesters from renewable resources. In that respect, Chen *et al.* prepared a degradable polymer from □-ANL. The resultant polyester exhibited good degradability under light or acidic/basic circumstances owing to the presence of a C=C bond.[346] Hirabayashi and Yokota utilized □-ANL in a different manner. They carried out radical copolymerization of □-ANL with styrene in order to obtain polymers containing lactone units in the backbone chains.[347]

2-butanone (methyl ethyl ketone, MEK) (Figure 23), is commonly used as an industrial solvent. Although it is the precursor of MEK peroxide; a commonly peroxide used in polymer industry,[348] its direct utilization as a monomer is scarce. Nonetheless, Glowacz-Czerwonka obtained polyurethane foams from melamine solution in reactive solvents based on MEK and 4,4'-diphenylmethane diisocyanate.[349]

4-hydroxypentanoic acid (4-hydroxyvaleric acid) (Figure 23) is a monomer belonging to hydroxyalkanoates family. It can be polymerized into polyhydroxyalkanoates (PHAs) by chemical and/or biological methods. Detailed scrutiny regarding PHAs based on lignocellulosic biomass is provided in Section 5.16.

LevA can be hydrogenated to GVL by employing either homogenous or heterogeneous catalysts. So far, the best results were obtained with noble metal-based catalysts, particularly with Ru-based catalysts. However, the utilization of expensive noble metal-based catalysts prevents scaling-up the GVL production process. Replacement of noble metals by more widely available cheap metals as well as recovery and recycling of non-precious metal-based catalysts are important issues in preventing GVL to become a commodity chemical. The challenges associated with industrial production of GVL were highlightened in detail by Wright and Palkovits.[350] Conversion of LevA to GVL holds great potential because GVL is a valuable fuel and green solvent as well as it is a precursor for other value-added chemicals as shown in Figure 3c.[22, 117, 118] Alonso *et al.* reviewed the upgrading of lignocellulose derived GVL to various chemicals and fuels, including polymers, fuel additives, and jet fuels.[22] Jedlinski *et al.* synthesized homopolymer of GVL with alkali metal alkoxides and alkali metal supramolecular complexes based initiator systems to obtain polymeric materials with potential biodegradability.[351] Its copolymerization with various monomers, including diglycidyl ether of BPA,[352, 353] -CL,[354] -butyrolactone,[355] and L-lactide[356] were also studied (Figure 24). Over basic catalysts, gas phase catalytic condensation of formaldehyde with GVL yields -methylene--valerolactone (MeGVL). There are several problems regarding this chemistry. First of all, it is very difficult to handle formaldehyde in the gas phase because it polymerizes very rapidly. MeGVL is also a reactive monomer so that it polymerizes on standing. Fortunately, dissolving GVL in formalin (37 wt% aqueous solution) works well and a MeGVL yield of more than 95% is obtained using barium-based catalysts supported on silica.[119] MeGVL is a new and an attractive acrylic monomer that imparts high thermal stability to polymers.[119] Naturally renewable MeGVL has been attracting interest in exploring the prospects of substituting the petroleum-based methacrylate monomers for specialty chemicals production. The cyclic ring in MeGVL, imparts significant enhancements such as enhanced resistance to heat, solvent and scratch in the materials properties of its polmers.[357] For instance, Tg of poly-MeGVL (Figure 24) is about 200°C, which is about 100°C higher than that of its acyclic counterpart, PMMA. However, the absence of an economically attractive catalytic process prevents the commercial development of MeGVL.[119] Particularly, its synthesis through LA-GVL-MeGVL pathway must be optimized accordingly for its large scale production. Achieving this pathway in one pot would be more desirable for cutting-off extra synthesis steps and thus, the production costs.

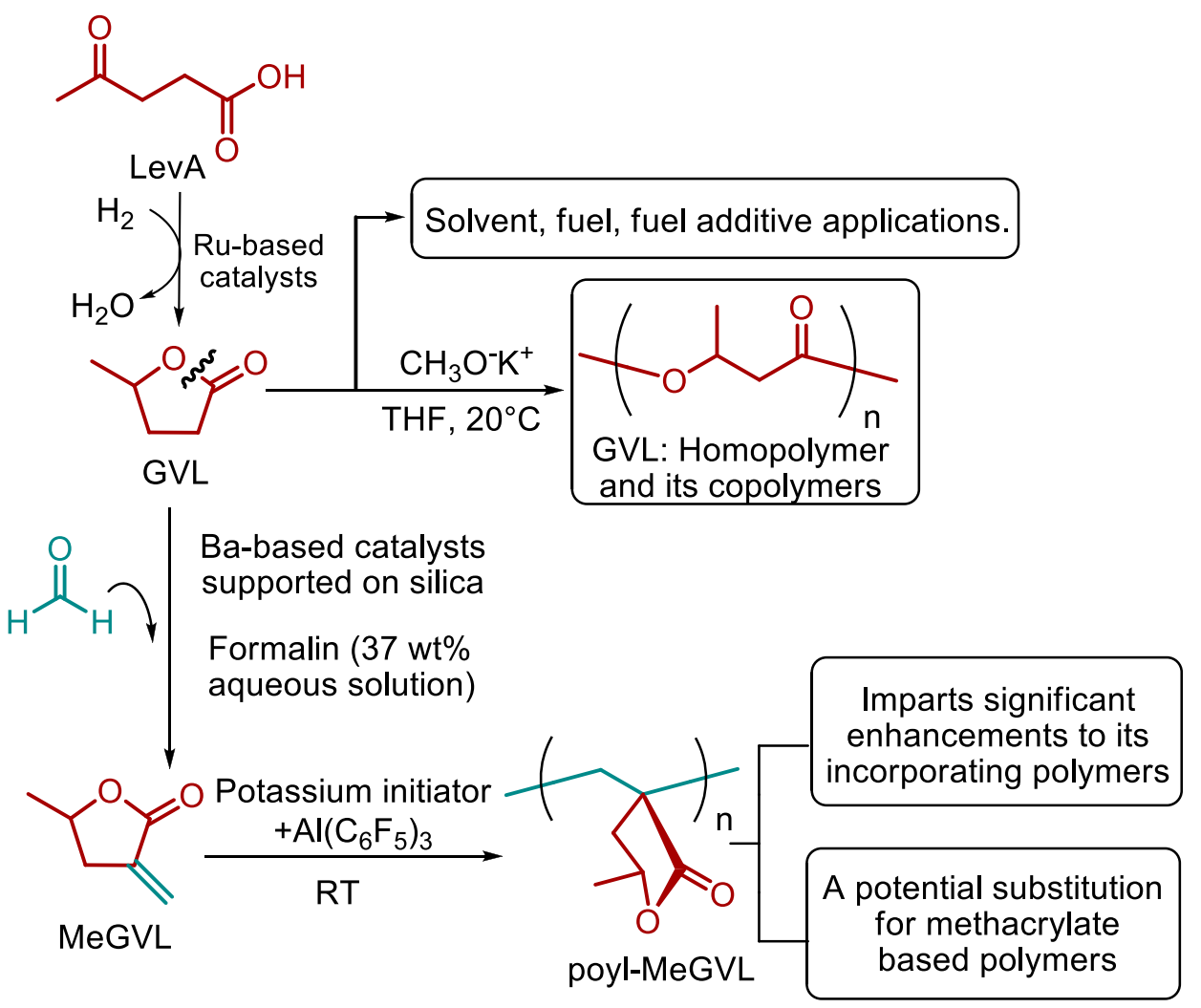

Figure 24. Applications of GVL and MeGVL.

Ring-opening of GVL is a new versatile route to produce polymer precursors with relatively high molecular weights (Figures 3c and 25).[358] In the presence of an amine, GVL ring can be opened to give the corresponding γ-hydroxy(amino)amide compounds. Chalid *et al.* synthesized novel GVL derived -hydroxy-amide monomers through the addition of various amino compounds. Yields of the final products ranged between 22-95%. It was shown that sterically less hindered amine compounds can effectively result in ring opening of GVL even without the use of a catalyst and solvent. Hence, GVL ring opening by amines, which do not have steric hindrance around their nucleophilic nitrogen centre, is a promising pathway for production of novel monomers.[122] This new family of monomers can be employed to produce polymers such as polyethers or polyurethanes.[22] For instance, GVL/1,2-aminoethanol adduct was polymerized with 2,4-toluene-di-isocyanate (TDI) at 140°C by using TEA as the catalyst and DMA as the solvent. A polymer with a molecular weight of 156 KDalton was produced in 97% yield.[358] Also, the hydrogenation GVL can be performed at 40°C to yield 1,4-pentanediol (1,4-PDO) over an inexpensive Cu catalyst, which was calcinated in the presence of $H_2$, with a selectivity of over 98% at complete conversion.[22] Although polycondensation of 1,4-PDO is rarely reported, it has a similar structure to 1,4-BDO so that it can be largely employed in synthesis of polymers. As such, Rothen-Weinhold *et al.* used 1,4-PDO in the synthesis of poly(ortho ester)s; hydrophobic and bioerodible polymers that have been investigated for pharmaceutical use. In the study, poly(ortho ester)s, prepared from 3,9-diethylidene-2,4,8,10-tetraoxaspiro[5.5]un decane, 1,4-PDO and 1,6-hexanediol glycolide, were used as agents for Bovine serum albumin (BSA) protein delivery.[359, 360] Acid catalyzed ring opening of GVL using a $SiO_2/Al_2O_3$ catalyst at 225-375°C results in the formation of an isomeric mixture of pentenoic acids.[120, 121, 123] Amongst these isomers, 4-pentenoic acid (4-PA) can be utilized in both radical and condensation reactions owing to its dual functionality (vinyl and carboxylic acid groups). An *et al.* employed this characteristic of 4-PA in the production of UV-induced thiol-ene crosslinked films. In the study, well-controlled pendant hydroxyl containing copolymers of 2-hydroxyethyl methacrylate (HEMA) and methyl methacrylate (MMA) were synthesized in the presence of benzyl -bromoisobutyrate initiator. The pendant –OH groups were then converted to pendant enes by their coupling with the carboxyl group of 4-PA. Under UV radiation, the resulting pendant ene groups can undergo thiol-ene polyaddition reactions with polythiols to form crosslinked films with a uniform network.[361] Lange *et al.* produced another ring-opened derivative of GVL; methyl penteonate, in more than 95% yield *via* a transesterification method.[127] The process basically relies on the large boiling point difference between GVL (207°C) and methyl pentenoate (127°C). In the process, methanol was slowly fed into the reaction flask containing GVL, and *para*-toluene sulfonic acid (pTSA) as the catalyst while boiling the reaction medium at 200°C. A mixture of methyl penteonate, methanol

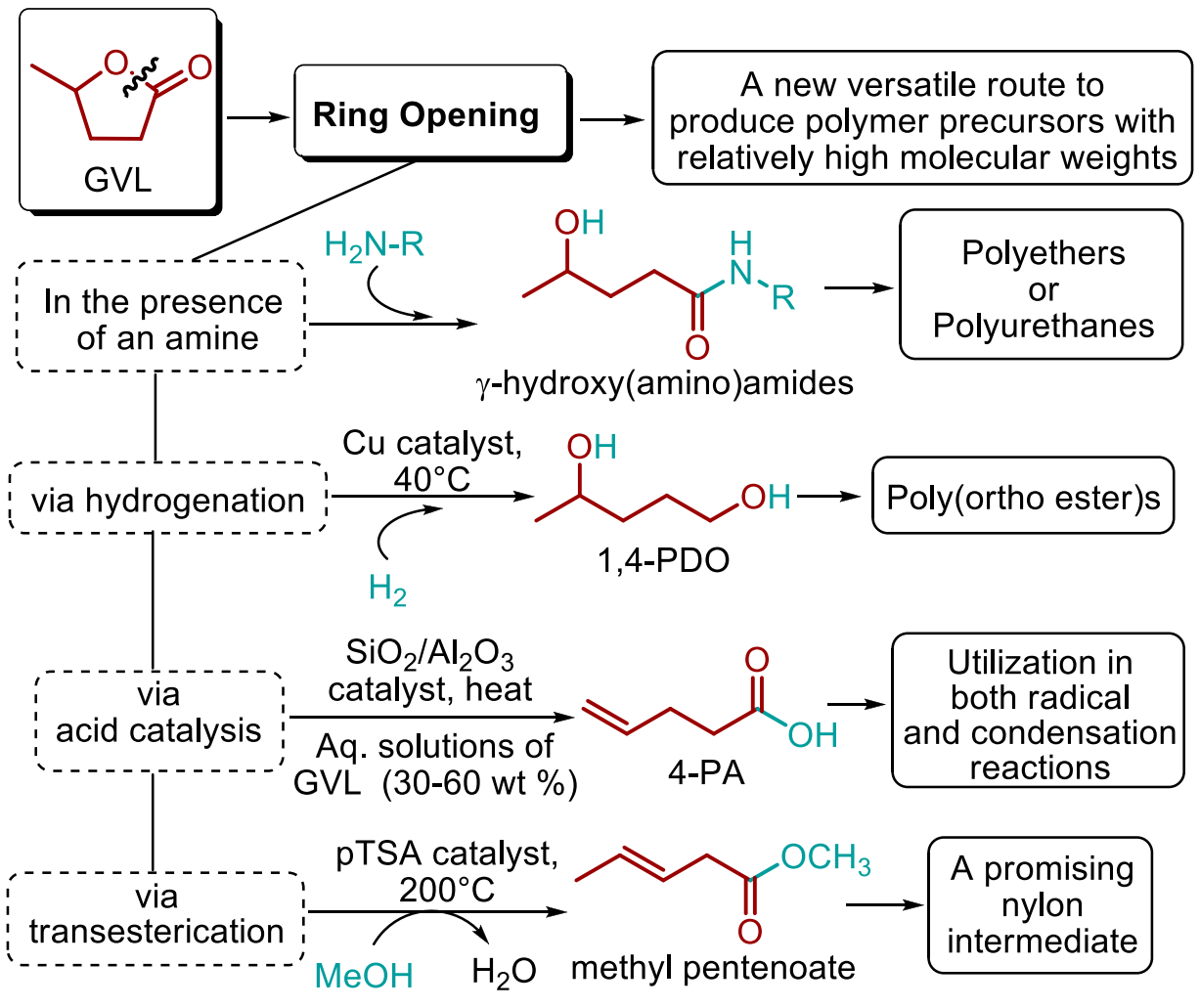

Figure 25. Ring opening of GVL as a new versatile route to produce polymer precursors.

and water was distilled of continuously, and then condensed at 90°C at the top of a rectification column. Finally, methyl pentenoate was collected over consecutive distillate fractions. Methyl pentenoate is considered as a valuable PA precursor because it can be converted to □-CL, caprolactam, or adipic acid by hydroformylation, hydrocyanation, or hydroxycarbonylation reactions, respectively. So, commercialization of the methyl pentenoate production through LevA-GVL-methyl pentenoate pathway may open new oppurtunities for production of conventional PAs (Figure 25).[22, 362]

Levulinic acid platform can also open novel synthetic routes for production of giant monomers such as butene and adipic acid (ADA) (Figure 3c). In this consideration, decarboxylation of GVL derived pentenoic acids on $SiO_2/Al_2O_3$ catalyst at 375°C result in the formation of butene isomers, including 1-butene (1-B).[121] In fact, GVL ring opening on acid catalysts at high temperatures first gives an isomeric mixture of pentenoic acid, and subsequently, an isomeric mixture of butenes. So, selective synthesis of 1-B starting from GVL is not feasible under these conditions. A better route for the production of 1-B would be its synthesis from n-butanol. Cobalt And The Naval Air Warfare Center recently teamed up for scaling up and optimizing the dehydration chemistry for the conversion of non-food feedstock based bio n-butanol to 1-B.[363] Polymerization of 1-B by using supported Ziegler-Natta catalysts results in its high molecular weight homopolymer; poly(1-butene) (PB-1). PB-1 exhibits excellent creep properties and its major applications are summarized in Figure 26.[270] 1-B is also used as a comonomer in the production of sine qua non polymers of today's modern society such as linear low density polyethylene (LLDPE), LLDPEs based on the metallocene catalyst technology (mLLDPE), very low density polyethylene (VLDPE) and isotactic polypropylene (i-PP).[270, 364, 365] Kraton Polymers Company employs 1-butene in the production of linear styrene-(ethylene-butylene)-styrene triblock copolymer series under the trend name of Kraton G1600 SEBS (Figure 26).[270]

GVL derived pentenoic acids can be carbonylated to produce ADA over palladium catalyst for 5 h under 60 bar of CO at 105°C (Figure 26). However, poor yields of 22-48% are obtained. For further improvement of the reaction yields, more information on the kinetics and the thermodynamic constraints of the process is required.[124] In fact, there are many other possibilities for production of ADA. There is a growing research in both industry and academia in the development of bio- and chemocatalytic routes for ADA from other biorefinery building blocks such as 5-HMF, D-glucose, compounds representative of lignin, and lignin-derived bio-oils. Van de Vyver and Roman-Leshkov stated that the production of ADA particularly from lignocellulosic biomass-derived chemicals could provide an even more sustainable route, instead of its petroleum-derived production route.[124] In the end, these bio-based routes for ADA production have to be cost-competitive with its current industrial manufacture process, which is based on catalytic oxidation of a mixture of cyclohexanol and cyclohexanone with nitric acid.

The global market size of ADA is projected to be 6 billion pounds by 2017.[124] Since it has such a huge market, many start-up companies such as Rennovia, Verdezyne, BioAmber, Celexion, and Genomatica have been engaged in developing bio-based routes to produce ADA. Some of them have already reached advanced pilot or demonstration scales. Especially, the Rennovia's process was suggested to be highly cost competitive with the conventional DuPont/Invista cyclohexane-based oxidation process.[124] The ultimate motivation of these companies for synthesis of bio-ADA is to produce 100% bio-based PAs. In this consideration, ADA is a critical monomer. About 85% of the total produced ADA is used for the production of PA-6,6.[366] Another important commercial polymer of ADA is PA-4,6. After the polyamides, esters and polyesters comprise the second most important class of ADA derivatives.[367] Esters of ADA can be obtained from alkylpenteonates,[127] or as another route, ADA can be readily reacted with alcohols to give either its mono- or diesters.[367] Moderately long-chain esters of ADA find large applicability as plasticizers, particularly for poly(vinyl chloride) (PVC).[367] As an example, BASF manufactures and markets di-2-ethylhexyl adipate and di-isononyl adipate derivatives of ADA under the trend name of Plastomoll® (Figure 26).[368]

### 5.10. 3-Hydroxybutyrolactone Platform Based Polymers.

3-hydroxybutyrolactone (3-HBL) is a versatile chiral building block (Figure 3d) having applications in the synthesis of a variety of pharmaceuticals, polymers and solvents. Enantiopure 3-HBL is a significant precursor for chiral drugs such as the cholesterol-reducing statins like Lipitor® (Pfizer) and Crestor® (AstraZeneca), antibiotics such as carbapenems and linezolid (Zyvox®), and the anti-hyperlipidemic medication Zetia®. Both enantiomers of 3-HBL can be used in the synthesis of L-carnitine; the nutritional supplement. Moreover, the functional groups of enantiopure 3-HBL can be derivatized to yield chiral building blocks such as lactones, THFs, amides, nitriles, epoxides, and solvents.[40, 109, 369, 370]

Owing to its versatile applicability, the U.S. Department of

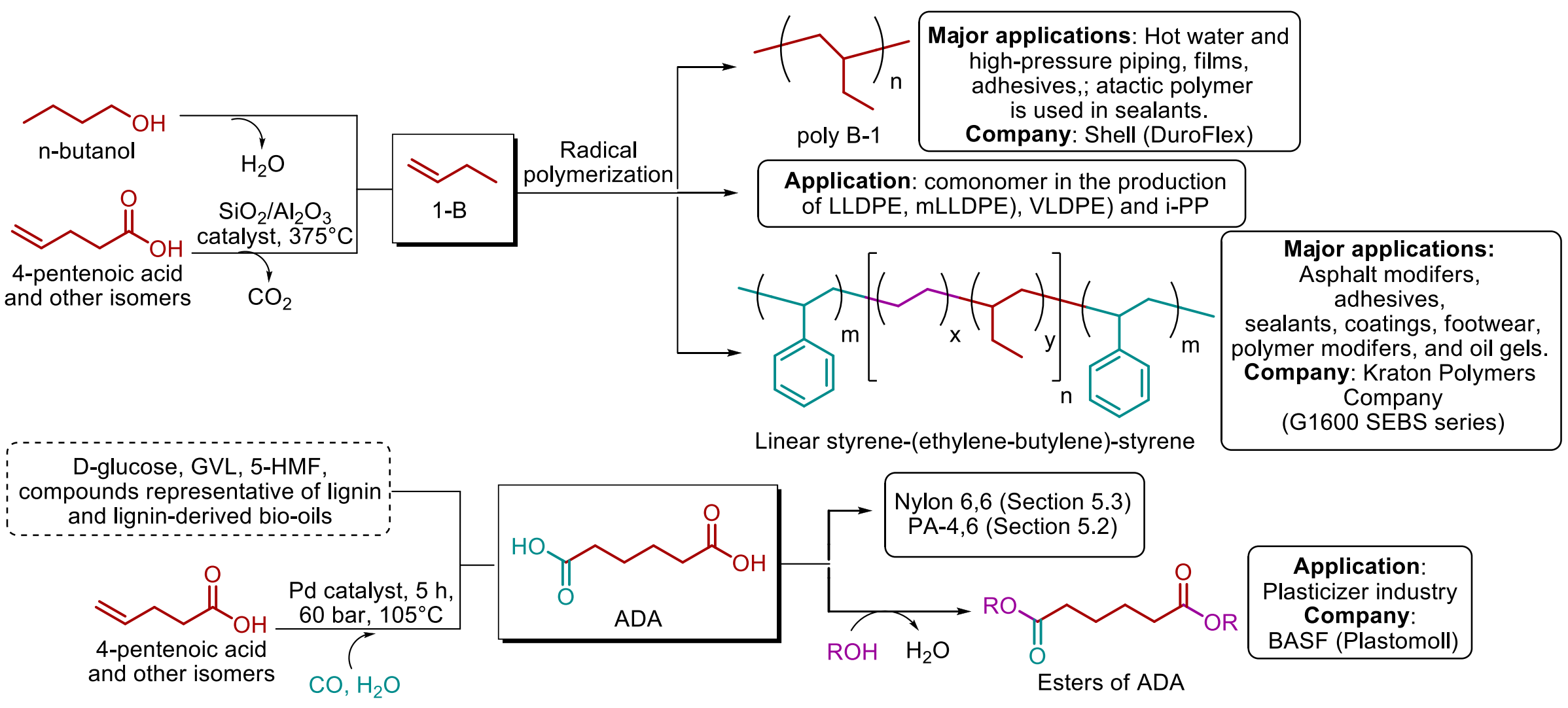

Figure 26. Major applications of 1-B and ADA.

Energy proposed 3-HBL as a top value-added chemical from biomass. Even so, there is not a facile and low-cost chemical pathway for its large scale synthesis. For instance, the commercial synthesis of (S)-3-HBL is performed through a continuous chemical synthesis process using high pressure hydrogenation of L-malic acid over a Ru-based catalyst. However, the process consists of expensive Pd catalyst and purification steps as well as hazardous processing conditions. The other various chemical and chemoenzymatic synthesis routes of 3-HBL also consist of multiple steps and expensive catalysts. That is why; 3-HBL currently has an expensive wholesale cost of ~$450 per kg and this alleviates 3-HBL from being a commodity chemical.[109]

Since chemical synthesis of 3-HBL is messy, a special attention should be given to its biological production routes. Even so, there are no known naturally occurring biosynthetic pathways for 3-HBL. Here, the design of a novel pathway through microbial engineering holds a great importance. Recently, Martin *et al.* presented such a promising platform pathway for 3-hydroxyacid synthesis, which provides a biosynthetic route to 3-HBL, for the first time. The pathway results in the complete biological production of 3-HBL and its hydrolyzed form, 3,4-dihydroxybutyric acid (3,4-DHBA), in recombinant *Escherichia coli* by using glucose and glycolic acid as feedstocks. It was also stated that direct production of 3,4-DHBA/3-HBL from glucose requires integration of the endogenous glyoxylate shunt with the 3,4-DHBA/3-HBL pathway. Engineering of this integration was described in detail in their recent article.[109, 369]

Once large scale production of 3-HBL *via* a biosynthetic route is achieved, it can be converted to a variety of fine chemicals by employing mainly three different chemistries; (i) functionalization of the –OH group, (ii) dehydration of the –OH group, and (iii) ring opening of the cyclic ester. Particularly, functionaliztion of the -OH group of 3-HBL can open new windows to novel acrylic monomers. As such, Murata *et al.* patented a process for converting 3-HBL to □-(meth)acryloyloxy-□-butyrolactones (B(M)AL-GBLs) by reacting 3-HBL with (meth)acrylic acid chloride, (meth)acrylic acid or (meth)acrylic ester in a simple and safe manner (Figure 27). These acrylic derivatives of 3-HBL are stated to be useful as a constituent monomer of paints, adhesives, sticking-agents, and resins for ink. Despite their expected use for various purposes, they have never been produced industrially due to a high risk of explosion in the chemical synthesis of their parent molecule; 3-HBL.[111] But now, as previously mentioned, achievements in the production of 3-HBL *via* fermentation routes can solve this problem.[109, 369] B(M)AL-GBLs have been finding applicability particularly in resist compositions and patterning processes.[371] As a different synthetic route, the -OH group of 3-HBL also provides functionality for its dehydration to the corresponding unsaturated lactone derivative; 2(5H)-furanone. The dehydration reaction can be easily performed in the presence of an acid, such as polyphosphoric acid, and under vacuum distillation (160°C).[110] Once obtained, 2(5H)-furanone can undergo a facile reaction with a primary amine by a Michael-type addition to give □,□-amino-gamma-lactone, which subsequently polymerizes to yield a polyamide with pendant R groups. In the end, many different types of polyamides can be produced starting from 3-HBL depending on the structure of the side R group (Figure 27).[268]

## 5.11. Sorbitol Platform Based Polymers.

Amongst similar polyols, sorbitol (SBL) (Figure 3b) is the least costly and the most commonly used sugar alcohol. It has the biggest market share and it is extensively used as sweetener, thickener, humectant, excipient, dispersant in food, cosmetic, toothpaste and other related industries. Owing to its large market volume, many companies have involved in its commercial production. Recently, Roquette Freres is the biggest SBL producer and it shares over 70% of the total market volume together with Cargil and SPI Polyols.[372, 373] Currently, no tecnical development is essentially needed for the use of SBL as a building block because its production is commercially practiced by several companies and the yields demonstrated are about 99%. It is suggested that the only necessary change requires a shift from batch process to continuous process.[40] SBL's industrial production begins with raw materials such as corn, cassava and wheat. Through enzymatic hydrolysis, these raw materials are first converted to glucose, and then the hydrogenation of glucose at 403−423 K with $H_2$ pressure ranging from 4.0 to 12.0 MPa provides the production of SBL. Also, there is a growing interest in one-pot conversion of biomass into SBL.[106, 372] Very recently, Zhu *et al.* reported efficient hydrogenolysis of cellulose into SBL in one-pot in the presence of a bifunctional, sulfonic acid-functionalized silica-supported ruthenium, catalyst. When the reaction was performed at 150°C for 10 h, a maximum yield of 61.2 % was obtained.[374] Such direct conversions of raw materials into SBL are more desirable for further decreasing current production prices of SBL. Nonetheless, cheap catalysts must be employed instead of noble metal-based catalysts as well as the reaction yields must be ~99%. In this consideration, research studies should be directed on the development of cheap and recyclable multifunctional catalyst systems for one-pot synthesis of SBL.

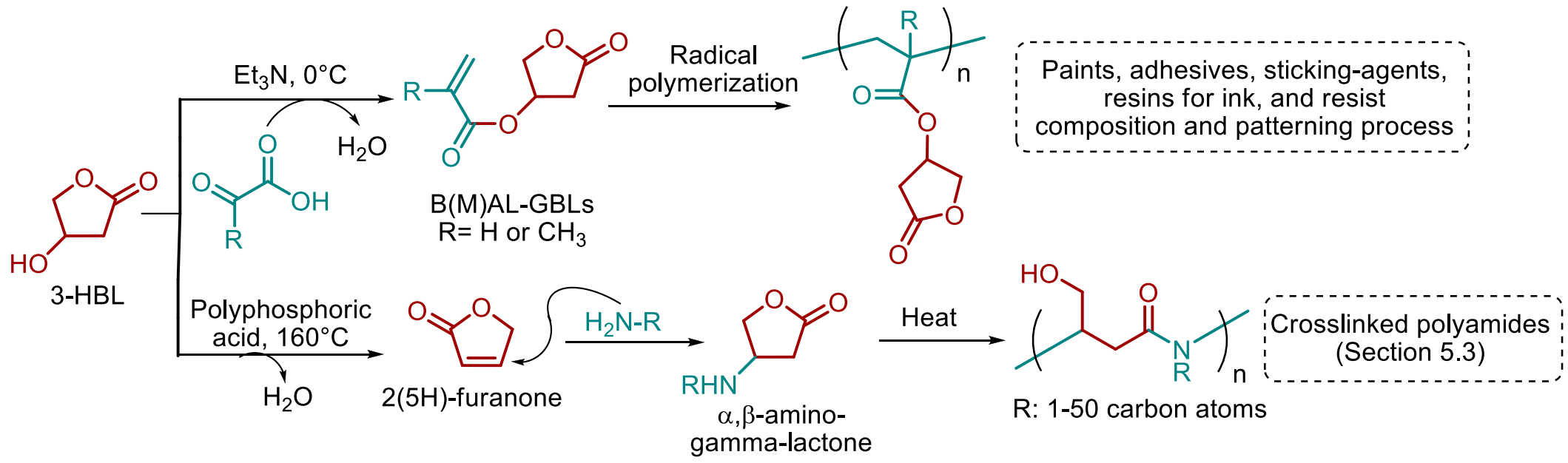

Figure 27. Polymer applications of 3-HBL derived B(M)AL-GBLs and 2(5H)-furanone.

SBL is an outstanding building block for achieving sustainable energy supply, chemicals and polymer production. It is industrially converted to vitamin C (ascorbic acid) by fermentation with or without chemical processes through sorbose and 2-ketogluconic acid as intermediates.[37] This process consumes almost 15% of world SBL production. As a different chemistry, 2-fold dehydration of SBL in the presence of sulfated copper oxide catalyst at 200°C results in the formation of isosorbide (IS).[107] In this process, cyclodehydration of SBL first forms 1,4-sorbitan and 3,6- sorbitan intermediates, and afterward the dehydration of these intermediates yields IS (Figure 3b). One-step conversion of lignocellulose feeds to IS is also possible with designed reaction systems.[37, 372] In a recent study, Beeck *et al.* reported the catalytic production of IS from a diverse range of lignocellulosic biomass feedstocks by combining the heteropoly acid; $H_4SiW_{12}O_{40}$, with the redox catalyst; commercial Ru on carbon.[375] Apart from these, SBL hydrogenolysis with multifunctional catalysts can constitute a major route for the synthesis of lower alcohols such as glycerol, propylene glycol, ethylene glycol, ethanol and methanol. It is worthy to note that these lower alcohols have enormous market volumes and they can be further utilized to obtain many other high value-added products. Hence, SBL hydrolysis is expected to be a major research field in which the development of cheap and recyclable multifunctional catalysts as well as mild and facile reactions deserves priority. Repeated cycling of dehydration and hydrogenation reactions can also open up a new field in SBL chemistry, particularly for the production of $H_2$ and alkanes. Detailed information regarding the transformation of SBL to biofuels, in the frame of chemical and industrial considerations, is provided in the review article of Vilcocq *et al.* [37, 372, 376]

Polymer production from SBL or its derivatives constitutes another important building block consideration. Copolymerization of SBL with other glycols would create a major oppurtunity in the unsaturated polyester resin market.[40] Roquette recently introduced sorbitol-based polymer clarifiers under the trend name of Disorbene®; a bis(3,4-dimethylbenzylidene) sorbitol (bis-DMBS) product.[377] Bis-DMBS can be produced by heating 3,4-dimethyl benzaldehyde and D-sorbitol in the presence of a bronsted acid catalyst (Figure 28).[378] As being a linear monosaccharide, SBL can potentially be employed for production of a wide range of polymer families through its selective functionalization (Figure 8-Section 5.1). The most recent research studies (Table 5) reveal that SBL can be employed in a variety of polymerization reactions. Hence, SBL-based SBL-based polymers can have a wide range of applicability ranging from biodegradable polymers for everyday applications to specialty products including biocomposites and biomedicines.

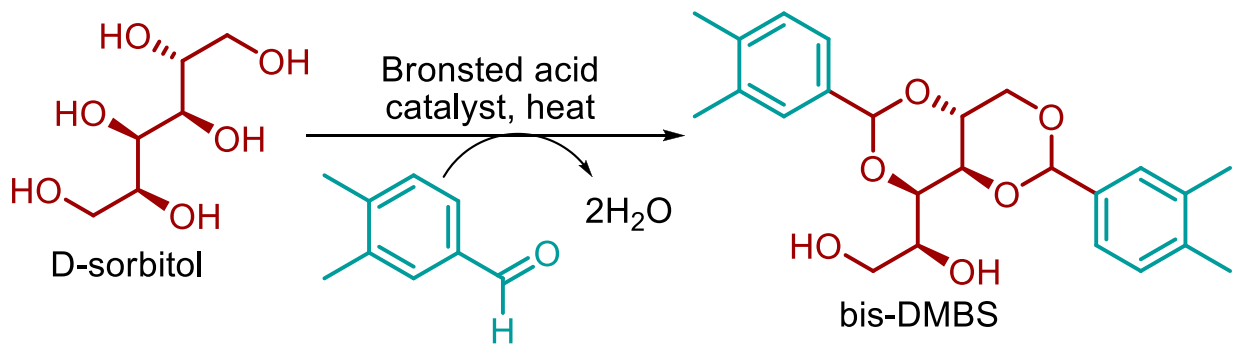

Figure 28. Synthesis of SBL derived bis-DMBS for polymer clarifier applications.

As a selective dehydration product of SBL, IS (Figure 29) imparts superior properties to its incorporating polymers by providing rigidity and chiral centers. [372] For instance, unlike the conventional PET having a Tg of 353 K, poly-(ethylene-co-isosorbide) terephthalate (PEIT) (Figure 29) can show a much higher Tg (up to 470 K) by increasing the ratio of IS to ethylene glycol. Hence, PEIT polyesters are quite useful for hot-fill bottles. Among aliphatic polyesters, poly(isosorbide oxalate) (Figure 29) also has a remarkably high Tg of 445 K. Moreover, it can exhibit good biodegradable properties.[37, 326] Recently, Naves

**Table 5. The most recent research studies on sorbitol-based polymers and their applications.**

| Monomers | Method | Polymer Structure | (Potential) Applications | Ref. |
|---|---|---|---|---|
| SBL, divinyl adipate | Enzymatic polycondensation *(Candida antarctica)* | 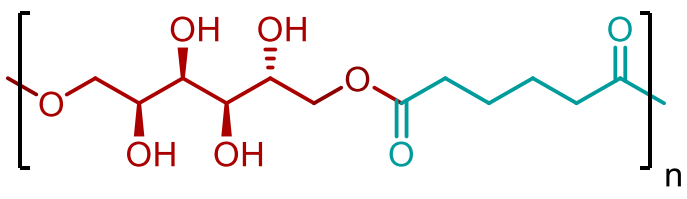 | Self-assembled nanostructures, super-soft elastomers, biomedicine | 379 |
| SBL, 1,2,3,6-tetrahydro phthalic anhydride, adipic acid and diethylene glycol | Melt condensation | 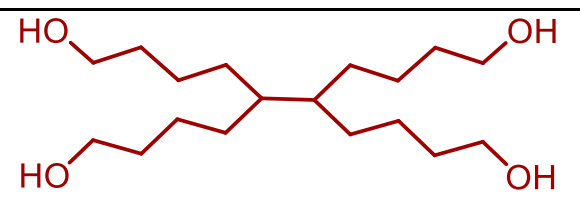 Representative structure | Polyurethane coatings (after reacting with polyisocyanates) | 380 |
| SBL polyglycidyl ether, quercetin or phenol novolac | Thermal polymerization and then, compression molding | The structure was not provided. | Biocomposites | 381 |
| SBL dimethacrylate, polyethylenimine | Michael addition | 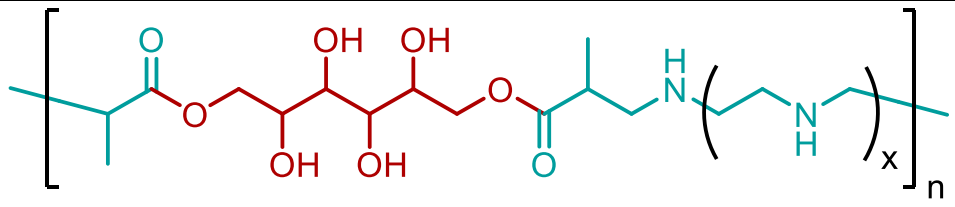 | Gene carrier, osmotically active transporter | 382 |
| SBL, citric acid (or tartaric acid), sebacic acid | Catalyst-free melt condensation | 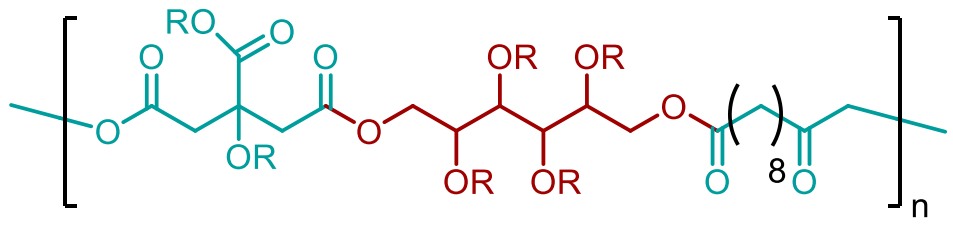 | Biodegradable polymers for various applications | 383 |
| SBL methacrylate | Surface-initiated polymerization | 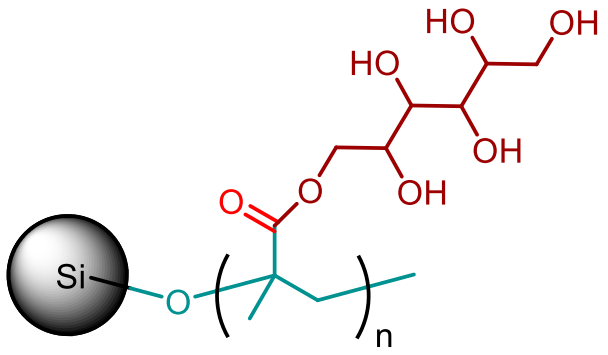 | Hydrophilic interaction chromatography | 384 |
| SBL, L-aspartic acid | Thermal polycondensation (no catalyst) | 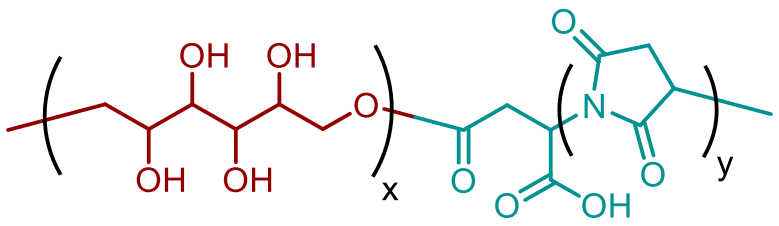 | Biocompatible gels, absorbents, controlled release matrices | 385 |

*et al.* reported enzymatic copolymerization of IS with diethyl adipate and different unsaturated diesters including diethyl itaconate, diethyl fumarate, diethyl glutaconate, and diethyl hydromuconate.[386] The study is specifically important for providing insights on unsaturated polyester production from fully renewable monomers. Another IS incorporating sustainable polyester production was reported by Gioia *et al.* The sustainability stems from a novel method which combines the chemical recycling of poly(ethylene terephthalate) with the use of IS and succinic acid derived from renewable resources.[387] Other IS based polymer families can also exhibit unique characteristics. As such, poly(isosorbide carbonate) (PIC, Figure 29) is highly transparent, heat-resistant and water-tolerant. As being a bio-based and harmless plastic, PIC is identified as a promising alternative to the polycarbonate derived from BPA. Hence, the commercial production of PEIT and PIC can boost up in the future since their petroleum derived counterparts; PET and PC, have an annual consumption capacity of about $5\times10^7$ and $3.5\times10^6$ t, respectively. Owing to the current interest in IS based polymers, various companies have been involving in the commercialization of IS based polymeric materials. Mitsubishi and Teijin chemical companies launched mass production of PIC in 2012 and 2011, respectively. Roquette also constructed a plant in 2007 for the manufacture of polymer-grade IS. The company also produces IS diesters, as phthalate-free plasticizers, under the trend name of Polysorb® ID (Figure 29). The diesters are produced as substitutes for conventional petroleum-based plasticisers such as dioctyl phthalate and particlularly for the production of a flexible polyvinyl chloride. Many other kinds of IS polymers have been studied in recent years and detailed information about them was provided in the review articles of Fenouillot *et al.*, and Rose and Palkovits.[37, 388, 389]

SBL derived sorbitans[108] (Figure 3b) also could potentially be employed in polyester production.[325] Besides, esterification of sorbitan with fatty acid methyl esters, which is followed by PEG-ylation, results in commercially avaliable polysorbates (Figure 29). These products are commonly sold as emulsifiers and solubilizers in cosmetic, pharmaceutical, and food industries under different trend names such as Tween, Span, Arlacel and Crill.[37, 390]

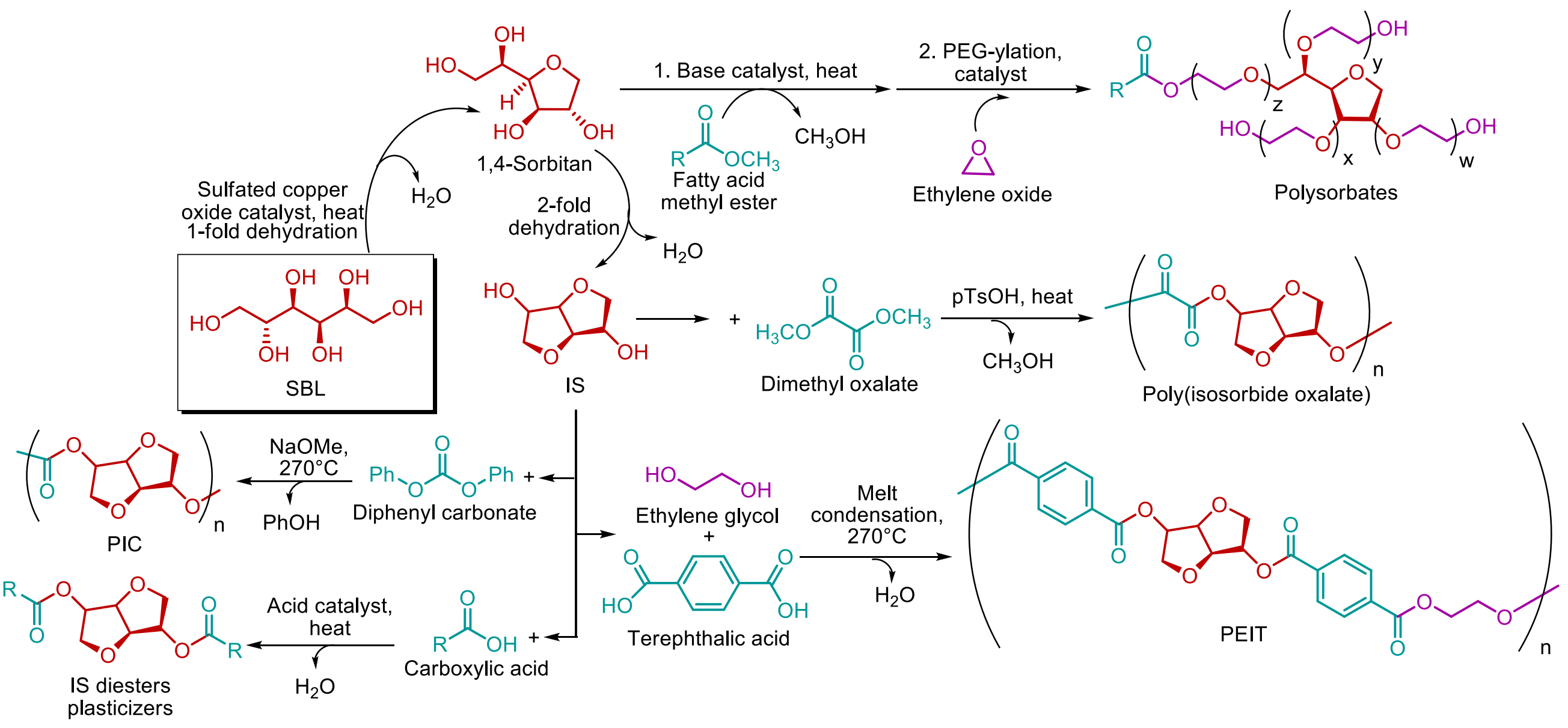

Figure 29. Common polymers based on SBL dehydration compounds; IS and sorbitan.

### 5.12. Glycerol Platform Based Polymers.

Microbial production of GLY has been known for about 150 years. Significant improvements have been made in the GLY production using osmotolerant yeasts on a commercial scale particularly in China. Microbial fermentation has been supplying more than 12% of the country's needs. More detailed information regarding GLY production by microbial fermentation was provided in the review article of Wang *et al.*[90] On the other hand, the fast growth in biodiesel industry has led to overproduction of GLY worldwide. About 90% of GLY has been recently produced as a by-product in the conversion of fats and oils to fatty acids or fatty acid methyl esters for biodiesel.That is why, the other production routes of GLY such as fermentation of sugar or hydrogenation of carbohydrates are not industrially important.[391] Instead of this, the main consideration has currently been focused on the valorization of overproduced GLY. Both industry and academia has been putting a lot of effort into conversion of GLY to other commodity chemicals and polymers. So, the global demand for GLY as a platform chemical is also expected to extend in parallel with its production capacity. In fact, this is confirmed by the recent report published by Transparency Market Research. According to the report, the global demand for glycerol (GLY) was 1,995.5 ktons in 2011. This demand is expected to reach 3,060.4 kilo tons growing at a compound annual growth rate (CAGR) of 6.3% from 2012 to 2018. In terms of revenues, the demand is predicted to reach $2.1 billion by 2018.[392]

There are more than 2000 different applications of GLY which arise from its unique properties. Its main application areas include cosmetics, pharmaceuticals, food, tobacco, plastics and resins. About 20% of the totally produced GLY has been utilized in plastic and resin applications.[391] GLY is commonly used as a polyol for production of alkyd resins; synthetic resins made from polyhydric alcohols and polybasic acids, and modified with resins, fatty oils, or fatty acids. The largest use of alkyd resins is their utilization as binders for surface coatings. Alkyd resins are the most versatile of coating binders and thus, they can be extensively used in all major categories of coatings. Other uses of alkyds include ink binders, caulks, adhesives, and plasticizers.[393] As main by-product of biodiesel, crude GLY is a promising and abundant feedstock for industrial microbiology. Without any pretreatment step, it can be used to produce polyhydroxyalkanoates with an economically competitive overall production process.[394] Many different copolymers of GLY including poly(glycerol sebacate), poly(glycerol methacrylate)s and poly(glycerol-co-□-caprolactone) were reported for various applications in the literature. It is out of scope of this article to summarize countless GLY based polymeric materials and their applications. Detailed scrutiny is provided in review articles on polyglycerols (PGs).[395-398]

Research efforts have been directed to make use of the large surplus of GLY by introducing a number of selective processes for its conversion into commercially value-added products (Figure 3a). Hence, GLY is considered a platform-central raw material in future chemical industry.[399, 400] Particularly, its hydrogenolysis opens an alternative route to the industrial production of its diol derivatives; 1,3-PDO, propylene glycol and ethylene glycol,[93-95, 399, 400] which are indispensible monomers of today's modern industry. Selective hydrogenolysis of GLY and the resultant diol yields are highly dependent on the used catalyst. Nakagawa *et al.* reported that conventional-unmodified catalysts generally results in formation of propylene glycol as the main product, whereas; Pt–$WO_3$ catalysts on Al-based supports and Ir–ReOx/$SiO_2$ catalysts are specifically favors production of 1,3-

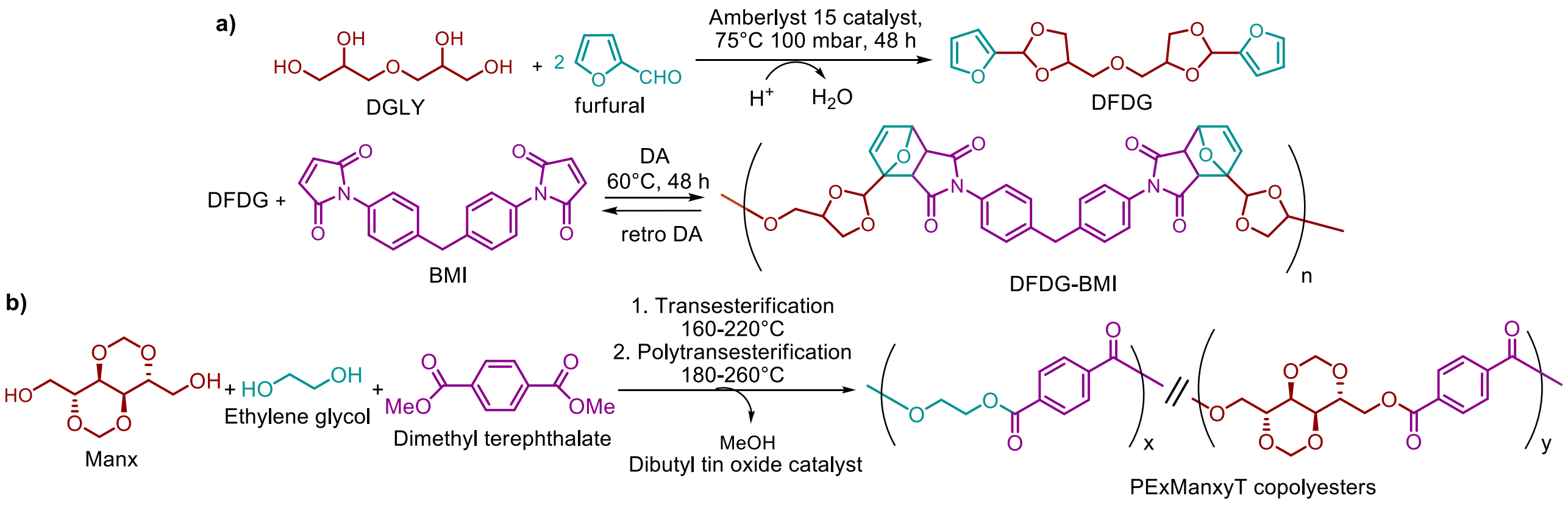

Figure 30. a) Thermo-reversible DA polymerization of DFDG and BMI. b) Production of mannitol based PExManxyT copolyesters.

PDO.[93] Further hydrogenolysis of GLY is also critically important since the resultant products are bio-based fuels; 1-propanol (1-PO) and propane.[98] As a common fuel, production of propanol is currently too expensive.[401] Its production routes from lignocellulosic biomass might decrease its current market price. As a promising study, Zhu *et al.* converted biomass-derived GLY to biopropanols in one-step over bi-functional catalysts. Especially, the conversion of GLY to 1-PO in the presence of Pt–HSiW catalyst supported over $ZrO_2$ is quite promising. The reaction proceeded with a conversion of 99.7% and a selectivity of 80.0%.[402] However, the above-mentioned chemical transformations of GLY employ expensive noble-metal catalysts. For industrial applicability, research studies should be directed on cheap and recyclable catalyst systems. Apart from these, Deng and Fong reported direct conversion of untreated lignocellulosic biomass to 1-PO by using an engineered strain of the actino bacterium, *Thermobifidafusca*.[403] Such direct conversions of raw lignocellulosic biomass to target compounds are more desirable. Of course, commercialization of such shortcuts will depend on the production costs and efficiencies.

As a different way of valorization, GLY can be converted to its short lenght oligomers (Figure 3a). GLY-derived oligomers have GLY like applications so that they are also widely used in cosmetics, food industry, and in polymer and plastics industries. An important GLY oligomer is its dimer; diglycerol (DGLY). Heterogeneous catalysts exhibit better selectivities for oligomerization of GLY to DGLY. As such, Amberlyst catalysts provide DGLY selectivities up to 85%, however; the conversion of GLY remains very low (35-40%). Hence, development of heterogeneous catalysts with very high GLY conversion efficiencies is essential for selective oligomerization of GLY.[91, 400] Once selectively obtained, DGLY can be extensively used in the manufacture of polyurethanes and polyesters. Its industrial applications include the production of plasticizer in polyvinyl alcohol films or starch-based biodegradable thermoplastic compositions.[404] Satoh *et al.* have recently employed DGLY in a different manner. They produced a novel monomer; difurfurylidene diglycerol (DFDG), by the acetetalization reaction of DGLY and furfural in the presence of Amberlyst 15 catalyst at 75°C under a reduced pressure of 100 mbar for 48 h. The new DFDG monomer can be fully bio-based since both DGLY and furfural are bio-derivable compounds. A bio-based linear polyimide with Mw of 5400 was then obtained as a result of Diels-Alder (DA) polymerization of DFDG with 4,40-bismaleimidodiphenylmethane (BMI) at 60°C for 48 h (Figure 30a). The polymaleimide is expected to be environmentally benign having feedstock recovery, thermo-responsive and remendable features.[405]

Production of hexitols from GLY is also possible. For instance, under aerobic condition, Khan *et al.* reported production of mannitol (Figure 3a) from GLY by employing resting cells of *Candida magnoliae*.[92] As shown in Figure 8, a large variety of polymers can be obtained through selective functionalization of hexitols. As a recent example, Lavilla *et al.* produced a novel carbohydrate-based bicyclic diol; 2,4:3,5-di-O-methylene-D-mannitol (Manx), by internal acetalization of D-mannitol. Manx monomer was then reacted in the melt with ethylene glycol and dimethyl terephthalate for production of random $PE_xManx_yT$ copolyesters (Figure 30b). The produced copolyesters have similar thermal stability, higher Tg and low crystallizability with respect to PET. These findings suggest that Manx, as a bio-based comonomer, is suitable for obtaining amorphous PET products with enhanced glass-transition temperature. It can be used in polymer applications requiring thermal stability and transparency.[406] More detailed information regarding the utilization of hexitols for the preparation of bio-based polymers was provided in Section 5.1.

GLY and its derivatives, such as hydroxyacetone (acetol)[104] (Figure 3a), can be valorized as potential platform compounds in the synthesis of renewable diesel or jet fuel.[407, 408] For the first time, Li *et al.* obtained diesel or jet fuel range branched alkanes by the hydroxyalkylation–alkylation (HAA) of 2-methylfuran (2-MF) with lignocellulose derived hydroxyacetone. The hydroxyacetone pathway showed higher HAA reactivity and diesel yield over the previous acetone route owing to the electron-withdrawing effect of the hydroxyl group.[407] Another GLY derivable hydroxy ketone compound is dihydroxyacetone (DHA) (Figure 3a).[99, 399, 400] DHA is industrially produced by microbial fermentation of GLY over *Gluconobacter oxydans.* It is an intermediate in the metabolism of glucose in humans as well as an FDA approved compound for topical use as the active ingredient in sunless tanning lotions. DHA-based polymers are expected to have advantageous biomaterial applications since DHA degradation product can enter the normal metabolic pathway. Owing to this motivation, Putnam and Zelikin patented polycarbonates, poly(acetal carbonate)s, poly(spiroacetal)s, polyesters and polyurethanes by using chemically protected DHA and/or its dimers.[409] In solution, monomeric DHA is in equilibrium with its hemiacetal dimer. Locking monomeric DHA in the dimer form allows the preparation of poly(carbonate acetal)s. Alternatively, through the conversion of the $C_2$ carbonyl into a dimethoxy acetal, DHA can be converted to 2,2-dimethoxypropane-1,3-diol carbonate (TMC(OMe)$_2$). The bulk ROP of TMC(OMe)$_2$ in the presence of stannous octanoate (Sn(oct)$_2$) results in formation of the polycarbonates (Figure 31).[410]

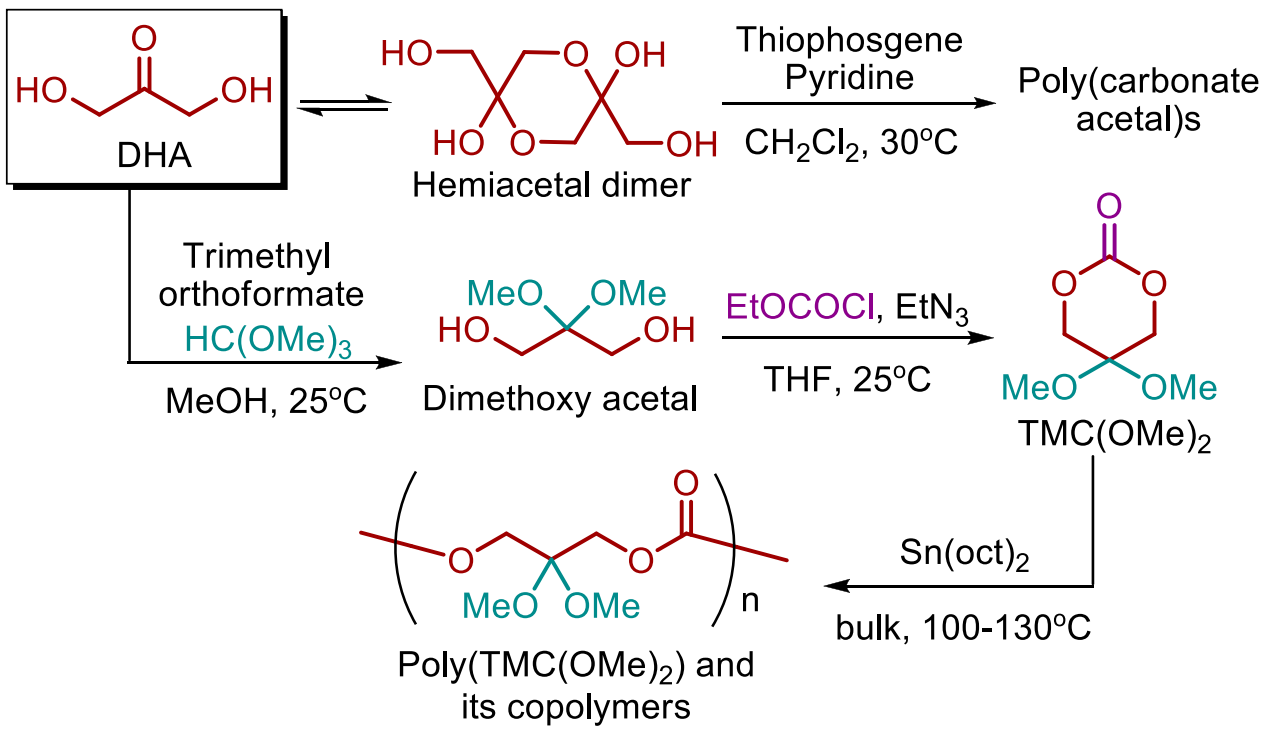

Figure 31. Conversion of DHA to poly(carbonate)s and poly(carbonate acetal)s.

Selective dehydration of GLY affords the production of extensively used vinyl monomers such as acrylic acid, acrolein, propene, allyl alcohol and 3-methoxy-1-propene (Figure 3a). Acrolein is commericially produced by oxidation of petroleum-derived propene with a Bi/Mo mixed oxide catalyst at temperatures above 300°C. Large scale production of acrylic acid also starts from propene. The process involves a two-step reactor system and metal oxide catalysts, in which propene is first oxidized to acrolein as the intermediate, and then further oxidation of acrolein yields acrylic acid (Figure 32). As shown in Section 5.4-Figure 12, bio-based production of acrolein and acrylic acid can be realized starting from 3-HPA as the feedstock. In the process, 3-HPA can be converted to propene through a dehydration reaction coupled with a decarboxlation process. Conversion of propene to acrolein and/or acrylic acid has been already commercially practiced as mentioned above. On the other hand, the conversion of 3-HPA to propene is yet to be realized. Besides, no known organisms can produce 3-HPA as a major metabolic end product from sugars. These issues currently alleviate 3-HPA from being a platform chemical for large scale production of propene and acrylic monomers. In this consideration, GLY, as being a cheap byproduct of biodiesel production, may become a better alternative. As such, acrolein can be produced from GLY *via* the dehydration of GLY on acidic solid catalysts. In the process, full conversion of GLY to acrolein is achieved by passing a mixture of GLY-$H_2O$ gases at 250-340°C over an acidic solid catalyst (Figure 32). Recently, the usages of sub- and supercritical water as the reaction media, and a biocatalysis pathway employing *Lactobacillus reuteri* have also been investigated.[399, 400] Nonetheless, these acrolein production routes from GLY do not satisfy the criteria of an economical process, and thus, they were not commercialized so far.[399] The GLY-based synthesis methods should be cost-competitive with comparison to the current propene-based process. Hence, a process development is required for further reducing the GLY-derived acrolein production costs. Once obtained, GLY-derived acrolein can be further oxidized to acrylic acid by employing the current industrial process (Figure 32). It is also possible to produce propene through GLY-acrolein-allyl alcohol-propanol-propene pathway.[98] However, this process is quite cumbersome and it does not seem to be a cost-competitive route for production of bio-propene. Common polymers of acrylics and propene were described in Sections 5.4 and 5.14, respectively. So, no further detail was provided in this section.

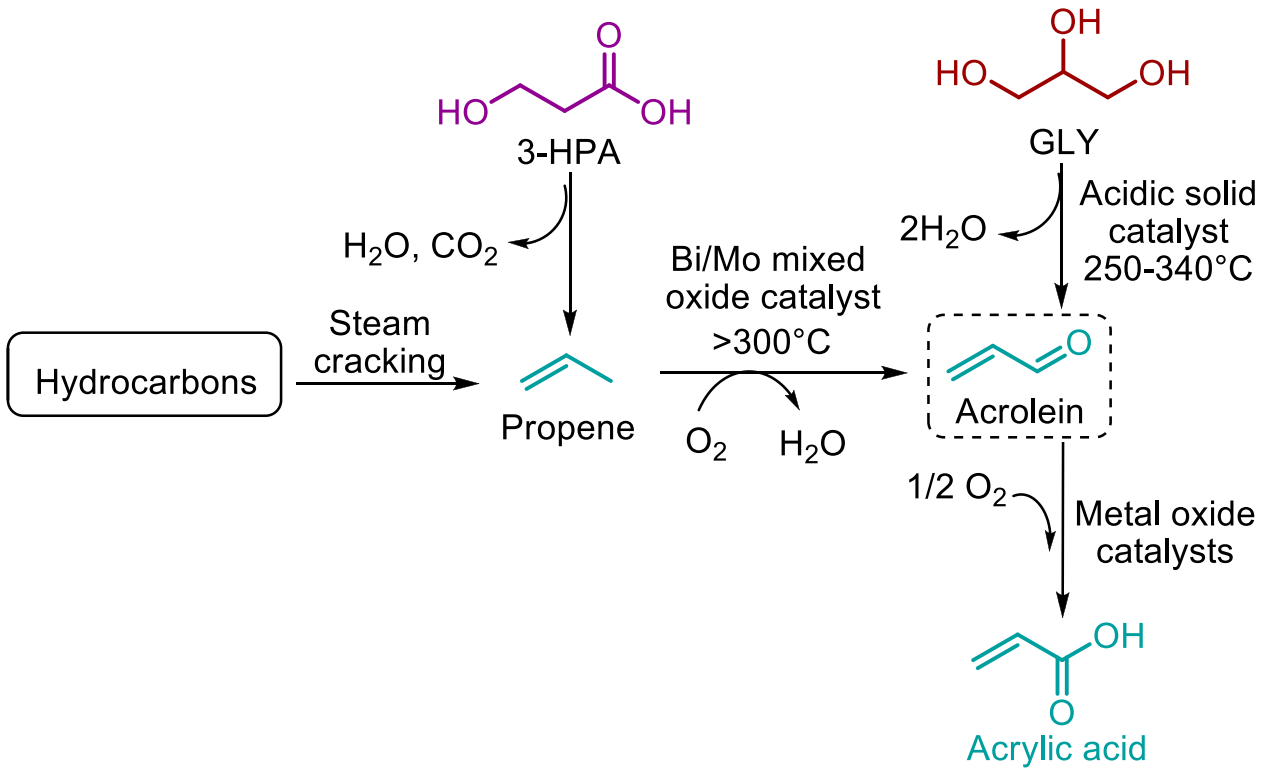

Figure 32. Production pathways of propene, acrolein, and acrylic acid from different feedstocks.

Liu *et al*. have shown that using iron oxide as catalyst first catalyzes the dehydration of GLY to acrolein and consequtively, the selective hydrogenation of acrolein to allyl alcohol (AlOH). The catalyst results in full conversion of GLY but AlOH yields remain between 20-25%.[411] Although the yields are low, the study provides promising insights for direct conversion of GLY to AlOH. GLY-derived AlOH can be copolymerized with other monomers. As an example, in the presence of oxygen, it is copolymerized with styrene. AlOH based copolymers are used as an intermediate for the production of flame-resistant materials, or as a nematocide, fungicide, or preservative. It was reported that its condensation with methyl glucoside polyethers, and subsequent bromination and addition of isocyanates, yields flame-resistant polyurethane foams. Apart from these, AlOH is also a building block for the synthesis of polymerizable allyl esters and ethers. Allyl esters are generally produced by reacting AlOH with the free acids, acid anhydrides, or acid chlorides in the presence of *p*TsOH. On the other hand, allyl ethers are obtained when AlOH is heated with monoalkanols in the presence of mineral acids (Figure 33). Detailed information regarding polymeric applications of AlOH derived allyl esters and allyl ethers was provided by Krähling *et al.*[412]

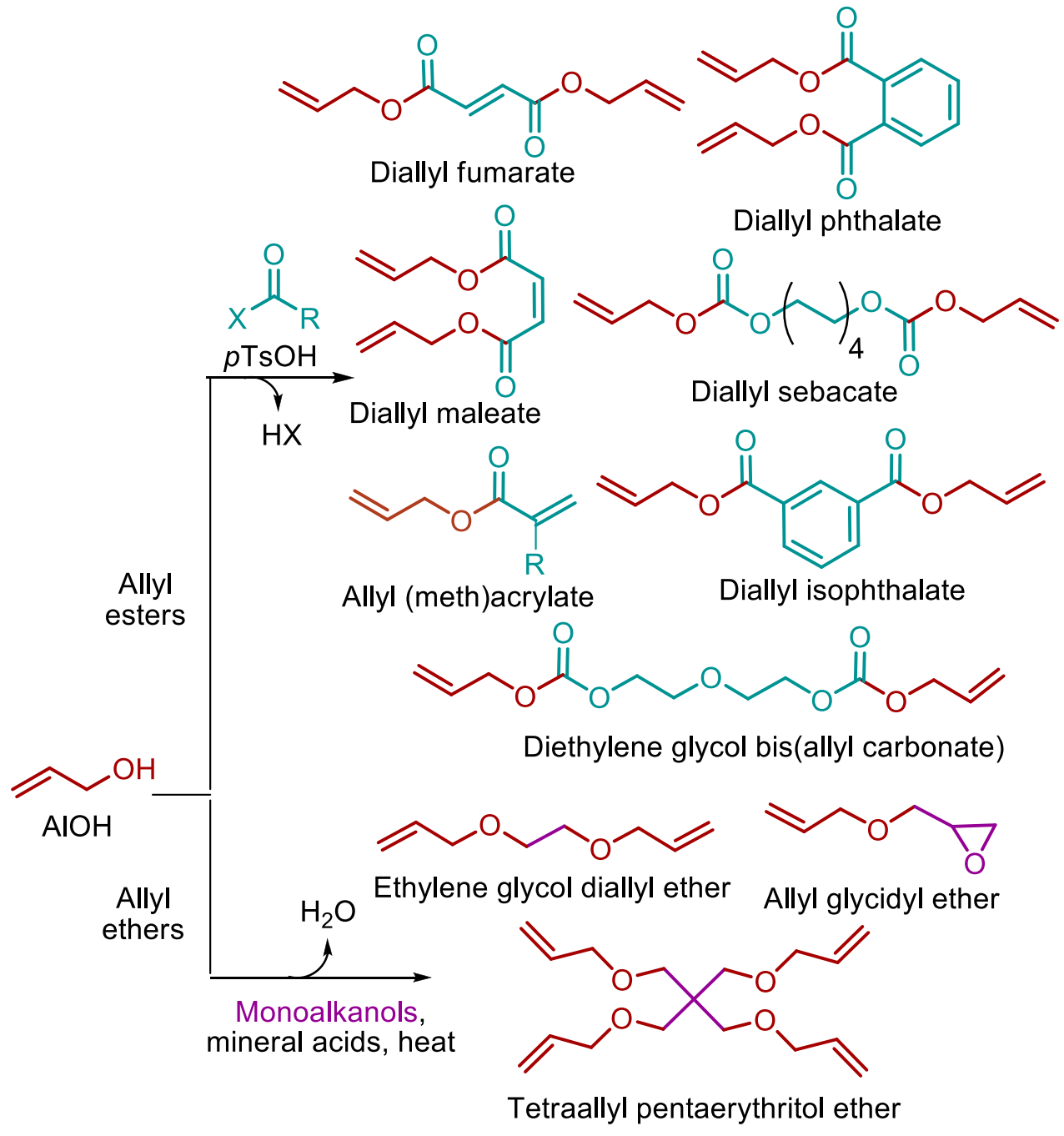

Figure 33. AlOH derived allyl ester and allyl ether monomers.

Ring formation reactions can be performed on GLY for production of industrially important cyclic monomers such as epichlorohydrin (ECH), glycidol and glycerol carbonate (GC) (Figure 3a). Through the use of GLY as a renewable feedstock, the Dow Chemical Company has announced a significant improvement in the production of ECH by introducing a solventless, glycerin to epichlorohydrin (GTE), process, which proceeds in two steps.[105] In the first step, GLY is hydrochlorinated to dichlorohydrins in the presence of a carboxylic acid catalyst under pressure at 120°C. Subsequently, the ring closure reaction of the dichlorohydrins in the presence of a base yields ECH (Figure 34). This process produces only one equivalent of waste chloride. On the other hand, the dominant commercial route employing propene as starting compound, retains only one of the four employed chlorine atoms in the final ECH product.[105] This shows that the GLY-based process is more efficient in terms of atom economy. Besides, the two step GTE route is less laborious with comparion to the propene-based process (3 steps). Owing to these advantages, a shift from the dominant process to the GLY-based process might be observed in the near future. As such, Solvay Chemicals has also launched a GLY-based Epicerol® technology for the production of bio-based ECH in its new production plant in China.[413] As a chemical intermediate, ECH has a wide range of applications (Figure 34). It is primarily used in the manufacture of epoxy resins. The most familiar epoxy resin; bisphenol A diglycidyl ether (BADGE), is produced by condensing ECH with BPA. ECH is also homopolymerized or copolymerized with other monomers to form elastomers. ECH based elastomers exhibit excellent physical properties and are resistant to oxygen, weather, fuel, oil and ozone. The reaction of ECH with alcohols, alcoholates, or the sodium salts of fatty acids gives products which are used as vinyl polymer plasticizers. It is used in the manufacture of ion-exchange resins, polyamines and polyquaternary ammonium salts for water treatment applications. Many ECH-based surface-active agents are produced by reacting ECH with a polyamine. Other main application areas of ECH include textile, paper, agricultural and pharmaceutical industries.[414, 415]

Another GLY derivable epoxide is glycidol (Figure 3a). Commercially, glycidol is produced through epoxidation of AlOH or reaction of 3-chloro-1,2-propanediol with bases. However, both of the processes suffer from drawbacks mainly including; (i) petroleum-based raw materials which are not sustainable, (ii) multistep synthesis which decreases the synthetic efficiency and thus, increases the production cost, (iii) production of a large amount of waste liquid and salt.[416] So, a bio-based process development for large scale production of glycidol is necessary. Such a process recently reported by Bai *et al.*[416] They demonstrated one-pot synthesis of glycidol from GLY and dimethyl carbonate (DMC) in the presence of reusable and recoverable $NaAlO_2$ catalyst under pressure (101.3 kPa) at 80-92°C. The selectivity of glycidol and the conversion of GLY were reported to reach 80.7% and 94.7%, respectively. As a different route, glycidol can be synthesized from the decarboxylation of GC. Zeolithe A was proved to be a very efficient catalyst for this reaction by providing a yield of 86% and a purity of 99% at a temperature of 180°C and a pressure of 35 mbar.[399, 400] These developments can potentially provide large scale synthesis of bio-glycidol starting from GLY in a two step process (Figure 34). Once obtained in large scale, bio-glycidol can have a variety of industrial uses owing to its bifunctionality, having both epoxide and alcohol functional groups. Like GLY, glycidol can be polymerized into PGs. Sunder *et al.* described in detail the synthesis of hyperbranched PGs (*hb*PGs) through the reaction of glycidol with trimethylolpropane.[417, 418] For applications ranging from cosmetics to controlled drug release, a variety of polyglycerols were commercialized. For instance, Hyperpolymers Company (Germany) has been providing small-scale production *hb*PGs with different molecular weights and special derivatives. Besides these, glycidol is a high-value compound in the production of epoxy resins, polyurethanes and polyglycerol esters (PGEs).[54, 399, 400] Esters of glycidol are also useful monomers. Recently, Geschwind and Frey reported the poly(glycerol carbonate)s which were synthesized from the polymerization of the glycidyl ethers; ethoxy ethyl glycidyl ether or benzyl glycidyl ether, with $CO_2$. In the study, an innovative route for the synthesis of poly(1,2-glycerol carbonate) was presented.[419]

A different GLY derivable important cyclic monomer is GC. By taking industrial feasibility into account, Ochoa-Gómez *et al.* stated that the synthesis of GC starting from GLY and/or $CO_2$-derivatives is the most attractive. Thus, the most suitable industrial process seems to be the transesterification of GLY with dimethyl carbonate in the presence of uncalcined CaO as catalyst. This process is highly promising because (i) CaO is very cheap and widely available, (ii) the reaction proceeds with a 100% GLY conversion and a GC yield of 95%, and (iii) the reaction is completed in just 1.5 h at 80 °C.[96] . Many other synthesis routes of GC by considering its reactivity and applications were

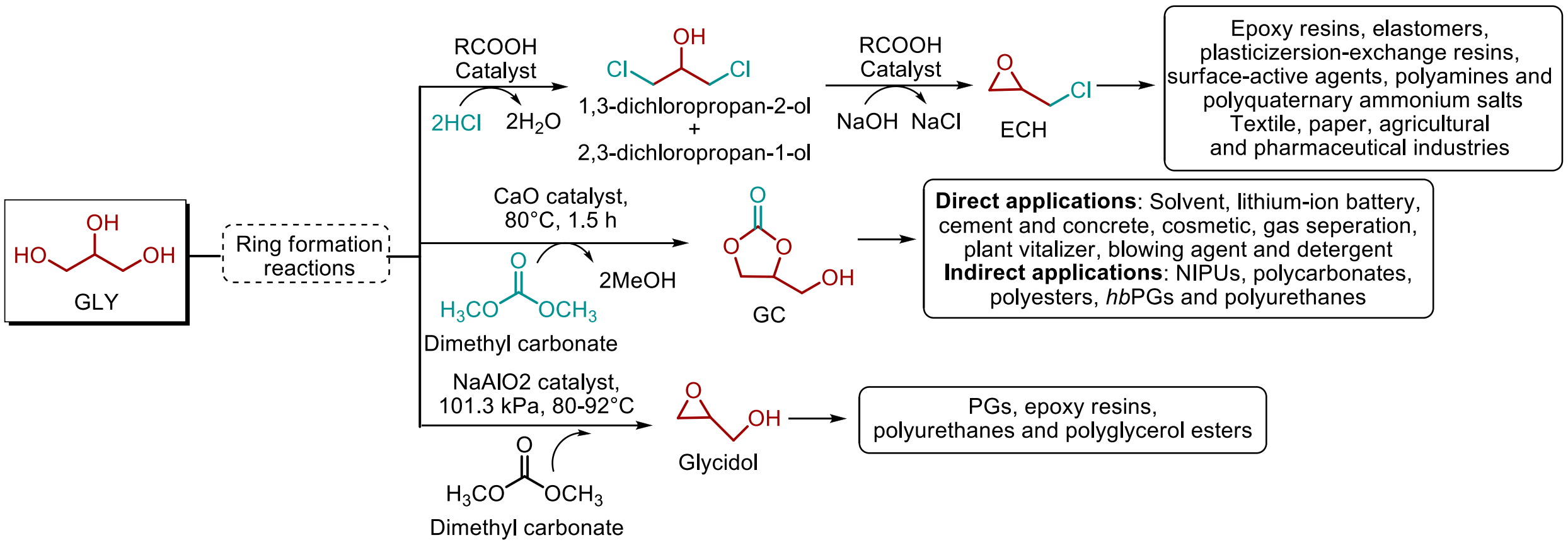

Figure 34. Common applications of GLY derived ECH, GC and glycidol.

discussed in detail by Sonnati *et al.*[420] Over the last 20 years, GC has attracted much interest because it is a very important compound in valorization of waste GLY. Owing to its wide reactivity, it has many direct and indirect applications (Figure 34). Unlike other cyclic carbonates, the reactivity of GC does not only stem from the dioxolane ring but it also stems from the pendant hydroxyl moiety. The wide reactivity of GC opens numerous oppurtunities for its utilization as a raw material for synthesis of industrially important chemical intermediates such as glycidol and ECH, as well as polymers such as non-isocyanate polyurethanes (NIPUs), polycarbonates, polyesters, *hb*PGs and polyurethanes. These indirect applications of GC hold enormous potential in manufacturing useful materials such as coatings, adhesives, foams, and lubricants.[96, 420-422]

GLY oxidation creates new oppurtunities for synthesis of value-added aldehyde and acid compounds (Figure 3a). Kim *et al.* demonstrated that electrochemical dehydrogenation process can be used to oxidize glycerol to glyceraldehyde (GAD).[101] Another GLY derivable value-added aldehyde is 3-hydroxypropionaldehyde (3-HPAl), which can function as a precursor for monomers such as 3-HPA, 1,3-PDO, acrolein and acrylic acid. Furthermore, it is transformed into polymeric derivatives and resins. 3-HPAl is produced through both bacterial fermentation and traditional chemistry. It was stated that biotechnological production has several advantages with respect to chemical process.[423] As a recent study, Krauter *et al.* reported very high 3-HPAl concentration and productivities from glycerol by employing *Lactobacillus reuteri*. In the study, 138 g/L glycerol was converted into 108 g/L 3-HPAl with an overall productivity of 21.6 g $L^{-1}$ $hours^{-1}$ in a single fed-batch biotransformation at 45°C.[424] Further oxidation of GAD results in the formation of glyceric acid (GLA)[103] and, subsequently, tartronic acid (TTA).[99, 400] The oxidation of both primary and secondary alcohol groups, on the other hand, gives the highly functionalized compound ketomalonic (mesoxalic) acid (KMA).[99] GLA has both medicinal and industrial value.[103] However, it is very expensive and thus, it has not been commercially produced yet. Nonetheless, it can be abundantly produced from GLY by employing acetic acid bacteria[425] and this pathway may open new oppurtunities for its mass production. The biotechnological production of GLA and its possible applications was reviewed by Habe *et al.*[426] GLA acts as an ABB'-type trifunctional monomer since it has one –COOH and two –OH groups. Hence, various GLA based polymers having different structures such as linear, hyperbranched, and network structures can be obtained through suitable polymerization reactions. Fukuoka *et al.* reported a novel hyperbranched poly(lactic acid) by polymerizing lactide in the presence of GLA. The produced branched polymer could potentially be used as a bio-based modifier for poly(lactic acid)s.[425] Catalytic oxidation of TTA in water over BiPt/C catalyst first results in formation of KMA and then, in polymerization of the produced KMA to its corresponding polyether; poly(ketomalonate) (PKM). Kimura stated that the decarboxylation of PKM can be easily performed and this proves that PKM has pendant carbon dioxide groups. When completely decarboxylated, poly(oxymethylene) is produced from PKM.[427]

### 5.13. Lactic Acid Platform Based Polymers.

Lactic acid (LA) (Figure 4a) is the most widely studied carboxylic acid from natural resources and it has extensive success in commercialization. Industrially, LA is mainly produced *via* the fermentation of glucose and sucrose by lactic acid bacteria. Through the fermentation process, the global production of LA is around 350 kt/year and it is expected that there will be a substantial growth in the next decade.[261] Since microbial production of LA has been extensively reviewed in the recent literature,[131-133, 428] it was not described in detail here. However, it is worthy to note that production of enantiopure L-lactic acid or D-lactic acid depends on the microbial strain used during the fermentation process. This is highly significant because LA stereo-chemistry greatly controls relevant physical properties of its final products.[236] According to Abdel-Rahman *et al.*,[131] it is very expensive when sugars, such as glucose, sucrose, and starch, are used as the feedstock for LA production. Hence, lignocellulosic biomass can have a great contribution to the expected substantial growth of LA production by considering its great availability, sustainability and low cost compared to refined sugars. Despite these advantages, the commercial use of lignocellulose for LA production is still problematic. The main problem stems from the costly pretreatment and enzymatic hydrolysis of lignocellulosic biomass for production of fermentative sugars (Section 3). The pretreatment is a *sine qua non* process in which the structure of lignocellulosic biomass is broken down to separate cellulose and hemicelluloses from lignin. Hence, highly efficient and cheap pretreatment technologies need to be sought. On the other hand, the costly enzymatic hydrolysis step for depolymerising cellulose and hemicelluloses to fermentative sugars can be by-passed. In this consideration, the development of genetically modified lactic acid bacteria, which can directly ferment cellulose and/or xylan, is a necessary option.[131] Fortunately, the commercial production of LA from lignocellulosic biomass does not seem to be too far away. As an encouraging example, industrial biotechnology company Direvo has recently fermented lactic acid on pilot scale by introducing a consolidated bioprocessing technology. The consolidated process allows the conversion of lignocelluloses in a single step without having to add enzymes.[429]

LA is primarily used for production of biodegradable polymers, and in food and beverage sectors. These sectors are followed by pharmaceuticals and personal care products.[430] In polymer industry, LA is mainly consumed in production of polylactic acid (PLA). There are two major chemical ways to prepare polylactic acid (PLA); polycondensation of LA or ROP of its cyclic dimer; lactide (Figure 35). For the preperation of high molecular weight PLA, the ROP is more widely used. It is difficult to obtain high molecular weight PLA *via* polycondensation due to water formation during the reaction. ROP proceeds in the presence of catalysts and it allows the control of molecular weight as well as stereotacticity of PLA with comparison to polycondensation.[261, 431] Several distinct forms of PLA exist due to the chiral nature of LA. Polymerization of racemates results in formation of amorphous poly-DL-lactide (PDLLA). On the other hand, poly-L-lactide (PLLA); the product resulting from polymerization of L,L-lactide, exhibits high crystallinity (37%), a glass transition temperature between 60 and 65°C, and a melting temperature around 175°C. Hence, the degree of crystallinity depends mostly on a defined ratio of D-and L-enantiomers and it determines most physical properties of the final products.[325]

Owing to its low price and availability in the market, PLA has one of the highest potential amongst the other biodegradable polyesters.[261, 326] Its market volume is projected to exceed $4,840.1 million by 2019.[430] Thus, many companies have been involved in its commercial production (Table 6). NatureWorks is currently the major supplier of PLA under the brand name of Ingeo. There are other PLA manufacturers in the USA, Europe, China and Japan. They have been developing various grades of PLA suitable for different industrial sectors.[325, 431] Until the late 1980s, the applicability of PLAs outside the medical field was restricted by high production costs. Later on, major breakthroughs in the process technology decreased the production costs and this allowed the commercial-scale production of biodegradable polymers from LA for nonmedical applications.[285] Currently, the produced PLA is mainly used in the packaging market, which is projected to be valued at $994.9 million by 2019. The textile segment accounts for the second-largest share in the PLA market and it is followed by agriculture, electronics, automobile and other segments.[430] PLA as a biodegradable and thermoplastic polyester has the potential to replace traditional polymers such as PET, PS, and PC for various applications. However, there are still many challenges to be addressed. Tg and brittleness are the most noticeable weakness of PLA and there are many research groups actively involved in overcoming and addressing solutions to these problems.[261, 431] Toughening of PLA by blending it with a variety of materials is a commonly employed approach in this consideration.[432] However, a completely renewable and biodegradable toughening agent should be used to preserve the renewability and biodegradability of the final product. Since PLA based polymers and their applications are extensively reviewed in the literature, it was not described in detail here. Detailed information regarding synthesis and production, thermal-chemical-mechanical-rheological properties, degradation and stability, and applications of PLAs are provided in the comprehensive books and review articles.[433-438]

**Table 6. Major manufacturers of PLA.**

| Company | Brand name | Location |
|---|---|---|
| NatureWorks LLC | Ingeo | USA |
| Purac | Purasorb | Netherlands |
| Futerro | Futerro | Belgium |
| Tate & Lyle | Hycail | Netherlands |
| Synbra | Biofoam | Netherlands |
| Toyobo | Vylocol | Japan |
| Teijin | Biofront | Japan |
| Uhde Inventa-Fischer | | Germany |
| Hiusan Biosciences | Hisun | China |

LA is transformed to other valuable chemicals *via* esterification, hydrogenolysis, dehydration, decarbonylation, oxidation, and reduction processes (Figures 4a and 35). The leading esterification products of LA are lactide[132, 133] and PLAs. Its esterification leads to the formation of lactates too, depending on the reaction conditions. A wide range of alkyl lactates (linear esters), which exhibit unique solvation properties, are easily produced from LA using either homogeneous or heterogeneous catalysts. The esterification of LA has been investigated with different alcohols including methanol, ethanol, 2-propanol, isobutanol, n-butanol and benzyl alcohol. The most prominent example is ethyl lactate. This short chain ester is amongst the most promising green solvents owing to its high boiling point, low vapour pressure, low surface tension and renewable origin.[132, 133, 137]

The dehydration and reduction of LA result in the formation of giant monomers; acrylic acid (Section 5.4) and propylene glycol, respectively.[54, 132, 133, 138] In 1958, LA was for the first time converted to acrylic acid *via* a direct dehydration reaction. A maximum yield of 68% of acrylic acid was obtained over a $CaSO_4/Na_2SO_4$ heterogeneous catalyst system at 400°C. Thus far, no important progress was reported so that the acrylic acid yields generally remained below ~70%.[132, 133] Mainly due to this reason, LA-derived acrylic acid production has not been put into commercial practice. Finding the right heterogeneous catalyst seems to be the key challenge since the main consideration is to increase the acrylic acid yield. In this respect, Dusselier *et al.* suggested the use of alkali modified zeolities which can eliminate formation of the side products including $CO_2$, acetaldehyde and coke.[133] The LA derived AA can be further functionalized. It is converted to acrylates with the help of efficient, cheap and reusable catalysts (Section 5.4). A different way of producing acrylates is their direct production from lactates through the dehydration reaction.[133, 142] The production of AA or its esters starting from LA might become an industrially feasible technology depending on its cost competitiveness with comparison to the current propene process. Acrylic polymers and their applications were described in detail in Section 5.4.

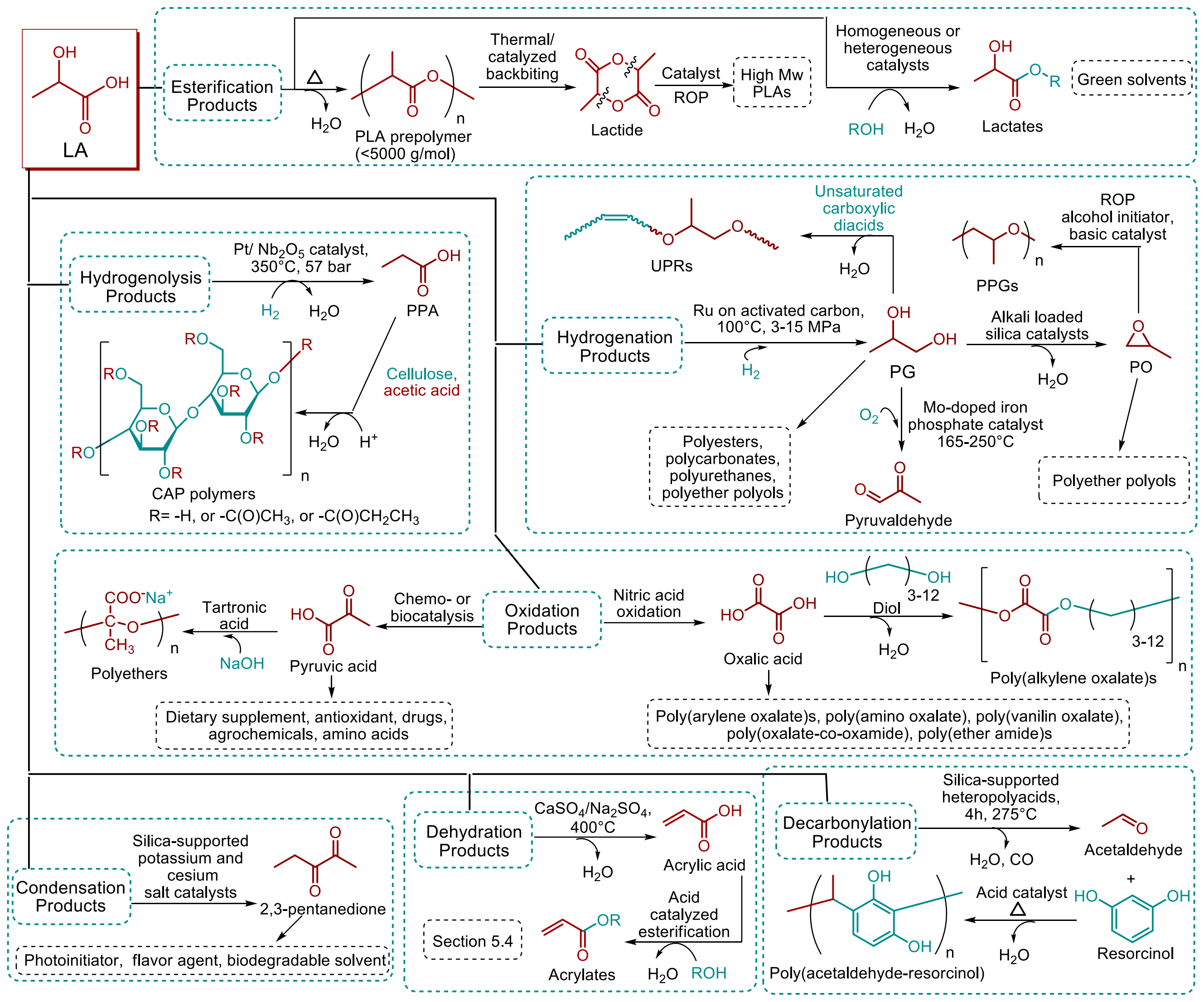

Figure 35. LA as a platform compound for production of fine chemicals and polymers.

Hydrogenolysis of the secondary alcohol group of LA generates propanoic acid (PPA). The reaction can take place over a bifunctional $Pt/Nb_2O_5$ catalyst at 350°C and 57 bar pressure. In the process, LA is first dehydrated on the acid support and subsequently, hydrogenated to PPA on the metal sites.[140] Nonetheless, the reaction yields very low amount of PPA (18%) and thus, a process development is required. Particularly, instead of expensive Pt-based catalysts, the development of bifunctional, cheap and recyclable catalysts which can provide high LA conversions as well as high PPA yields should be considered. Once obtained in large scale, LA-based PPA can be employed in the production of cellulose-acetate-propionate (CAP) polymers (Figure 35). Different formulations of CAP polymers are currently produced by Eastman Company for printing ink and nail care applications.[439] In addition to these, CAP has been studied as composite, adsorbent, membrane and coating material for a wide range of applications.[440-443]

Corma *et al.* stated that the direct hydrogenation of LA or lactates to propylene glycol (PG) can be an alternative green route to the current petroleum-based process, which is based on hydroperoxidation chemistry or the chlorohydrin process involving the use of hypochlorous acid.[54] Particularly, ruthenium on activated carbon was reported to be an effective catalyst to perform the hydrogenation process by providing 95% LA conversion and PG selectivities higher than 90%.[54] Nevertheless, the industrial manufacture of bio-PG is currently focused on glycerol mainly due to the large and cheap surplus of glycerol in the market, which is obtained as a byproduct in the production of biodiesel from vegetable oils (Section 5.12). Companies such as Dow Chemicals, Cargill/Ashland, ADM, and Senergy have been producing bio-based PG from glycerol.[444] Annually, 1.5 million tons of PG is manufactured by employing the above-mentioned petroleum-based process.[37] Such a high production capacity of PG simply stems from its large applicability. Nearly 45% of PG is consumed in the manufacture of unsaturated polyester resins (UPRs) (Figure 35), in which it is reacted with saturated and unsaturated carboxylic diacids such as isophthalic acid and maleic anhydride. The resulting resins are dissolved in styrene or another polymerizable monomer, which is combined with filler, chopped glass, a peroxide polymerization initiator, and other additives. Then, they are cured to give a hard, cross-linked, thermoset composite. The end-use applications for such UPRs include laminates, automotive plastics, fiber glass boats, construction, coatings, art objects, insulation, electrical components, pipes and tanks. Polymeric applications of PG are not limited to polyesters, but they also consist of polycarbonates, polyurethanes and polyether polyols.[270, 325, 445] PG has many other diverse applications since it is conferred as ‘‘Generally Regarded As Safe’’ (GRAS) by the FDA. It is commonly used as cosmetic and food additive, antifreeze agent, brake fluid, lubricant, and aircraft de-icing fluid.[37, 445] Besides, PG is employed as an intermediate for production of other valuable compounds including propylene oxide and pyruvaldehyde (Figures 4a and 35).[139, 141] Between 60% and 70% of the total volume of PO is used in the manufacture of polyether polyols. The Dow Chemical Company is the world's largest producer of PO and polyether polyols. Commercial polyether polyols are based on di-, tri-, or polyhydric alcohols which are usually copolymerized with PO/ethylene oxide. They are mainly consumed in the production of polyurethanes. Such polyurethanes can exhibit a wide range of hardness, rigidity, and density characteristics. Polyols with molecular masses higher than 3000 are used for the manufacture of flexible polyurethane foams for applications requiring flexibility such as furniture and automobile seating, bedding, and carpet underlay. On the other hand, polyols with lower molecular masses lead to the formation of rigid foams for applications such as thermal insulation.[446, 447] A different PO based polymer family is polypropylene glycols (PPGs) (Figure 35). They are produced by ROP of PO in the presence of an alcohol initiator and a basic catalyst. A variety of PPGs can be obtained by using different alcohol initiators and/or by copolymerizing PO with other monomers. Depending on the functionality of the initiator (i.e. mono-, di-, or multi-functional initiators), PPGs having different architectures such as linear, branched, or even linear-dendritic hybrid systems can be obtained for diverse applications.[446, 448-451]

Other considerable transformations of LA include its decarbonylation to acetaldehyde, condensation to 2,3-pentanedione, and oxidation to pyruvic and oxalic acids (Figures 4a and 35).[25, 132-136, 452] Regarding common applications of these fine chemicals, acetaldehyde is used in the manufacture of novolak resins of resorcinol (Figure 35). Acetaldehyde modified novolaks exhibit certain characteristics, i.e. special solubility or compatibility properties in rubber compounding applications, which may not be provided by straight resorcinol-formaldehyde resins.[453] 2,3-pentanedione has applications as a flavor agent, a biodegradable solvent, a photoinitiator, and an intermediate for other high-value compounds.[132] As an oxidation product of LA, pyruvic acid has been increasingly used as a dietary supplement and antioxidant, as well as a precursor for the synthesis of drugs, amino acids and agrochemicals.[133] Moreover, this alpha-keto acid can be anionically polymerized to form the corresponding polyether (Figure 35).[454] Very recently, Boamah *et al.* employed pyruvic acid in a different manner. They prepared pyruvic acid modified low molecular weight chitosans as potential lead adsorbent materials.[455] The other oxidation product of LA; oxalic acid, can be used in polycondensation reactions. Nonetheless, its diester[456] or dichloride (oxalyl chloride)[457] derivatives are more commonly used instead of oxalic acid in the literature since they provide more facile polycondensation reactions.

### 5.14. Acetone-Butanol-Ethanol (ABE) Platform Based Polymers.

Biomass based fuels currently seem to be the most suitable alternative for production of renewable, sustainable, and economically viable fuels such as bio-ethanol. In this context, lignocellulosic biomass resources have certain advantages. First of all, lignocellulosic biomass is the most abundant sustainable raw material worldwide and it has widespread avaliability. Furthermore, it has non-competitiveness with food supplies and high mitigation effects on GHG emissions. Owing these characteristics, various companies have been engaged in the commercial production of bio-ethanol from lignocellulosic biomass resources (Table 7). Although the production of bio-ethanol from lignocellulosic sugars is now a commercialized technology, the process of conversion is more complicated than that of starch based biofuels. This mainly stems from the rigid and complex molecular polymeric structure of lignocellulosic biomass. Hence, the production of lignocellulosic ethanol is still a challenge with many opportunities for progress. The current status and challanges as well as the future prospects of bio-ethanol production from lignocellulosic biomass were not comprehensively presented in this article since the detailed scrutiny was provided in the corresponding books and review articles.[143, 199, 458-463] Renewed attention has also been paid to butanol and acetone production from lignocellulose through acetone-butanol-ethanol (ABE) fermentation process (Figure 4c). In fact, butanol as a higher alcohol is a more promising gasoline substitute compared to ethanol. That is why; production of bio-butanol has been receving high interests from both small biofuel start-ups, and large oil and chemical companies such as British Petroleum, Chevron, DuPont and DSM.[144, 464-466]

ABE compounds are promising precursors to many value-added chemicals (Figure 4c). They are converted to other value-added chemicals through different chemical processes such as oxidation, reduction, dehydration, and chlorination reactions. For instance, Ozkar and Finke developed a system for room temperature and acid-assisted hydrogenation of acetone to 2-propanol in the presence of Ir(0) nanoclusters. The system exhibits exceptional

**Table 7. Existing, planned, and under construction facilities to produce bio-ethanol from lignocellulosic biomass resources**

| Company | Plant Location | Feedstock |
|---|---|---|
| Abengoa | USA | Corn stover, wheat straw, milo (sorghum) stubble, switchgrass, |
| Abengoa | Spain | Wheat straw, cereal |
| Agroethanol AB | Sweden | Wheat |
| ALICO, Inc. | USA | Yard and citrus wastes |
| American Process Inc. | USA | Woody biomass |
| BioEthanol Japan | Japan | Wood construction waste |
| BioFuels Energy Corp. | USA | Grass and tree trimmings |
| BlueFire Ethanol | USA | Green waste |
| Borregaard Industries Ltd | Norway | Wood |
| British Sugar | England | Sugar beet |
| Broin jointly with US DoE, DuPont and Novozymes | USA | Corn fiber and stover |
| Colusa Biomass Energy Coporation | USA | Rice straw and hulls |
| China Resources Alcohol Corporation | China | Corn stover |
| DINS Sakai | Japan | Waste construction wood |
| DuPont-BP Biofuels | England | Sugar beets |
| ICM Inc. | USA | Corn stover, switchgrass |
| Inbicon | Denmark | Wheat straw |
| Iogen | USA | Wheat straw, barley straw, corn stover, switchgrass and rice straw |
| Iogen | Canada | Wheat, oat and barley straw |
| Lignol | Canada | Softwood and hardwood |
| Mascoma | USA | Paper sludge, wood chips, switch grass and corn stover |
| Pacific Ethanol Inc. | USA | Wheat straw, corn cob, woody biomass |
| Poet | USA | Corn fiber, corn stover |
| Range Fuels | USA | Timber and forest residue |
| Sekab | Sweden | Forestry products |
| Tereos | France | Sugar beet, wheat and sugar cane |
| Verenium | USA | Sugarcane bagasse and specially bred energy cane |
| Western Biomass | USA | Ponderosa pine wood chips, waste |

activity by providing up to 100% selectivity at 100% conversion, and up to 188000 total turnover catalytic lifetime.[145] In terms of oxidation processes, ethanol and butanol can be selectively converted into acetic[152] and butyric acids,[147] respectively. Acetic acid can also be produced through fermentation of $C_5$ and $C_6$ sugars. As an example, ZeaChem Inc. has recently achieved a commercially scalable acetogenic process for fermentation of the cellulosic sugars to acetic acid without $CO_2$ as a by-product.[326] Cheung *et al.* pointed out that the global production of acetic acid was approximately $10.6 \times 10^6$ tons per annum in 2008. More than 65% of this production is converted into vinyl acetate or cellulose based polymers (Figure 36).[467] For the preparation of vinyl acetate monomer (VAM), acetic acid is reacted with ethylene and oxygen either in the liquid phase in the presence of Pd/Cu, or in the gas phase on heterogeneous catalysts containing palladium.[468] It should be noted that raw materials for VAM; ethylene and acetic acid, are nowadays based on fossil resources. As shown by Amann and Minge, this could be switched to bio-ethanol based renewable, sustainable and $CO_2$-neutral production process.[469] The polymerization of VAM can be carried out by employing different types of standard polymerization techniques like emulsion, suspension and solution polymerization. Through these techniques, VAM is used to produce poly(vinyl acetate) (PVAc) as well as a wide range of PVAc copolymers. Particularly, ethylene-vinyl acetate copolymers are one of the most important classes. Such polymers are abbreviated as EVA (high ethylene and low vinyl acetate content) or conversely VAE depending on the ethylene and VAM content. Another around 30% of the produced VAM is converted to poly(vinyl alcohol) (PVA) through saponification or transesterification process of PVAc. The remaining 20% is valorized in other ways, such as in the production of poly(vinyl butyral) (PVB). The major manufacturers of VAM and VAM-based polymers include, but are not limited to, BP, Wacker Chemie, Celanese Chemicals, Dow Chemical Corp., DuPont, Millenium, Kuraray, Nippon Gohsei, Showa Denko, Dairen Chemical, and Shanghai Petrochemical.[469] In addition to VAM-based polymers, acetic acid is also utilized in the synthesis of cellulose based polymers; predominantly cellulose acetate. Detailed scrutiny regarding properties, applications, and degradation behaviors of cellulose acetate-based materials is presented by Fischer *et al.*[470] and Puls *et al.*[471] Major applications of the above-mentioned acetic acid derived polymers were summarized in Figure 36.[270, 468-471]

Without a doubt, the most important process that can be performed on ABE alcohols is dehydration. Giant monomers of today's polymer industry such as ethylene,[149, 325, 472] propylene,[146, 472] butadiene,[148] 1-butene,[363] and isobutene[473] have been produced through dehydration of ethanol, butanol, and 2-propanol as shown in Figure 4c.

Bio-derived ethanol has been particularly dehydrated to ethylene. In fact, in the early 20th century, ethylene was mainly produced from ethanol. Later on, the process was shifted to the current petrochemical route in which ethylene is exclusively produced *via* steam cracking of hydrocarbons at 750-950°C. The main reason of this shift was the unbeatable cheap price of oil. Nonetheless, due to the current cost competitive production of bio-ethanol and public interest in polymeric products derived from renewable resources, several companies like Dow, Braskem and Solvay have engaged in bioethanol-to-ethylene projects. The ethanol dehydration reaction recently is conducted in the vapor phase through fixed bed or fluidized bed reactors over solid acid catalysts like alumina or silica-alumina. Over these catalysts and at around 400°C, full conversion of ethanol and 99.9% selectivity of ethylene could be achieved.[325] This conversion is quite valuable because ethylene dominates the petrochemical market by

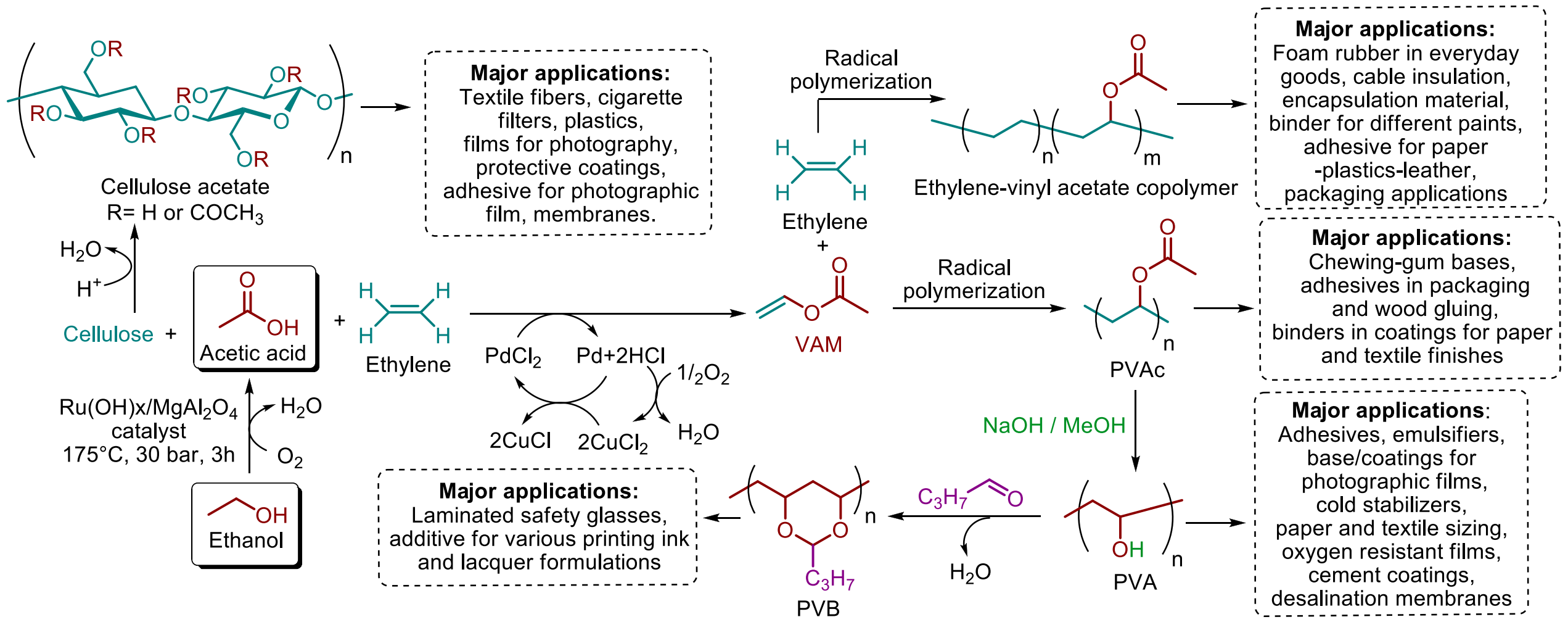

Figure 36. Acetic acid based commercial polymers.

being the largest petroleum chemical. In 2010, the worldwide production capacity of ethylene was approximately $160\times10^9$ lbs/year. To meet the demand for the enormous quantity of ethylene consumed worldwide, large quantities of renewable materials are needed. This, in fact, is another factor for the current shift towards the dehydration of bio-ethanol process.[325, 472] As an intermadiate chemical, ethylene is predominately used in the manufacture of many most important commodity polymers like high- and low-density polyethylene, poly(vinyl chloride), polystyrene and poly(ethylene terephthalate). These polymers make up almost two thirds of the current plastics market. Polyethylene (PE); the world's most widely used plastic, is produced by polymerization of ethylene under pressure and temperature, and in the presence of a catalyst. Ethylene can be either used as a monomer for high-density polyethylene (HDPE), or as a comonomer for linear low density polyethylene (LLDPE) (Figure 37). Moreover, it can be copolymerized with a wide range of monomers such as propylene, styrene, vinyl acetate, and acrylic acid to manufacture different materials with desired properties. With increases in oil prices, microbial PE or green PE is now avaliable in the market owing to the commercial production of bio-ethylene. Braskem is the first and largest producer of bio-PE with 52% market share, followed by Dow Chemical and Solvay with market shares of 12% and 10%, respectively. Bio-PE has exactly the same chemical, physical, and mechanical properties as petrochemical polyethylene, which means that all the applications of current fossil-based PE can be replaced by bio-PE. Currently, bio-PE is widely used in engineering, agriculture, packaging, and many day-to-day commodity applications due to its low price and good performance.[261, 270, 326, 431, 472, 474, 475]

Ethylene is also a valuable precursor for production of other related commodity compounds. Once bio-ethylene is obtained, no process development is required for its conversion to ethylene oxide (EO) and ethylene glycol (EG). Ethylene has been already commercially oxidized to ethylene oxide (EO), which is then hydrolyzed to produce ethylene glycol (EG) (Figure 4c). In 2011, the worlwide demand for EO was estimated to be higher than $28,000 million. From 2013 to 2018, this demand is expected to grow at a CAGR of over 6%, and pass $40,000 million in 2018. Currently, the global EO demand has been supplied by the vapor-phase oxidation of ethylene over silver-based catalysts. Several companies such as Dow Chemical Company, Shell, SABIC, Scientific Design, Japan Catalytic, BASF and China Chemical have been implementing this technology. However, depending on whether pure oxygen or air is used for oxidation, differences may exist in their technological details.[150, 476, 477] The produced EO is utilized as a decent disinfectant, sterilizing agent, and fumigant. Its ROP can proceed anionically or cationically but both of them generally give lower molecular weight products. For the production of extremely high molecular weight polymers, the process must involve a coordinate anionic reaction, in which EO is coordinated with a metal atom of the catalyst and subsequently, attacked by an anion.[478] Its major polymer derivative is polyethylene glycol (PEG), or also known as polyethylene oxide (PEO) (Figure 37). PEGs exhibit very high solubility, low toxicity, unique solution rheology, complexation with organic acids, low ash content, and thermoplasticity.[478] It is feasible to produce various PEG copolymers having different architectures such as linear, branched, star shaped, and com-like PEGs. Furthermore, highly diverse PEG-based polymers or polymeric systems can be developed *via* PEGylation method; the process of covalent attachment of one or more PEG chains to another molecule. For instance, various commercially avaliable polysorbates have been produced through PEG-ylation of sorbitan, which is followed by acylation with fatty acid (Section 5.11).[37, 390, 479] Owing to these characteristics, PEGs have myriad applications ranging from industrial to medical uses (Figure 37).[478] It is worthy to note that PEG is the most used polymer in the field of polymer-based drug delivery. There are many PEG-stabilized drug delivery systems in the market, which have gained regulatory approval from the US and/or the EU. The overwhelming utilization of PEG in biomedical applications are highlightened and discussed delicately in the review article of Knop *et al.*[480]

Approximately 60% of the global production of EO is converted to EG (Figure 4c) *via* the hydration of ethylene oxidation technology. Currently, this is the major industrial process for production of EG.[481, 482] The process simply involves non-catalytic liquid-phase hydration of EO. However; co-products such as diethylene glycol (DEG) and triethylene glycol (TEG) form together with EG. In order to ensure higher EG selectivities (roughly 90%), a large excess of water (20–25 mol water/mol EO) is generally used. This necessary operation, on the other hand, increases capital investment for unit operations.[158, 482] For the purpose of further increasing the EG selectivity as well as decreasing the production costs, Shell Global Solutions has implemented OMEGA; "Only MEG Advantage" technology. In the OMEGA process, EO is reacted with $CO_2$ to produce ethylene carbonate, which is then hydrolyzed to yield EG and $CO_2$. This robust operation gives over 99% EG yield without having a considerable by-product and it also uses nearly 30% less wastewater and 20% less steam.[483, 484] Although these are different industrial approaches for manufacture of EG, maximization of the EG yield and selectivity is a standard practice because EG is the most important one amongst the above-mentioned by-products. In 2010, its worldwide consumption capacity was pointed out to be about 20 million metric tons. Such a giant consumption capacity simply stems from its large applicability in diverse industrial areas such as energy, plastics, automobiles, and chemicals.[158, 482] Approximately 50% of the EG is utilized as an antifreeze. It is also consumed as a solvent, a coolant, a heat transfer fluid, a hydrate inhibitor and a fuel cell component. Besides its direct applications, nearly 40% of EG is directed into the polymer industry (Figure 37), particularly into the manufacture of polyethylene terephthalate (PET) polyesters for fiber and bottling applications.[481, 482] Another commercially avaliable EG based polyester; poly(ethylene succinate), was described in Section 5.2. Apart from the EG based polyesters, it is also feasible to convert EG into other types of polymers, like polyurethanes (PU). As a recent example, Pasahan *et al.* prepared PU films by reacting EG with glucose and diphenylmethane diisocyanate. The PU films were then used to design a novel polymer electrode for the detection of epinephrine. The PU film containing 5 % glucose by weight can be used as a membrane for EP detection in the presence of a high concentration of ascorbic acid.[485]

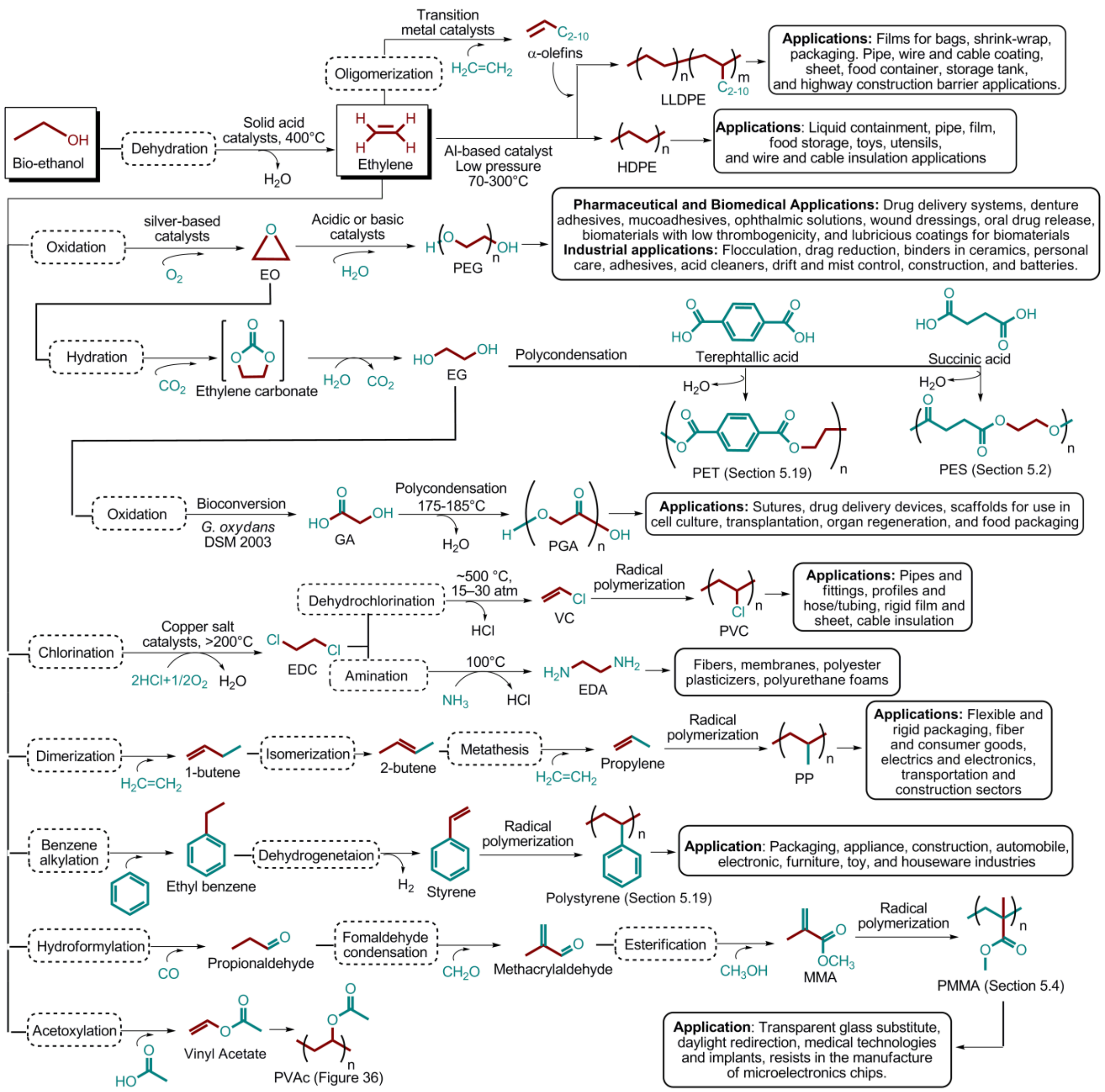

Figure 37. Bio-ethanol derived ethylene as a platform chemical for synthesis of commodity monomers and their corresponding polymers.

EG is also a fine chemical intermediate. It can be completely reduced to generate $H_2$-rich fuel gases, or selectively oxidized to glycolic acid (GA) or glyoxal (Figure 4c).[156, 157, 482] Partial oxidation of EG to GA can be performed on gold-based catalysts in aqueous solutions.[482] However, gold-based catalysts are too expensive, and it is difficult to control the oxidation process so that further oxidation results in formation of oxalic acid. Since the selective oxidation is perfectly done microbial species, bioconversion route could be a better option. As such, Wei *et al.* introduced an integrated bioprocess for the GA production from EG by employing G. *oxydans* DSM 2003. In the integrated process, simultaneous separation of GA was achieved by using anion ion-exchange resins to eliminate the effect of end product inhibition.[486] EG-derived GA can be polycondensed to form its homopolymer; poly(glycolic acid) (PGA) (Figure 37). This is the simplest process to synthesize PGA, however; it gives low Mwt product. Nonetheless, high Mwt PGA is obtained by ROP of cyclic dimer of GA; glycolide.[487] PGA is a low-cost tough fibre forming polymer but it is also biodegradable mainly by simple hydrolysis, which initially limited its use. Later on, this disadvantage of PGA turned out to be an advantage. The Davis & Geck used PGA to develop the first synthetic absorbable structure, which was marketed under the tradename of Dexon. Nowadays, PGAs are routinely used for suture manufacture. PGA based commercial sutures are produced by either copolymerization of PGA with other monomers, such as lactic acid (Vicryl®, Polysorb®), ε-caprolactone (Monocryl®), trimethylcarbonate (Maxon®), or a mixture of monomers, like ε-caprolactone and trimethylcarbonate (Monosyn®) or *p*-dioxanone, and trimethylcarbonate (Bisyn®). Major applications of PGAs are not limited to sutures but it also cover drug delivery devices, scaffolds for use in cell culture, transplantation, organ regeneration, and food packaging applications. For instance, Kureha Corporation commercialized PGA as a film having good gas barrier properties for food packaging.[37, 270, 301] The other selective oxidation product of EG; glyoxal, is used as a cross-linking agent in polymer industry. It has been employed to crosslink functionalized polymers such as cellulose, polyacrylamides, poly(vinylalcohol), keratin, and other polycondensates.[488]

Chlorination of ethylene to 1,2-dichloroethane, or ethylene dichloride (EDC), has also been industrially practiced. EDC currently belongs to those chemicals with the highest production rates since it is the starting material for the production of vinyl chloride (VC), and subsequently of poly(vinyl chloride) (PVC) (Figures 4c and 37). The conversion of EDC to VC process is carried out through either direct or oxychlorination methods, in which chlorine or hydrogen chloride is used, respectively. Industrially, both of the methods are practiced together and in parallel because most EDC plants are connected to vinyl chloride (VC) production units. The hydrogen chloride generated from VC production is re-used in the oxychlorination of ethylene.[151, 153, 489] In 2012, approximately 37.4 million t of PVC was produced from VC. With such a huge production capacity, PVC is the third-most widely produced plastic, after polyethylene and polypropylene. The top manufacturers are Shin-Etsu Chemical, Formosa Plastics, Solvay, LG Chem, ChemChina, Ineos and OxyVinyls.[490] Among these companies, Solvay is the leader in the production of bio-PVC, in which the company employs bio-ethylene as the starting material.[325, 431] Depending on the intended end use, each PVC-company produces a range of PVC polymers varying in molecular mass and morphology. Moreover, commercial PVC is never used alone. Heat stabilizers, lubricants, plasticizers, fillers, and other additives are always mixed with PVC in order to make its processing possible. These additives consequently influence physical and mechanical properties of PVC. For instance, unplasticized (rigid) and plasticized (flexible) PVC can have very distinct applications.[491] The most important application areas of PVC were reported to be pipes and fittings (42 %), profiles and hose/tubing (18 %), rigid film and sheet (17 %) as well as cable (8 %).[490] Although there are myraid applications of PVC in these areas, its usage in daily life is still debated due to the carcinogenic nature of VC monomer. Strict limits have been set for the quantity of residual VC in PVC polymers, which are produced for food contact applications.[491] Hence, replacement of PVC with its alternative and more reliable polymers is more desirable.

Another commercial derivative of EDC is ethylenediamine (EDA). This diamine is mainly produced by reacting EDC with aqueous or liquid $NH_3$ at about 100°C (Figure 4c).[154, 492] Major application sectors of EDA include detergents, resins, crop protection agents, paper chemicals, lubricants, and pharmaceuticals. Having diamine functionality makes EDA a useful monomer in different polycondensation reactions (Figure 37). Such as, EDA-polyester condensates, which are hydroxymethylated with formaldehyde, can be used as plasticizers. As nitrogen containing polyol components, condensates of EDA, epoxides, and urea were proposed for the production of polyurethane foams. Also, the incorporation of EDA into diisocyanate–polyester prepolymers forms useful polymers for the manufacture of elastic polyurethane fibers.[492] In the recent literature, EDA cored poly(amidoamine) (PAMAM) has emerged as an alternative monomer in synthesis of polyamide thin film composite (TFC) membranes owing to its dendritic and hydrophilic nature.[493] Furthermore, research studies were also reported on plasma polymerized EDA (PPEDA) films for various applications.[494]

The oligomerization of ethylene is a common method for synthesis of alpha-olefins, such as 1-butene, 1-hexene, and other higher alpha-olefins (Figure 4c). A wide variety of catalysts based on nickel, chromium, iron, cobalt, and aluminum can be employed in that respect. As previously mentioned in this section, α-olefins are predominantly copolymerized with ethylene for the commercial production of LLDPE (Figure 37).[472] Vinyl acetate, methyl methacrylate (MMA), and styrene are the other ethylene derivable giant monomers. Amongst these monomers, vinyl acetate is obtained *via* the reaction of ethylene with acetic acid, and it is mainly consumed in the production of PVAc and PVA (Figure 36). MMA is currently commercially produced by condensation of acetone and hydrogen cyanide (ACH process). However, hydrogen cyanide is highly toxic and the ACH process results in formation of large quantities of ammonium bisulphate as a corrosive byproduct. Thus, many alternative routes have been put into practice for replacing the problematic ACH process. As a promising method, BASF employs ethylene, carbon monoxide and formaldehyde as raw materials (Figure 37).[495] In this respect, a sustainable and renewable route for synthesis MMA could be realized once petroleum based ethylene is replaced with bio-ethanol derived ethylene. As the other giant monomer, styrene is produced in industrial scale *via* the dehydrogenetaion of ethyl benzene, which in turn prepared by alkylation of benzene with ethylene. Hence, the combination of lignin derived benzene (Section 5.19) with bio-ethanol derived ethylene could potentially provide a sustainable pathway for mass production of styrene.[472] As shown in Figure 37, the development of bio-based routes for production of MMA and styrene are quite important because these monomers are particularly converted into their homopolymers, which have myriad of applications.[270]

After ethylene, propylene is the second most demanded product in the petrochemical industry.[261] Mathers pointed out that the high demand for propylene production would be alleviated from its petroleum-based indutrial route to biomass resources.[472] In that respect, bio-based alcohols have a flaming potential as key intermediates in the synthesis of bio-propylene. For instance, Braskem has been planning to produce propylene from bio-ethanol. The process involves bio-ethanol conversion to ethylene, and then to propylene through dehydration, dimerization, isomerization, and metathesis processes (Figure 37). Other strategies might include fermentation of biomass to yield 2-propanol, butanol and methanol, and then conversion of these alcohols to bio-propylene (Figure 4c). Furthermore, fermentation and direct cracking of cellulose, lignin, and sugar for propylene production have been actively pursued in industry.[146, 261, 472] About two thirds of the globally produced propylene is consumed in the manufacture of polypropylene (PP) (Figure 37).[472] According to the new market study of Ceresana, the worldwide PP production capacity is about 62 million tonnes and this is expected to increase more than 23.5 million tonnes by 2019. Commercial PP is primarily consumed in flexible and rigid packaging, which accounts for more than 50% of the demand. This is followed by fiber and consumer goods (12% each), electrics and electronics as well as the transportation and construction (6% each) sectors.[496] Several kinds of PPs, such as atactic PP, elastomeric PP, isotactic PP and syndiotactic PP, are avaliable in the market under different trend names and each of them have their unique applications.[270]

### 5.15. Xylose/Furfural/Arabinitol Platform Based Polymers.

Xylose and arabinose, as being $C_5$ sugars of lignocellulosic biomass, lead to the production of many value added chemicals as shown in Figure 4d. Ring-opened derivatives of these sugars, such as xylitol, arabinitol, and xylaric acid,[170, 172, 173] are useful intermediates for synthesis of various kinds of linear polycondensates (Section 5.1). For instance, Zamora *et al.* synthesized and characterized new polyesters based on L-arabinitol and xylitol as PET and poly(butylene terephthalate) (PBT) analogs.[497] In another study, Garcia-Martin *et al.* reported a new series of AABB-type polyamides derived from L-arabinitol and xylitol.[498] Very recently, Pohjanlehto *et al.* employed xylaric acid in the preparation of a series of lignin-xylaric acid polyurethanes. The polyurethanes were prepared through the network-forming reaction between lignin and esterified xylaric acid. In the reaction, MDI, PEG, and di-n-butyltin dilaurate were used as a coupling agent, a compatibilizer and a catalyst, respectively. The content of bio-based starting materials in the polyurethane products was reported to be as high as 35%.[499]

Industrially, most of the xylose is consumed in the production of furfural through its dehydration using mineral acids as homogeneous catalysts. Besides, many studies have been reported to search new sustainable ways for furfural production by employing heterogeneous acid catalysts and different extracting techniques as well as by tuning the temperature, pressure and solvent.[171, 177] This conversion holds important industrial oppurtunities because xylose derived furfural can be further transformed to furan and many other furan derivatives such as furfuryl alcohol, hydroxy furans, furoic acid, 2(5H)-furanone, furfuryl amine, difurfuryl diamines, furanacrylic acid, furylidene ketones, methyl furan, 2-hydroxymethyl-5-vinyl furan and 2,5-bis(hydroxymethyl) furan (Figure 4d).[174, 175, 177, 179-181, 183-186] Moreover, the corresponding reductions of the parent furan ring of these products result in formation of THF and THF derivatives including 2-methyltetrahydrofuran, tetrahydrofurfuryl alcohol and 2,5-bis(hydroxymethyl) tetrahydrofuran (Figure 4d).[176, 178, 179, 187] Common polymers of these furfural derived compounds were discussed in Section 5.3.

In addition to furans and THFs, maleic anhydride and maleic acid are the other two furfural derivable industrially important compounds (Figure 4d).[182, 188] MA is mostly converted to fumaric acid, which is often preferred instead of MA to produce polyesters and copolymers. Nonetheless, small amounts of MA is utilized for maleinate resins and for copolymers.[500] Unlike MA, MAnh is a major monomer of the current polymer industry. Particularly, it is employed in the manufacture of glass-reinforced or unreinforced unsaturated polyester resins (UPRs). Moreover, MAnh is used in a myraid of applications through its copolymerization with other molecules having vinyl functionality. Typical copolymers of MAnh and their end uses are summarized in Table 8.[500-502]

### 5.16. Polyhydroxyalkanoates.

Polyhydroxyalkanoates (PHAs) are a family of naturally-occurring diverse biopolyesters produced by various

**Table 8. Major polymers of MAnh and their corresponding applications**

| MAnh-Based Polymers | Major applications |
|---|---|
| UPRs and alkyd resins | Myraid of applications in marine, construction, corrosion, transportation and electrical industries |
| MAnh – styrene copolymers | Engineering plastics, paper treatment, floor polishes, emulsifiers, protective colloids, antisoil agents, dispersants, stabilizing agent, adhesives, detergents, cosmetics, polymer-protein conjugates |
| MAnh – acrylic acid copolymers | Detergent industry |
| MAnh – diisobutylene copolymers | Dispersing agent |
| MAnh – butadiene copolymers | Sizing agent |
| MAnh – $C_{18}$ □-olefin copolymers | Emulsification agent and paper coating |
| Polyaspartic acid | Section 5.5 |

microorganisms.[431, 503, 504] As shown in Figure 38, each monomer unit of PHAs harbors a side chain R group which is generally a saturated alkyl group. Although it is less common, the R group can also take the form of unsaturated alkyl groups, branched alkyl groups, and substituted alkyl groups. PHAs be classified as either short- (3-5 carbon atoms), medium- (6-14 carbon atoms), or long-chain lenght depending on the total number of carbon atoms within a PHA monomer.[504, 505]

PHAs are produced by large scale microbial production. Chen and Patel pointed out that over 30% of soil inhabiting bacteria can synthesize PHA. Moreover, many bacteria species having habitats in activated sludge, in high sea, and in extreme environments are also capable of producing PHA in their cell environment.[506] In a nutshell, the microbial production process mainly involves strain development, shake flask optimization, lab and pilot fermentor studies, and finally industrial scale up. Various factors, such as growth rate of the bacteria, time duration for reaching high final cell density, the final cell density, PHA percentage in cell dry weight, substrate to product transformation efficiency, price of substrates, and robust extraction and purification of PHA, are critical for efficient microbial production of PHAs.[503] Depending on the specific PHA required, the PHA feedstocks can include cellulosics, vegetable oils, organic waste, municipal solid waste, and fatty acids.[431] Amongst these feedstocks, the main contribution of lignocellulosic biomass would be to supply glucose as a carbon source for PHA fermentation. The fermentation of glucose has been well studied and thus, it is commonly employed both in laboratory and industry for PHA production.[507] In addition to glucose, other lignocellulosic monomer sugars, such as xylose, arabinose, mannose, galactose and rhamnose, and some lignocellulose derivable compounds, such as glycerol, lactic acid and levulinic acid, can be employed as PHA feedstocks.[507-509] Furthermore, it is possible to produce PHAs directly from cellulose.[510] Detailed scrutiny regarding recent trends and future perspectives of microbial PHA production from different feedstocks, including lignocellulosic biomass, are provided in the comprehensive books and articles.[503-505, 507-509, 511, 512]

PHAs are completely biodegradable, biocompetible and thermoplastic polymers. They are enantiomerically pure, non-toxic, water-insoluble, inert and stable in air. Moreover, they exhibit piezoelectricity, and they have good processibility as well as high structural diversity. Their thermal and mechanical properties depend on their composition so that the Tg of the polymers varies from −40°C to 5°C, and the melting temperatures range from 50°C to 180°C.[431, 505] They can be blended with synthetic or natural polymers for tuning their physical properties and biodegradability. PHAs having longer side chains are elastomeric, on the other hand, short side chain PHAs show similar characteristics of polypropylene.[285] Owing to these characteristics, many companies (Table 9) have engaged in commercial manufacture PHA polymers[431, 503] with a specific motivation towards producing low cost and 100% bio-polymers which can replace conventional polymers like polyethylene and polypropylene in certain applications.

Depending on the bacterial species and growth conditions, it is possible to produce homopolymers, random copolymers, and block copolymers of PHA.[506] Polyhydroxybutyrate (PHB), poly(3-hydroxybutyrate-co-3-hydroxyvalerate (PHBV), poly(3-hydroxybutyrate-co-4-hydroxybutyrate) (P3HB4HB), poly(3-hydroxybutyrate-co-3-hydroxyhexanoate) (PHBHHx) and medium-chain length (mcl) PHAs are successfully commercialized so far.[503] Some specific examples of PHB include poly-3-hydroxybutyrate (P3HB), poly-4-hydroxybutyrate (P4HB), polyhydroxyvalerate (PHV), polyhydroxyhexanoate (PHH) and polyhydroxyoctanoate (PHO) (Figure 38, Table 9).

PHAs and its related technologies have an industrial importance ranging from fermentation, materials, energy to medical fields. As shown in Table 9, recent commercial applications of PHAs mainly include packaging, fiber and nonwoven production, blending with other conventional polymers, biomedical products and drug delivery systems. In packaging industry, PHA based materials are suitable for short term usage due to their biodegradability. P&G, Biomers and Metabolix have developed PHA based packaging films generally for shopping bags, containers and paper coatings, disposable items, cosmetic containers and cups, and compostable bags and lids.[503] PHA can also be processed into fibers like conventional PAs.[503] Thermoplastic processing of PHB was already optimized for production of fibers which have physical properties suitable for the production of scaffolds or for other medical applications.[513] It is convenient to blend PHA with other polymers. As such, BASF has been blending PHAs with its commercially avaliable product Ecoflex[503] to further improve use of the product. In the literature, blending different kinds of PHAs among themselves or blending PHAs with other low cost materials, such as cellulose and starch, have also been reported to tune their physical properties and biodegradability.[503, 514, 515] PHA and its copolymers are widely preferred as biomedical materials. Biomedical applications of PHAs include but not limited to sutures, suture fasteners, meniscus repair devices, rivets, bone plates, surgical mesh, repair patches, cardiovascular patches, tissue repair patches, and stem cell growth.[431] Tepha Inc. specializes in manucaturing PHA based biomedical materials. It markets P4HB for medical application under the name of PHA4400 (Section 5.2).[503] Furthermore, PHAs have become candidates for use as drug carriers owing to their biocompatibility and controlled degradability.[431] So far, only PHB and PHBV have been studied for controlled drug release. Hence, this application area still remains to be exploited.[503]

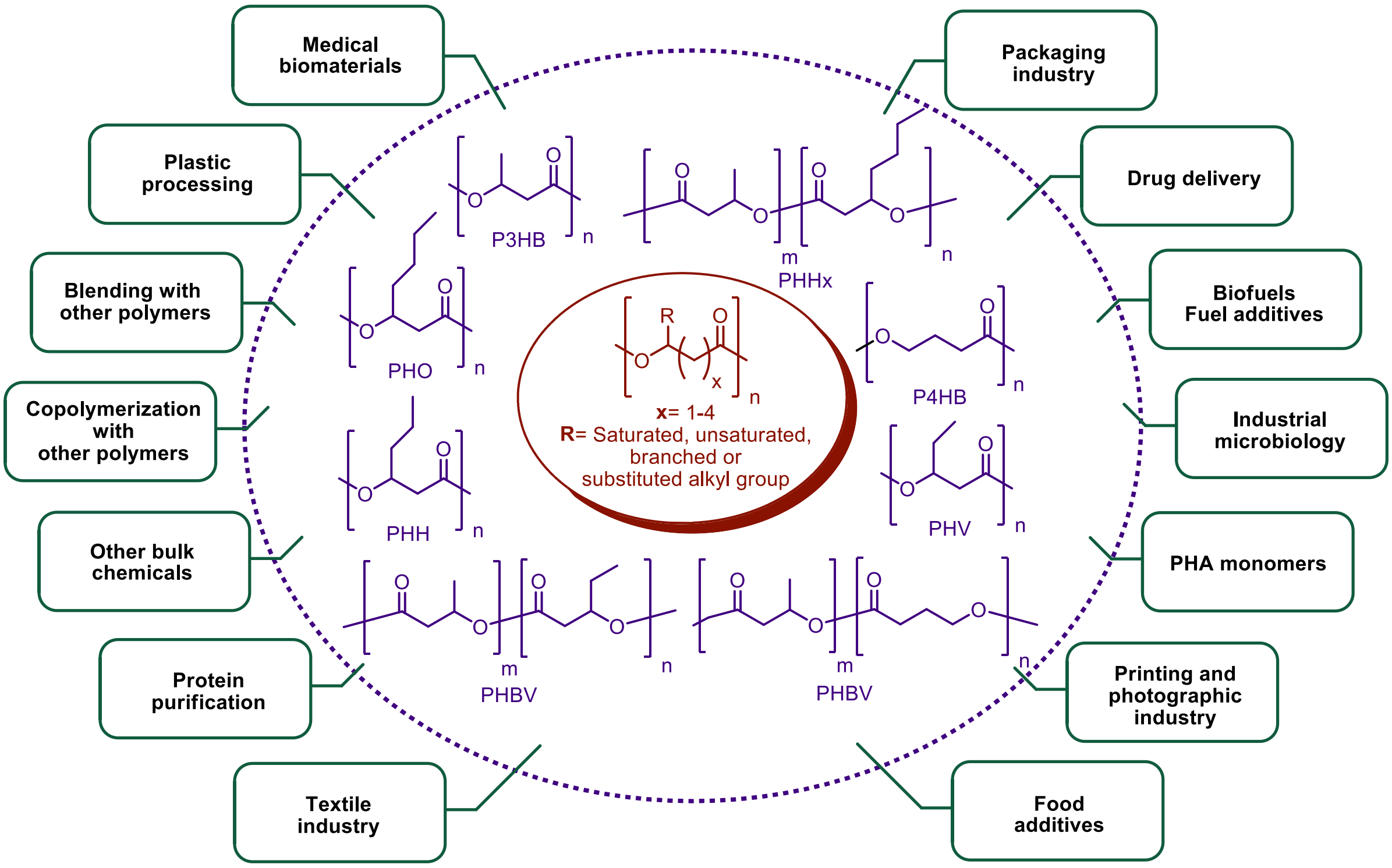

Figure 38. General structure, specific examples and common applications of PHAs.

Apart from the above-mentioned commercial uses, applications of PHAs have been rapidly expanding (Figure 38). Advancements in PHA research greatly contribute to progress of industrial microbiology. In this context, the performances of industrial microbial strains can be improved by the PHA synthesis operon which is utilized as metabolic regulator or resistance enhancer.[503, 516, 517] As another important point, PHAs are degraded into its constituent monomers. These monomers can be produced *via* various routes including chemical synthesis, acidic hydrolysis of PHA, and in vitro and in vivo enzymatic depolymerization of PHA.[506] So far, more than 150 PHA monomers have been identified.[431] PHA derived monomers can be utilized as chiral building blocks for synthesis of novel polymers, particularly for chiral polyesters.[506] In addition, chemical modifications, which can not be easily achieved by bioconversion processes, can add valuable attributes to PHAs.[518] For example, through transformation of PHAs into PHA diols, block copolymerization of PHAs with other polymers becomes feasible.[503] PHAs are also suitable candidates for production of other bulk chemicals, such as heat sensitive adhesives, latex, smart gels,[503] biofuels or fuel additives,[503, 519] and healthy food additives.[503] Moreover, PHA granule binding proteins are used to purify recombinant proteins and stained PHAs can find applications in printing and photographic industry.[503, 520, 521]

### 5.17. Rubber Polymers.

The huge market volume of rubber has been substantially expanding particularly due to the growth of the automotive industry and the growing demand for tires.[522] Hence, the market has been experiencing a tightening supply of both natural and synthetic rubber. In this context, the production of isoprene, isobutene, and butadiene; the three most important monomers for manufacture of renewable rubber, would create a very positive impact in the field.[261] In fact, the three largest tyre manufacturers; Goodyear, Michelin, and Bridgestone, have already been engaged in the production of renewable rubber from biomass resources. Recently, the collaboration of Genencor and Goodyear has resulted in development of a microbial process for bio-based isoprene production. By adopting genes from *Enterococcus faecalis*, *Methanosarcina mazei*, *S. cerevisiae*, and *Populus alba*, the companies synthetically constructed a pathway towards isoprene production in *E. coli*. The successful metabolic engineering of the isoprene pathway has resulted in the production of 60 g of isoprene per liter of sugar.[243, 289] Moreover, Michelin teamed up with Amyris, and Bridgestone announced that they are working with Ajinomoto for developing isoprene-based tyres.[523]

**Table 9. Major suppliers of various types of PHAs[431, 503]**

| Company | Brand Name | Country | PHA Type | Applications |
|---|---|---|---|---|
| Metabolix | Mirel | USA | Several PHA | Packaging |
| TEPHA | TephaFLEX, TephaELAST | USA | Several PHA | Medical bio-implants |
| PHB Industrial S/A | Biocycle | Brazil | PHB | Raw materials |
| Mitsubishi Gas Chemicals | Biogreen | Japan | PHB | Packaging |
| Tianjin Green Bio-Sciences | GreenBio | China | P3HB4HB | Raw materials and packaging |
| Tianan Biological Materials | Enmat | China | PHBV, PHB | Thermoplastics, fiber, nonwovens |
| Bio-On | Minerv | Italy | PHA (unclear) | Raw materials |
| P&G | Nodax | USA | PHBH | Packaging |
| ICI (1980s to 1990s) | Biopol | UK | PHBV | Packaging |
| Biomer | Biomer | Germany | PHB | Packaging and drug delivery |
| BASF | | Germany | PHB, PHBV | Blending with Ecoflex |
| ADM | | USA | Several PHA | Raw materials |
| Meridian | | USA | Several PHA | Raw materials |
| Kaneka (with P&G) | Aonilex | Japan | Several PHA | Packaging |
| Biomatera | Biomatera | Canada | Several PHA | Packaging and biomedical products |
| PolyFerm | VersaMer | Canada | mcl PHAs | Bioplastic materials |
| Zhejiang Tian An | | China | PHBV | Raw materials |
| Jiangmen Biotech Ctr | | China | PHBHHx | Raw materials |

Lignocellulosic biomass can also play a very active role in the development of bio-based rubber industry (Figure 39). Recently, Zhang *et al.* patented methods for production of bio-isoprene from lignocellulosic biomass by using genetically engineered strain of saprophytic bacterium.[200] Another interesting possibility concerns the conversion of lignocellulosic bio-ethanol (Section 5.14) into butadiene.[148, 325] In this consideration, two possible routes have been reported. The first route is to transform ethanol to butadiene at 400–450°C in the presence of metal oxide catalysts. Alternatively, ethanol and acetaldehyde can be converted at 325–350°C. These processes are even used on industrial scale in Eastern Europe, China, and India.[325] Furthermore, cellulose derived sugars can be fermented to give isobutanol and then, the dehydration of isobutanol over alumina catalysts at 250–400°C results in the formation of isobutene.[524, 525] Gevo has been producing bio-based isobutanol from sugars by fermentation through its Integrated Fermentation Technology (GIFT), which combines genetically engineered yeast with a continuous separation process.[326] As an alternative method, lignocellulosic biomass derived bio-butanol can be selectively dehydrated into isobutene in one-step over zeolite catalysts.[473]

About two thirds of the total rubber consumed worldwide is synthetic and one third is natural. Primarily owing to its heat and mineral oil resistance, synthetic rubber (SR) is superior to natural rubber (NR).[526] Hence, many different synthetic rubbers having various monomer compositions have been developed (Figure 39). In 2002, worldwide SR capacity was $11.2\times10^6$ t. 43% of this production was belonging to styrene-butadiene rubber (SBR). Other important butadiene based rubber polymers such as polybutadiene, or butadiene rubber (BR), and acrylonitrile-butadiene rubber (NBR) occupied 26% and 4% of the worldwide production capacity, respectively. There are different types of polybutadiene, which include 1,2-polybutadiene, cis-1,4-polybutadiene, and trans-1,4-polybutadiene. These structurally different polybutadienes are prepared by using different catalysts.[270, 526] Homopolymer of isoprene is another important rubber polymer. It has mainly been extracted from rubber tree (NR) or it has also been manufactured synthetically (IR).[307, 527] Higher 1,4-cis configuration of polyisoprene most closely exhibits the properties of natural rubber. Hence, a nearly pure cis-1,4 structure allows the production of synthetic natural rubber (IR) (Figure 39).[528] This is achieved by coordination, anionic, free- radical, or cationic polymerization of isoprene through the use of various catalysts.[270] Another widely used rubber polymer is butyl rubber (IIR); a copolymer of isobutylene and a small amount of isoprene.[529] Halley and Dorgan stated that butyl rubber will soon be available from renewable resources.[524] The usefulness of butyl rubber is greatly extended by halogenation, which provides higher vulcanization rates and improves the compatibility with highly unsaturated elastomers. In this consideration, butyl rubber can be either chlorinated or brominated (Figure 39).[529]

### 5.18. Other Lignocellulosic Biomass Derivable Polymers.

More classes of final chemicals can be produced from lignocellulosic feedstocks than petroleum resources owing to their compositional variety.[25] Although industrially important chemicals are presented under different platforms (Figures 2-5), it is beyond the scope of this review article to cover all of the lignocellulosic biomass derivable compounds. Nonetheless, some of the important ones are presented in Figure 4b.

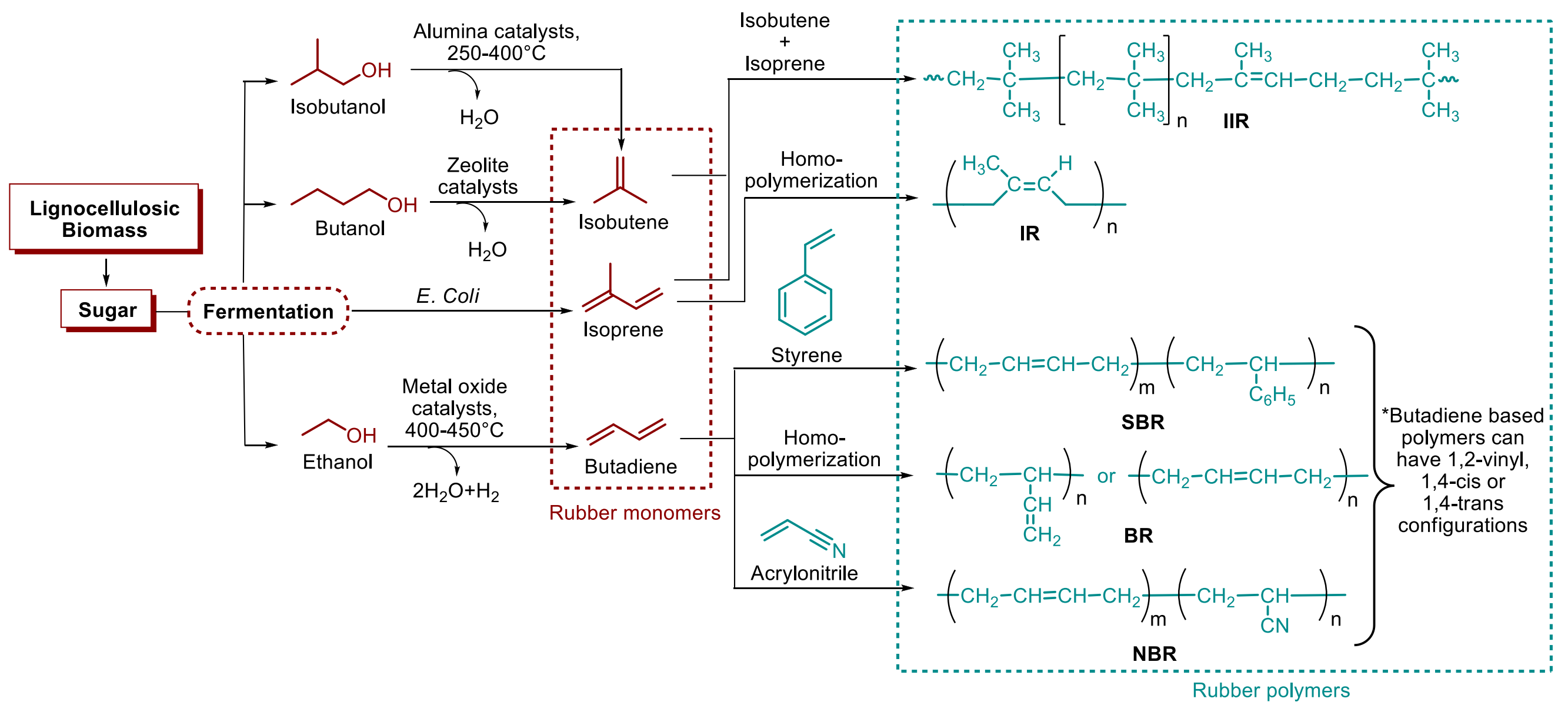

Figure 39. Lignocellulosic biomass derivable rubber monomers and polymers.

Citric acid can be obtained through aerobic fermentation of lignocellulosic glucose with *A. niger*. Its main applications include food, beverages, pharmaceuticals, detergents, buffering and chelating agents.[159] It is used as a multifunctional monomer to react with different aliphatic diols to give biodegradable polyester elastomers. Tran *et al.* reviewed various other citric acid based elastomers, including crosslinked urethane-doped polyesters (CUPEs), poly (alkylene maleate citrates) (PAMCs), poly (xylitol-cocitrate) (PXC) and poly(poly(ethylene glycol) maleate citrate) (PPEGMC).[530]

Another interesting monomer is limonene and it can be produced from citrus wastes. Particularly, the wastes from orange juice industry can be valorized in the production of this terpene derivative.[1] For removal of limonene from citrus wastes, Pourbafrani *et al.* reported a dilute acid hydrolysis process at 150°C for 6 min. In the process, the hydrolysis which is followed by explosive pressure reduction (flashing) resulted in drastic decrease of limonene in the hydrolyzates. 99% of the limonene content can be removed from citrus wastes by this method. The removed limonene can then be recovered through its condensation process.[164] Once removed from citrus wastes, limonene can be widely employed in the synthesis of polymers (Figure 40).[261] For instance, Singh and Kamal reported radical homopolymerization of limonene in the presence of benzoyl peroxide as an initiator at 85°C in xylene. The resultant polylimonene has a weight-average molecular of 43 000 g/mol and a Tg of 116 °C.[531] Copolymerization of limonene can also be carried out with different monomers including styrene[532] and N-phenylmaleimide.[533] Furthermore, limonene is oxidized to form mono- as well as di-functional epoxides.[54]

Byrne *et al.* performed catalytic copolymerization of limonene monoxide with carbon dioxide. The resulting thermoplastic polylimonene carbonates show polystyrene like properties.[1, 534] Instead of carbon dioxide, limonene monoxide can alternatively be copolymerized with dicarboxylic acid anhydrides, such as succinic anhydride, to give limonene-based polyesters.[1, 85] Limonene dioxide; the difunctional epoxide derivative of limonene, can be further converted into limonene dicarbonates. This process enables chemical fixation of 34 wt% carbon dioxide. Curing of this novel limonene-derived monomer with polyfunctional amines, such as citric aminoamides, yields a wide variety of crosslinked green polyurethanes without requiring the use of isocyanates (Figure 40).[1, 535]

Erythritol can be produced by several osmophilic yeasts such as *Aureobasidium, Candida, Moniliella, Pichia, Pseudozyma, Trigonopsis*, *Trichosporon*, *Tri-chosporonoides*, and *Yarrowia*.[169] Having –OH functionality allow erythritol to be used in polycondensation reactions. Barrett *et al.* produced various soft elastomers based on the polycondensation of erythritol with a dicarboxylic acid, including glutaric, adipic, pimelic, suberic, azelaic, sebacic, dodecanedioic, and tetradecanedioic acids. The elastomers exhibit a wide range of physical and mechanical properties. Young's modulus, ultimate tensile stress, and rupture strain values of the elastomers are in between 0.08–80.37 MPa, 0.14–16.65 MPa, and 22–466%, respectively.[536]

2,3-Butanediol (2,3-BDO) is produced as a natural fermentation product of many microorganisms, including many species of *Klebsiella*, *Bacillus*, and lactic acid bacteria.[161, 236] LanzaTech has been developing 2,3-BDO as a platform chemical for production of bio-based butadiene monomer.[236] Besides, several studies were reported on the copolymerization of 2,3-BDO with 2,5-FDCA and other FDCA derivatives in the recent literature. The common motivation of these studies is to produce fully bio-based polyesters.[258, 537, 538]

Another notable compound is L-lysine; an essential amino acid for humans. In 2005, about 1.5 million tons of L-lysine was produced by employing Gram-positive *Corynebacterium*

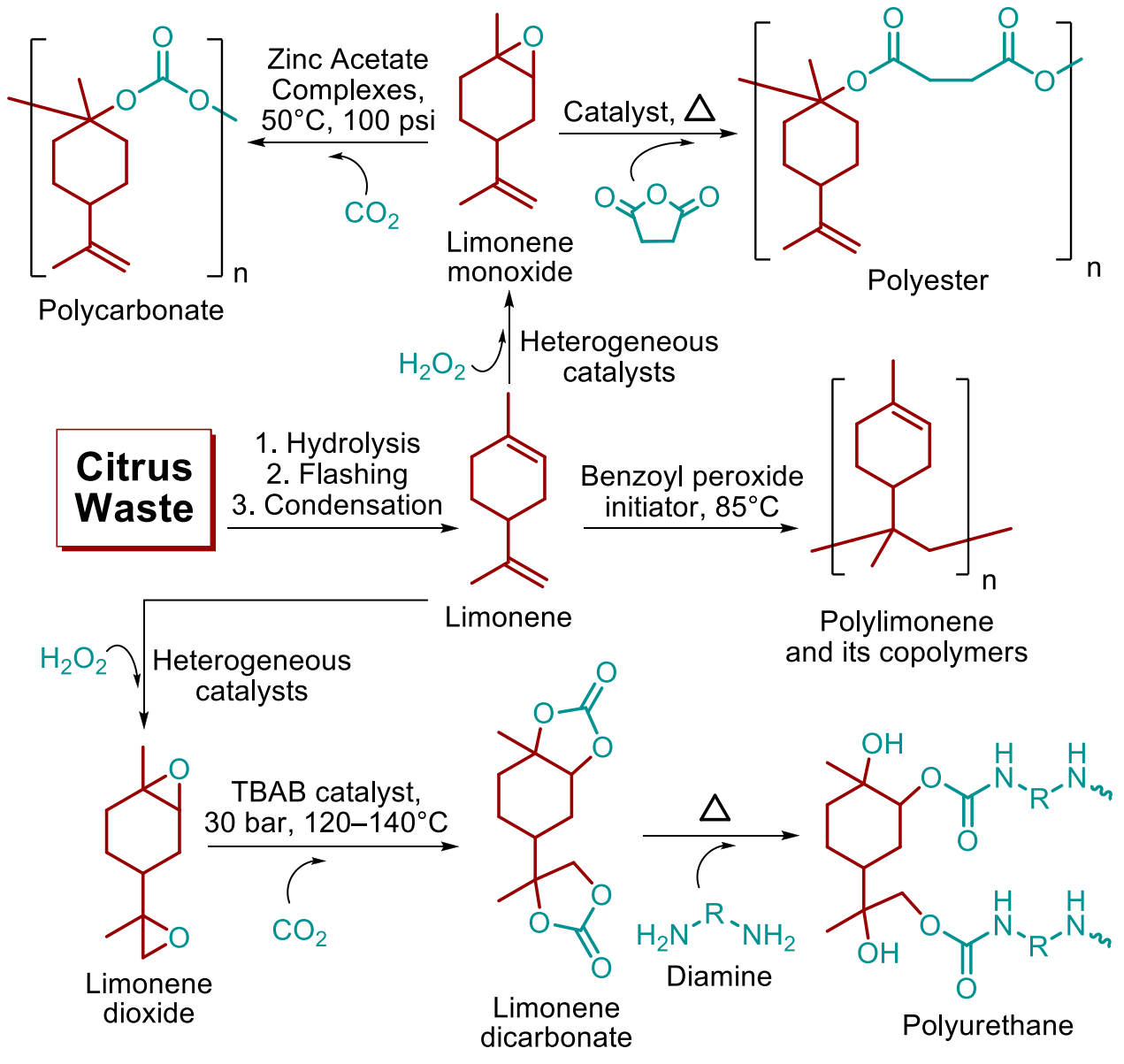

Figure 40. Limonene based polymers.

*glutamicum*.[167, 168] The most well known polymer of L-lysine is its homopolyamide; ε-polylysine (ε-PL). Naturally occurring ε-PL is nontoxic, water soluble, biodegradable and edible. Furthermore, it exhibits excellent antimicrobial activity and heat stability. Owing to these characteristics, ε-PL has been attracting interest especially in food, medicine and electronics industries. Its derivatives also offer a wide range of applications including emulsifying agent, dietary agent, biodegradable fibers, highly water absorbable hydrogels, drug carriers, anticancer agent enhancer, and biochip coatings. Production and medical applications of ε-PL are elegantly reviwed by Shukla *et al.*[539] Additionally, the design of novel functional monomers based on amino acids and their corresponding polymerization methods are described in Section 5.5.

Lastly, levoglucosan is another important lignocellulosic biomass derivable compound. It is obtained through fast pyrolysis of lignocellulosic biomasss.[166] Ring-opening multibranching polymerization of levoglucosan, and other anhydro sugars, results in formation of hyperbranched carbohydrate polymers. Satoh and Kakuchi reported levoglucosan incorporating water-soluble hyperbranched carbohydrate polymers with controlled molecular weights and narrow polydispersities. Such polymer architectures are expected to have useful applications in medical and medicinal fields.[540]

### 5.19. Lignin Derivable Polymers.

Lignin, which can be found in all vascular plant, is the second most abundant natural polymer coming after cellulose. Even so, it has received little attention regarding its valorization. For instance, approximately 50 million tons of lignin is avaliable worldwide from pulping processes but only about 2% of this lignin is used commericially. The remaining 98% is not isolated but burned on site. Borregard Lignotech, Westvaco and Tembec are currently the major producers of isolated lignin. These companies have been producing around 1.1 million tonnes of isolated lignin per year either through the acid sulphite or the kraft pulping processes.[45, 506, 541]

Lignin is considered as the major aromatic resource of the bio-based economy. A wide variety of bulk and fine chemicals, particularly aromatic compounds, as well as fuels can be obtained from lignin. Lignin derivable compounds are reviewed in detail by Zakzeski *et al.*,[45] Huber *et al.*,[9, 542] Bozell *et al.*[543] and Stöcker.[24] Representatives of lignin derivable aromatic compounds are also provided in Figure 5. In terms of polymer industry, the enormous quantity of lignin renders excellent oppurtunities for production of aromatic monomers and polymers. However, the full potential of lignin for commodity polymers is recently underutilized mainly due to the difficulty in obtaining aromatic chemicals from lignin.[472] Hence, new technological developments are required to breakdown the chemical structure of lignin into commodity chemicals such as benzene, toluene, xylene, phenols, hydroxybenzoic acids as well as coniferyl, sinapyl, and p-coumaryl compounds (Figure 41). According to Mathers, this difficulty could be overcome by employing either heterogeneous of homogeneous catalysts, which can break the complicated network of C-O and C-C bonds in lignin without leaving substantial amounts of tar or gasification.[472]

Once lignin is isolated and degraded into aromatic compounds, the subsequent processes do not require too much improvement. There is already a mature technology for conversion of these compounds into commodity monomers and polymers. As shown in Figure 41, the combination of the new and current technologies would provide novel oppurtunities for conversion of lignin to commodity polymers like polyethylene terephthallate (PET), polystyrene, Kevlar, unsaturated polyesters, polyaniline and many more aromatic polymers. Since common applications and properties of these polymers are well documented in the literature,[270] they are not reviewed here in detail. Nonethless, it should be noted that alternatives to petroleum-based aromatic polymers could be fully realized by valorization of lignin. For instance, two key components PET; ethylene glycol and p-terephthalic acid, must be produced from renewable biomass for production of fully bio-based PET. The Coca Cola Company has been using bioethanol-derived bio-ethylene to generate a ca. 30% plant-based PET. It is also necessary to use bio-based p-xylene as the raw material for the p-terephthalic acid to produce 100% plant-based PET. Pepsi Cola has been producing 100% plant-based green PET bottles by employing $C_5$ and/or $C_6$ sugars derived bio-terephthallic acid.[544] Whereas, the processes for production of bio-terephthallic acid from $C_5$/$C_6$ sugars require intermadiate steps and compounds. Hence, direct conversion of lignin to terephthallic acid may become more feasible depending on the development of new lignin valorization technologies.

## 6. Conclusion.

Industrial production of a wide range of chemicals and synthetic polymers heavily relies on fossil resources despite of their dwindling resources and frightening environmental effects

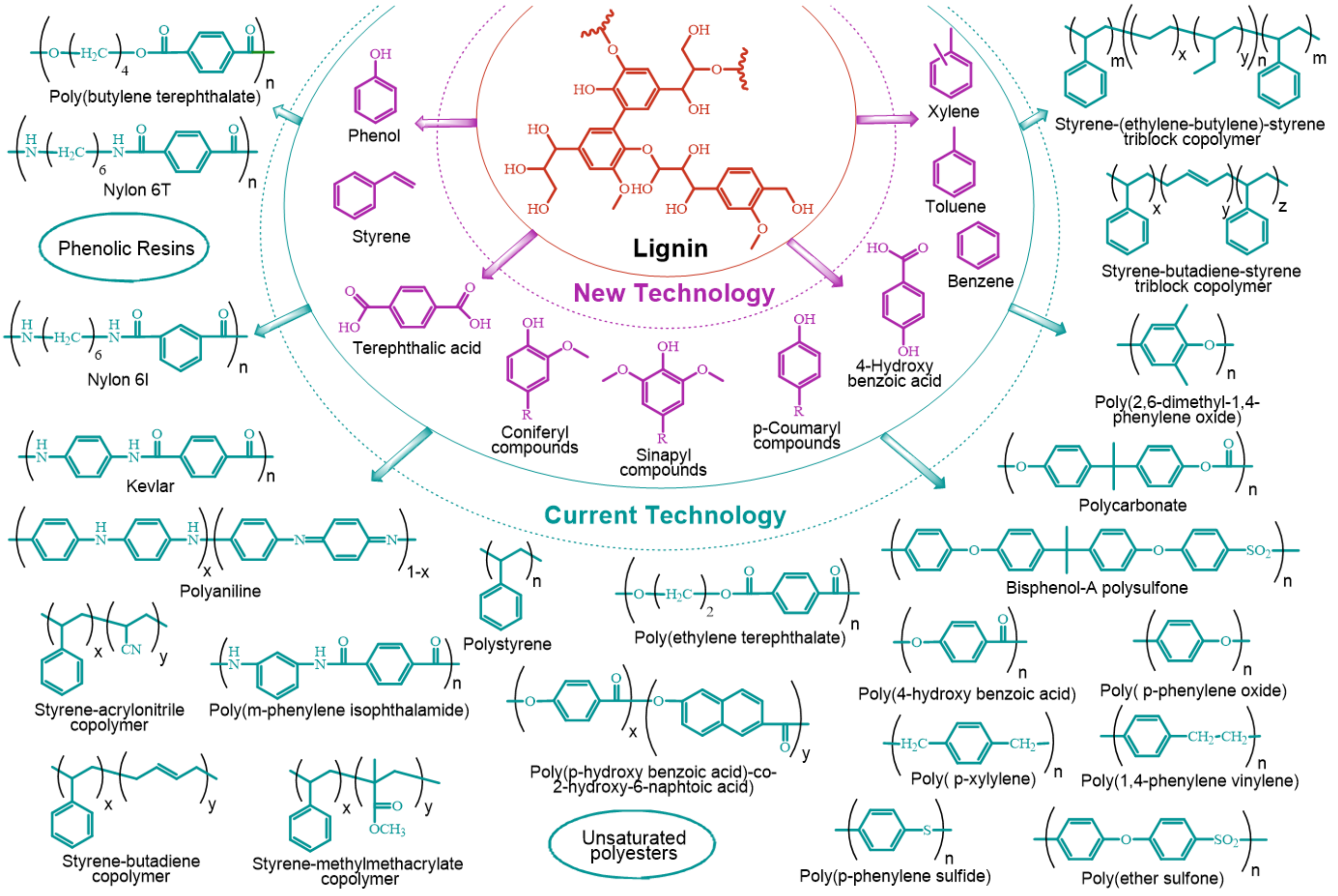

Figure 41. Valuable aromatic monomers and polymers obtainable from lignin *via* development and integration of new and current technologies.

Hence, alternative sources are sought to produce valuable chemicals and polymers from renewable natural resources for decreasing the current dependence on fossil resources as well as alleviating their corresponding environmental threats. In this context, lignocellulosic biomass, as being the most abundant bio-renewable biomass on earth, has a critical importance mainly owing to its worldwide avaliability and non-competitiveness with food supplies. Besides, it is significantly cheaper than crude oil and it can be produced quickly and at lower cost than other agriculturally important feedstocks. Thus, lignocellulosic biomass is considered as an ideal renewable feedstock to produce biofuels, commodity chemicals and polymers by exhibiting great economic significance and environmental friendliness. Owing to these characteristics, the ongoing research and industrial activities in the field of lignocellulosic biomass for the production of commodity chemicals and polymers were reviewed. Initially, the structure and sources of lignocellulosic biomass were described and subsequently, different pre- and post-teratment methods for the degradation of lignocellulosic biomass into its components were summarized. More than 200 lignocellulosic biomass derivable value-added compounds were presented with their references. Finally, the scope of the polymers that can be produced mainly in the frame of these compounds was depicted to reveal the potential of lignocellulosic biomass as being an alternative platform to petroleum resources. As shown throughout this article, the development in the valorization of lignocellulosic biomass still remains a big challenge together with many oppurtunities. Thus, extensive research is currently being undertaken all over the world to convert lignocellulosic biomass to value-added chemicals and polymers at high selectivities, and yields at economical costs. One of the most important goals is to fractionate lignocellulose into its three major components; cellulose, hemicelluloses and lignin. Various pretreatment approaches have been developed to increase accessibility and biodegradability of these components for enzymatic or chemical action. Once these components are isolated, target compounds can be obtained through either chemocatalytic or microbial production processes. That is why, future developments in the valorization of lignocellulosic biomass are directly correlated to improvements in the fields of chemical and microbial synthesis. Owing to the recent advancements in these fields, the number and diversity of lignocellulosic biomass based commodity and specialty chemicals have been rapidly increasing. Furthermore, biorefinery and biofuel technologies have been developed to refine lignocellulosic biomass in analogy to petrochemistry for producing green fuels, chemicals and polymers. The number of biorefinery-related pilot and demonstration plants has been increasing. As such, Lignol, Verenium and Mascoma are promising companies which aim to undertake the development of biorefining technologies for the production of advanced biofuels, biochemicals and biomaterials from non-food cellulosic biomass feedstocks. The world's largest chemical firms including DuPont, BASF, SABIC, Dow Chemical, LyondellBasell, and Mitsubishi Chemical have also been actively engaged in valorization of lignocellulosic biomass. Bio-production of various platform chemicals such as ethanol, butanol, lactic acid, levulinic acid, sorbitol, glycerol, 1,3-propanediol, itaconic acid, succinic acid, and 2,5-FDCA has already been achieved. Through research and development, many more lignocellulosic biomass derivable chemicals still await the realization of their commercial production. This, of course, includes a number of monomer building blocks, which are utilized to produce many conventional and novel polymers. So, current commodity polymers are expected to be replaced by their bio-derived counterparts in the near future. For instance, the applications of petroleum derived polyethylene can be easiyl reproduced by bio-polyethylene. It was shown that bio-polyethylene has exactly the same chemical, physical, and mechanical properties as petrochemical polyethylene. As a different future prospect, existing conventional polymers may also be replaced by their new bio-based alternatives. As such, Avantium's 100% bio-based polyethylene-furanoate (PEF) can potentially replace PET in certain applications. However, the ultimate commercial success of bioplastics will depend on three factors; economics, performance, and environmental factors. The first factor seems to be more significant since bio-polymers have been proven to exhibit similar performance and more environmental friendliness with comporison to their petroleum-based counterparts. Economical considerations can be improved *via* continued research and development as well as government and private sector investment. Fortunately, the recent trends as mentioned throughout this article suggest that we are on the path of establishing a worldwide bio-based economy and lignocellulosic biomass may have a great contribution in this context.

## Notes and references

[a] *Department of Chemistry, Boğaziçi University, Bebek, 34342 İstanbul, Turkey*

[b] *School of Engineering and Materials Science, Queen Mary University of London, Mile End Road, E1 4NS London, United Kingdom*
*Tel: 020 78826534; E-mail: r.becer@qmul.ac.uk*